\documentclass[a4]{article}

\usepackage{tikz}
\usepackage{pgfplots}
\pgfplotsset{compat=1.15}
\usepackage{mathrsfs}
\usetikzlibrary{arrows}
\usepackage{bm}
\usepackage[all]{xy}
\usepackage{graphicx}
\usepackage{verbatim}
\usepackage{wrapfig}
\usepackage{ascmac}
\usepackage{makeidx}
\usepackage{amscd}
\usepackage{url}
\usepackage{comment}
\usepackage{enumerate}
\usepackage{here}
\usepackage{latexsym}
\usepackage{array}
\usepackage{booktabs}
\usepackage{multirow}
\usepackage{mathrsfs}
\usepackage{geometry}
\usepackage{fancyhdr}
\renewcommand{\title}[1]{

\begin{center} \Large \bf #1 \end{center}
}

\renewcommand{\author}[2]{
 \begin{center} #1  \vspace{3mm} \\
  #2 \\
 \end{center}
\addvspace{\baselineskip}
}

\usepackage{amssymb}
\usepackage{amsmath}

\usepackage{amsthm}
\newtheorem{theorem}{Theorem}[section]
\newtheorem{proposition}[theorem]{Proposition}

\theoremstyle{definition}

\theoremstyle{remark}
\newtheorem*{rem}{Remark}

\makeatletter
\@addtoreset{equation}{section}

\makeatother

\def\propagator{\includegraphics[width=0.1\textwidth]{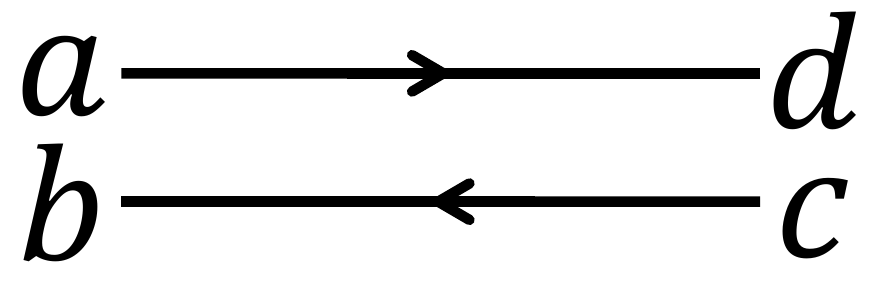}}

\def\vertex{\includegraphics[width=0.1\textwidth]{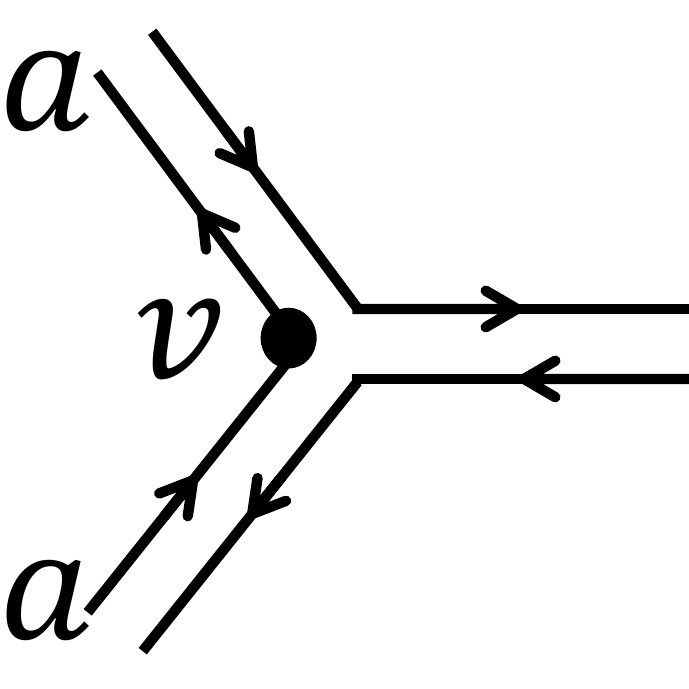}}

\def\vertexs{\includegraphics[width=0.1\textwidth]{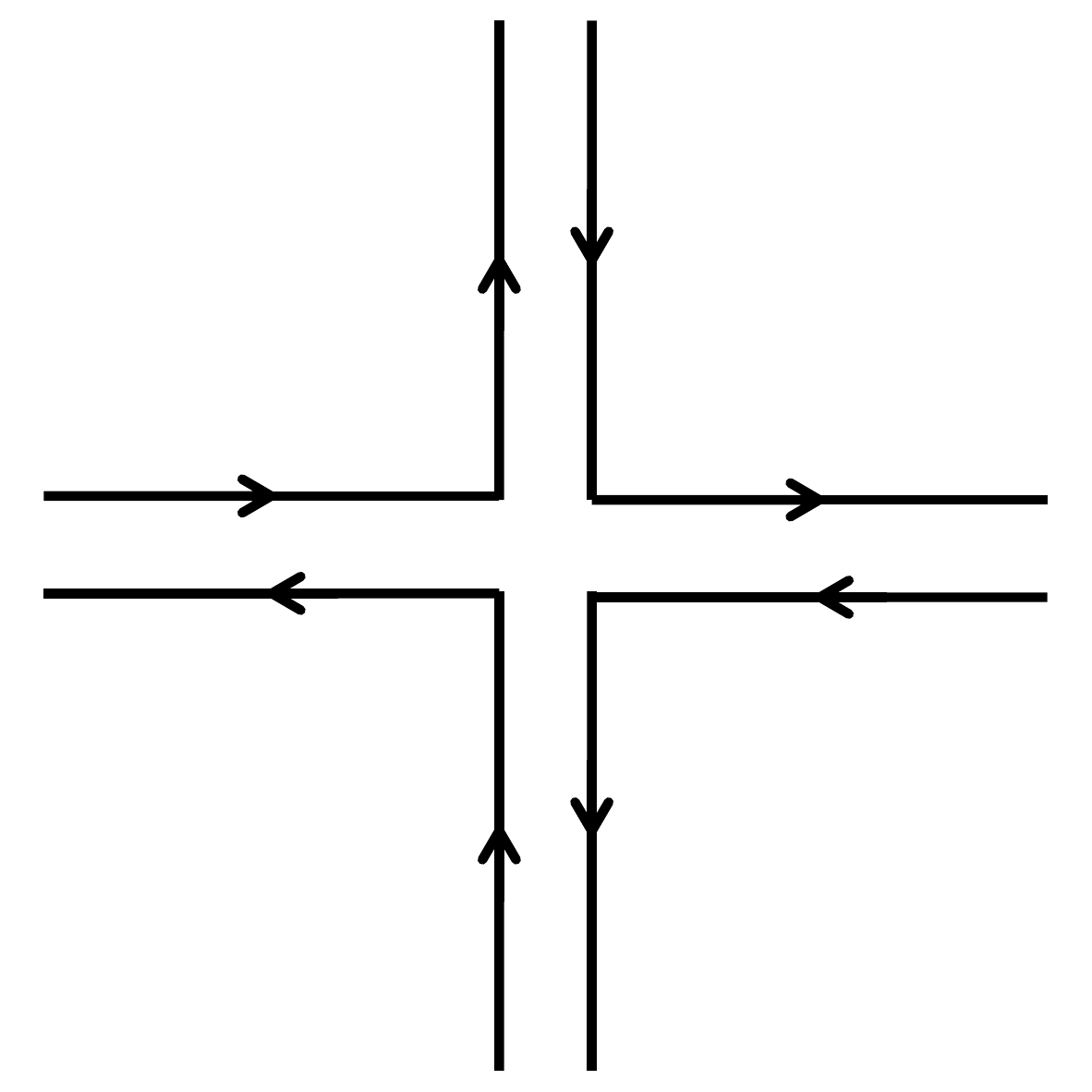}}

\def\one{\includegraphics[width=0.15\textwidth]{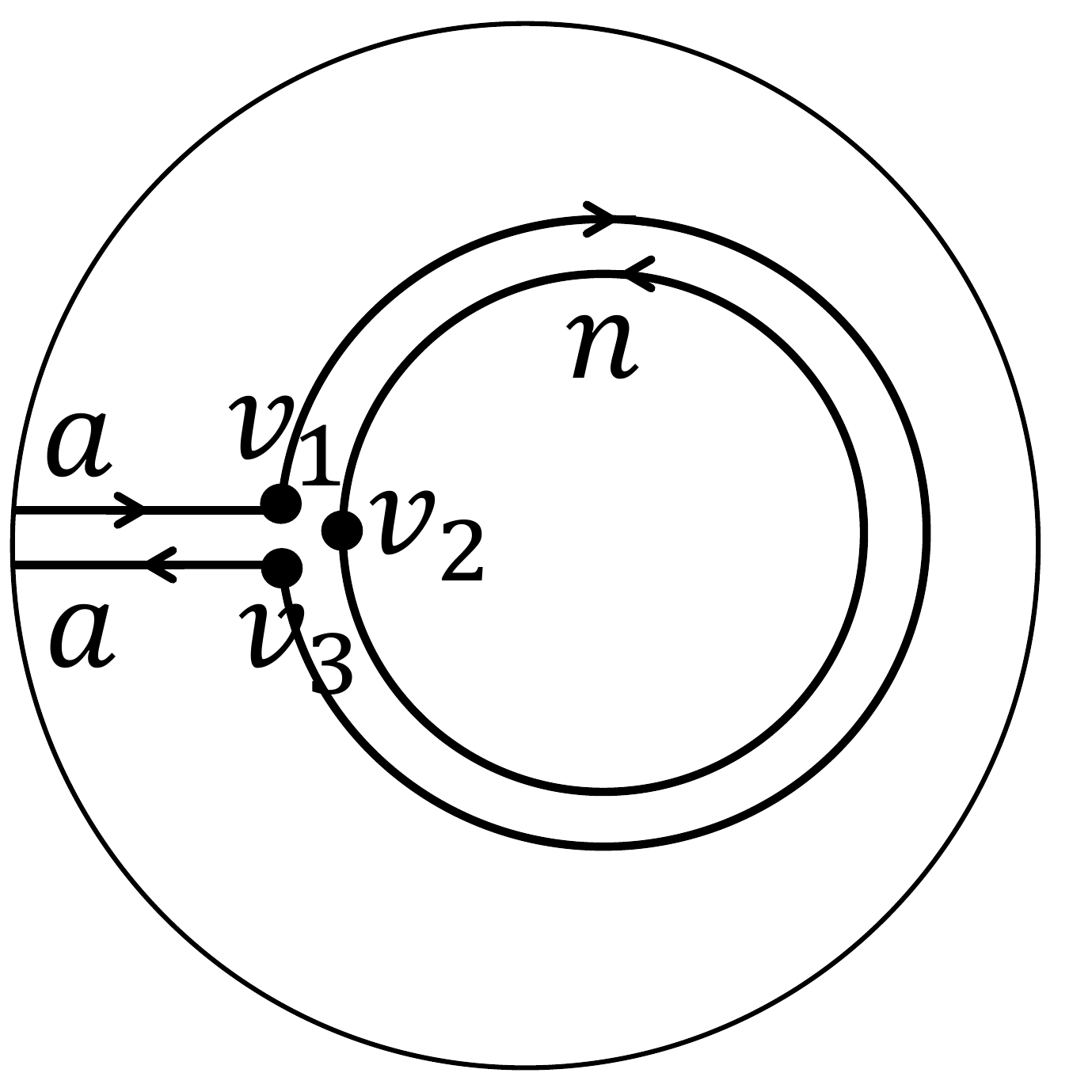}}

\def\six{\includegraphics[width=0.15\textwidth]{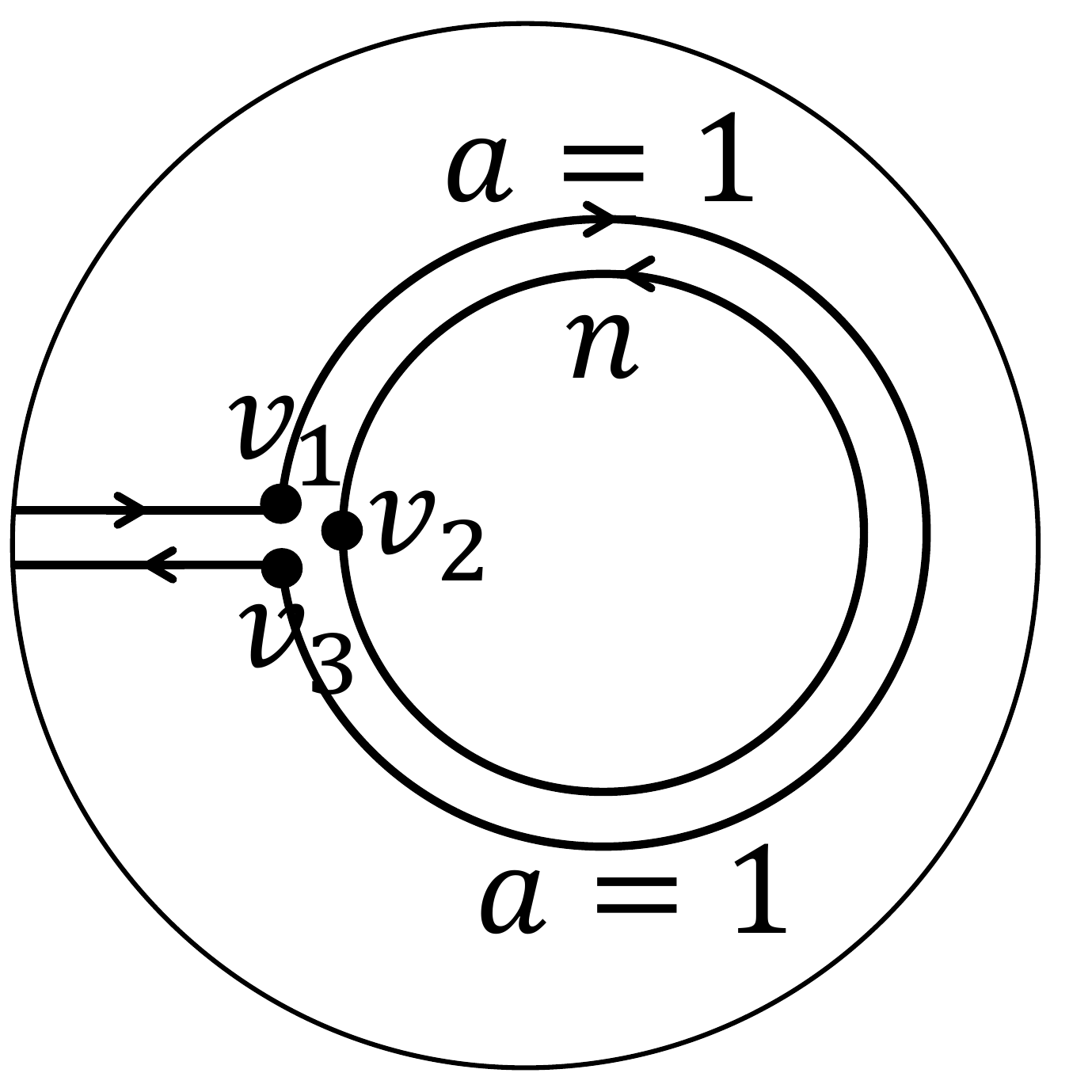}}

\def\two{\includegraphics[width=0.15\textwidth]{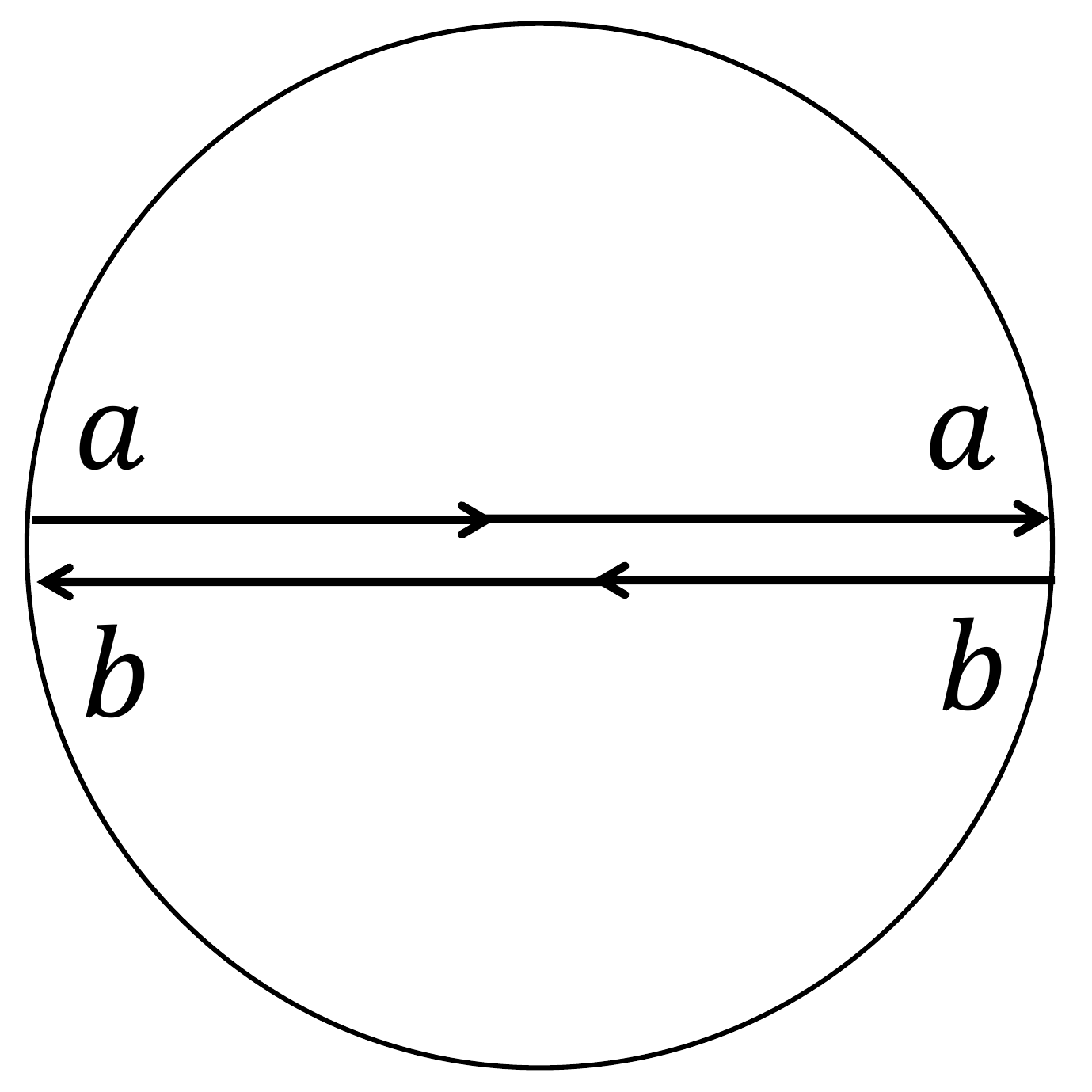}}

\def\three{\includegraphics[width=0.15\textwidth]{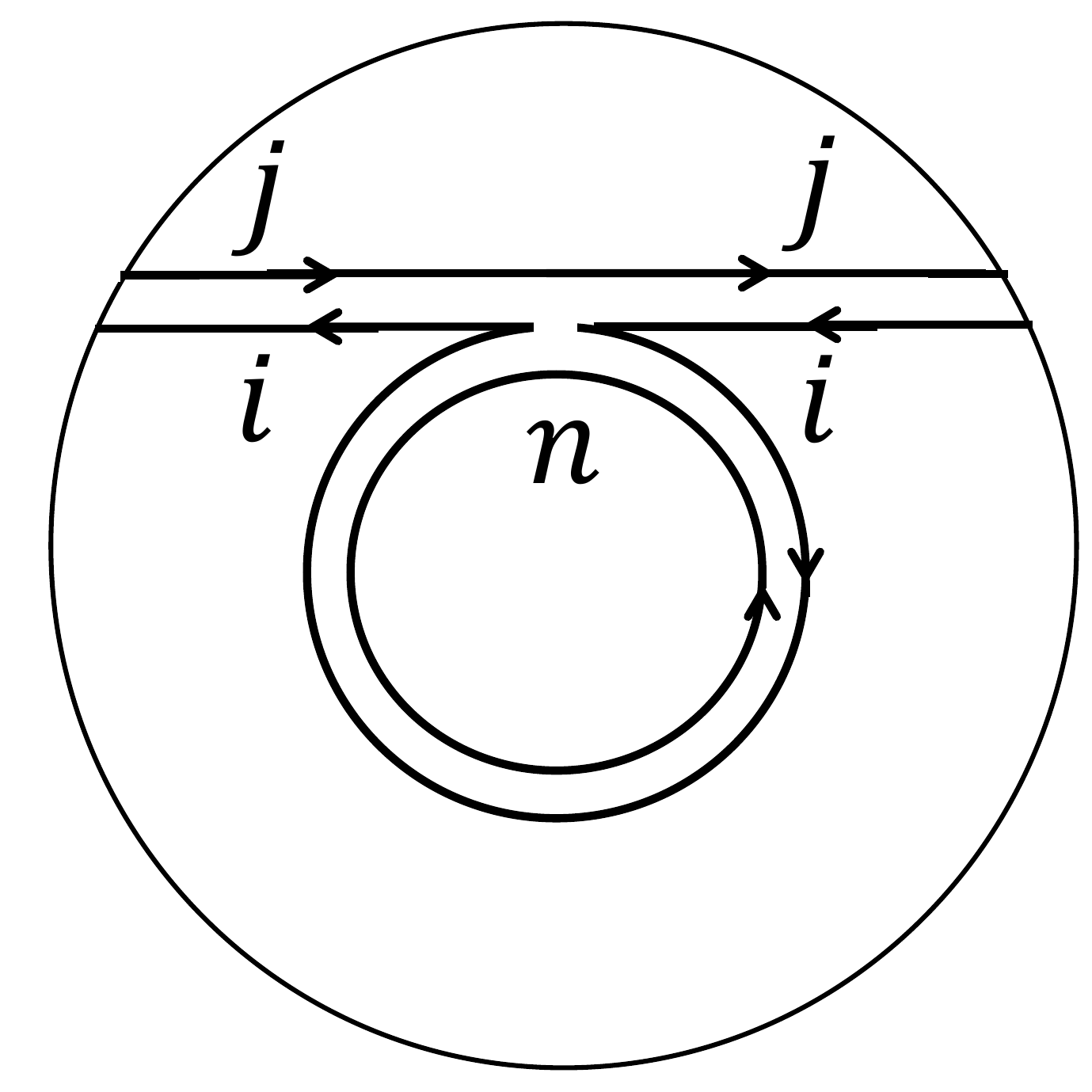}}

\def\four{\includegraphics[width=0.15\textwidth]{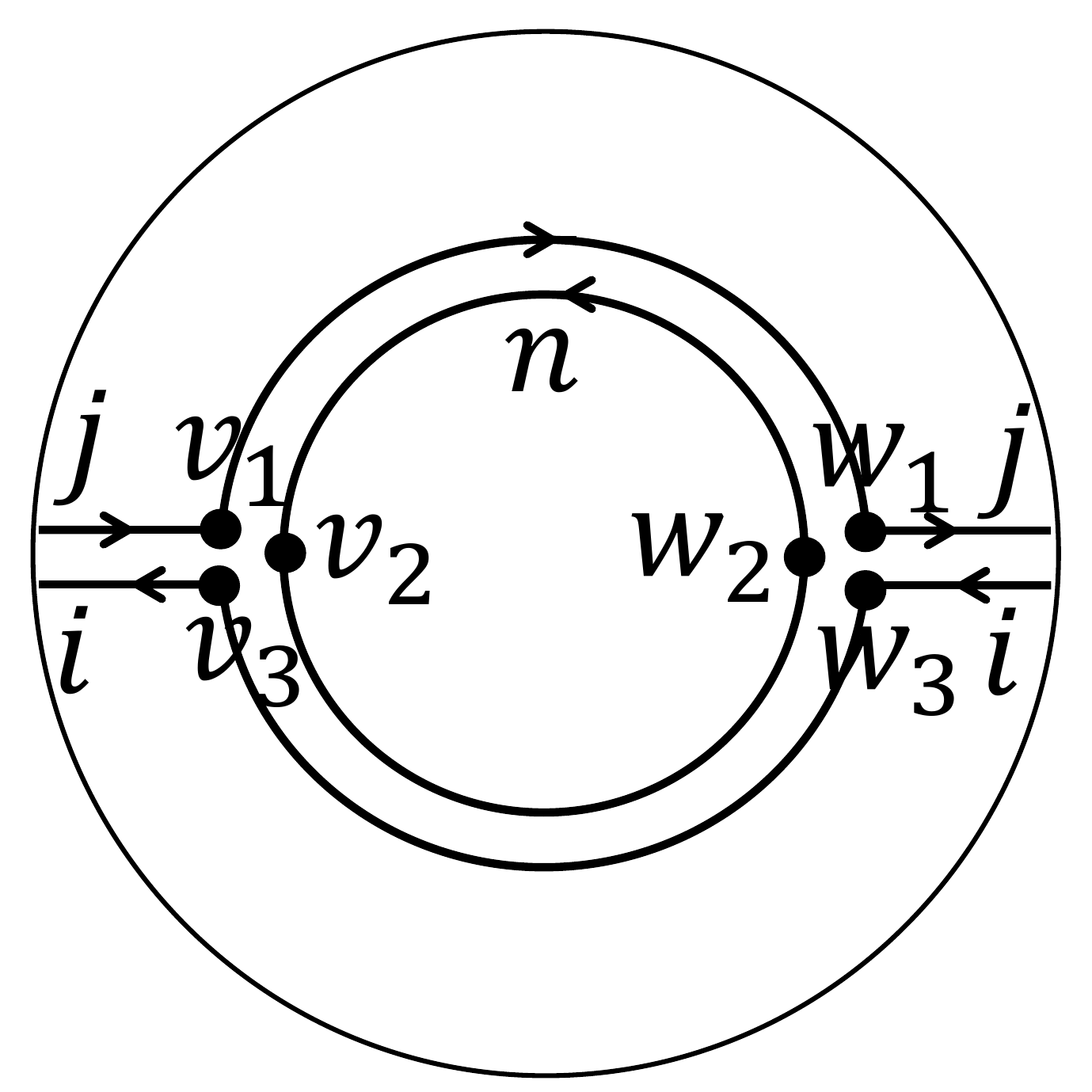}}

\def\five{\includegraphics[width=0.15\textwidth]{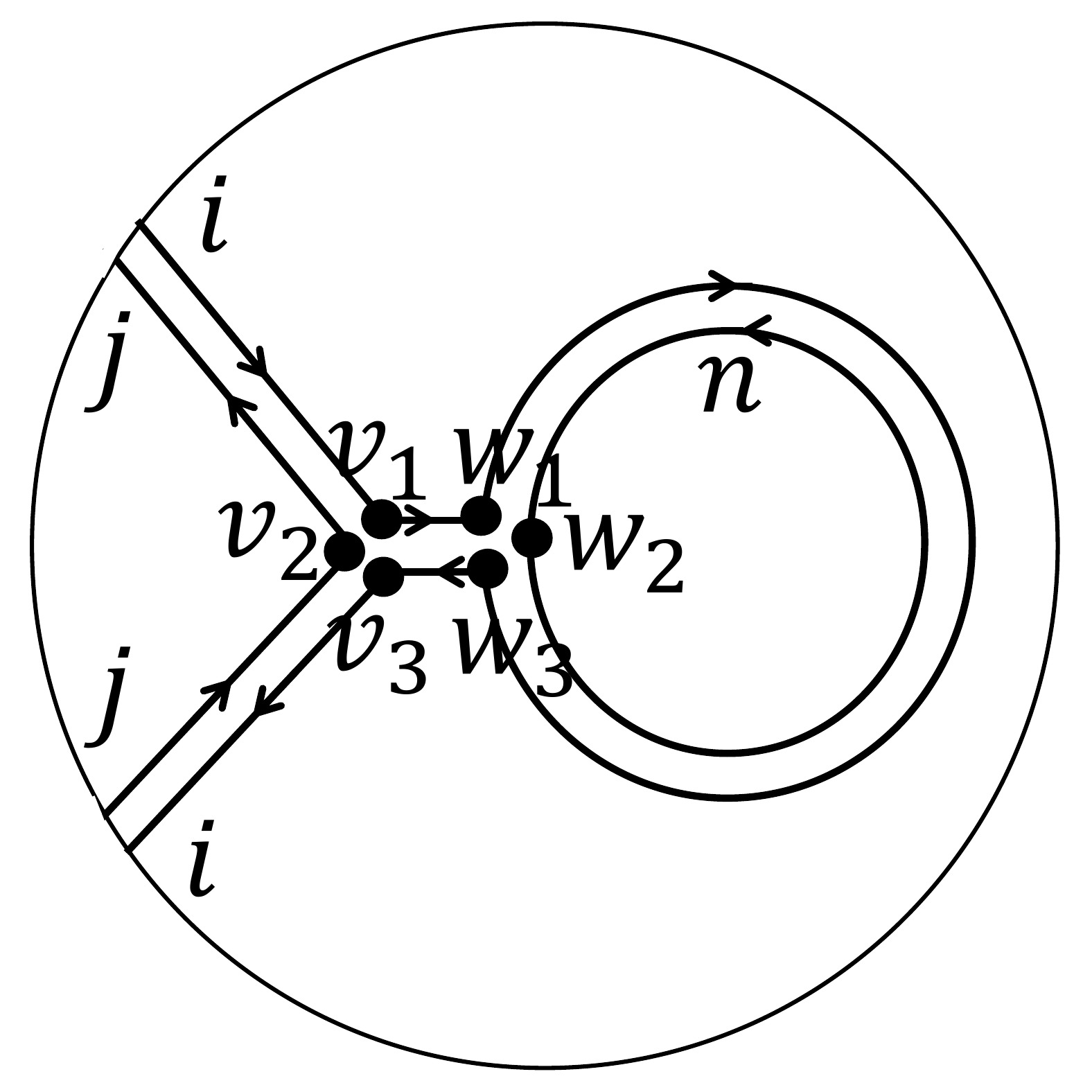}}

\def\ten{\includegraphics[width=0.15\textwidth]{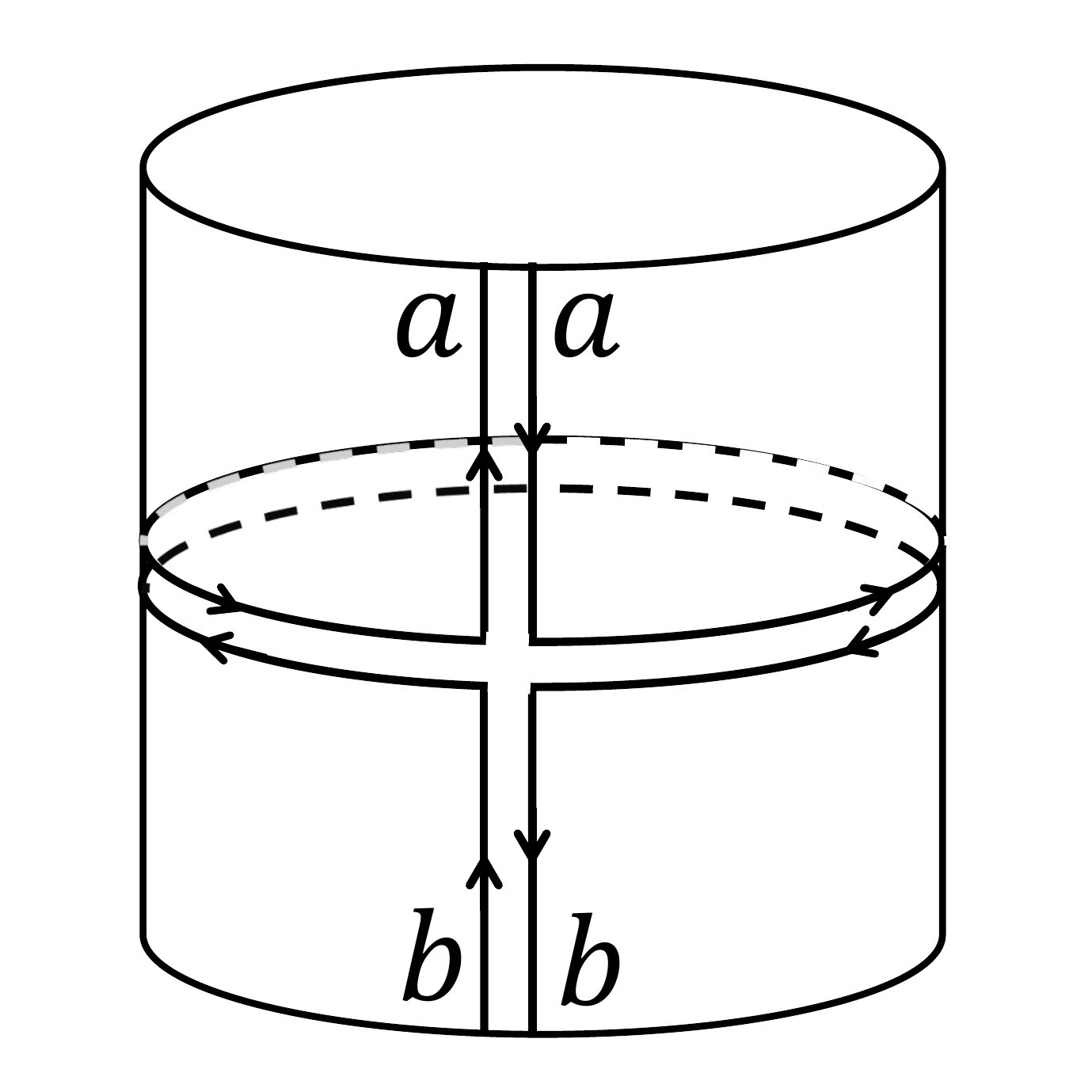}}

\def\seven{\includegraphics[width=0.15\textwidth]{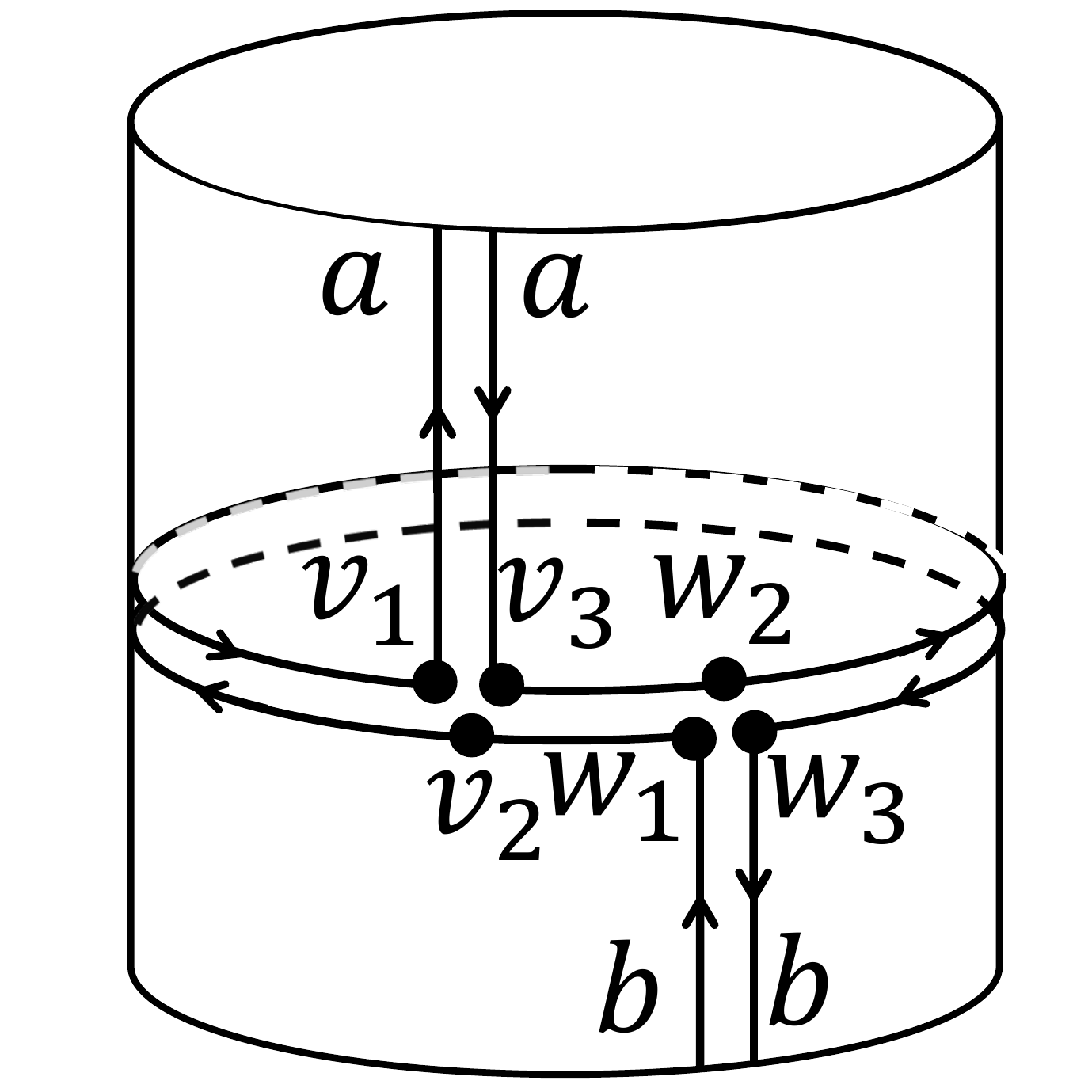}}

\def\eight{\includegraphics[width=0.15\textwidth]{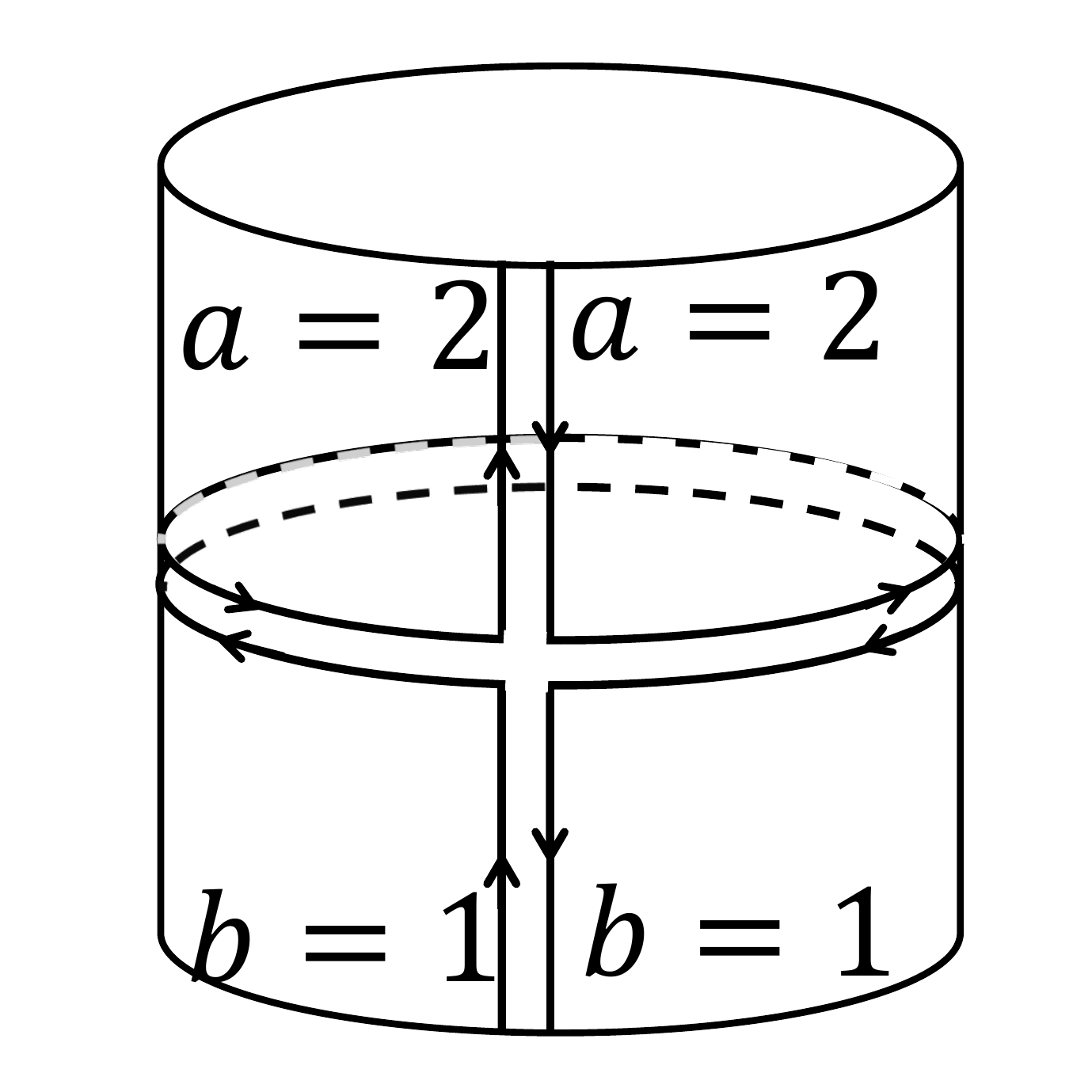}}

\def\nine{\includegraphics[width=0.15\textwidth]{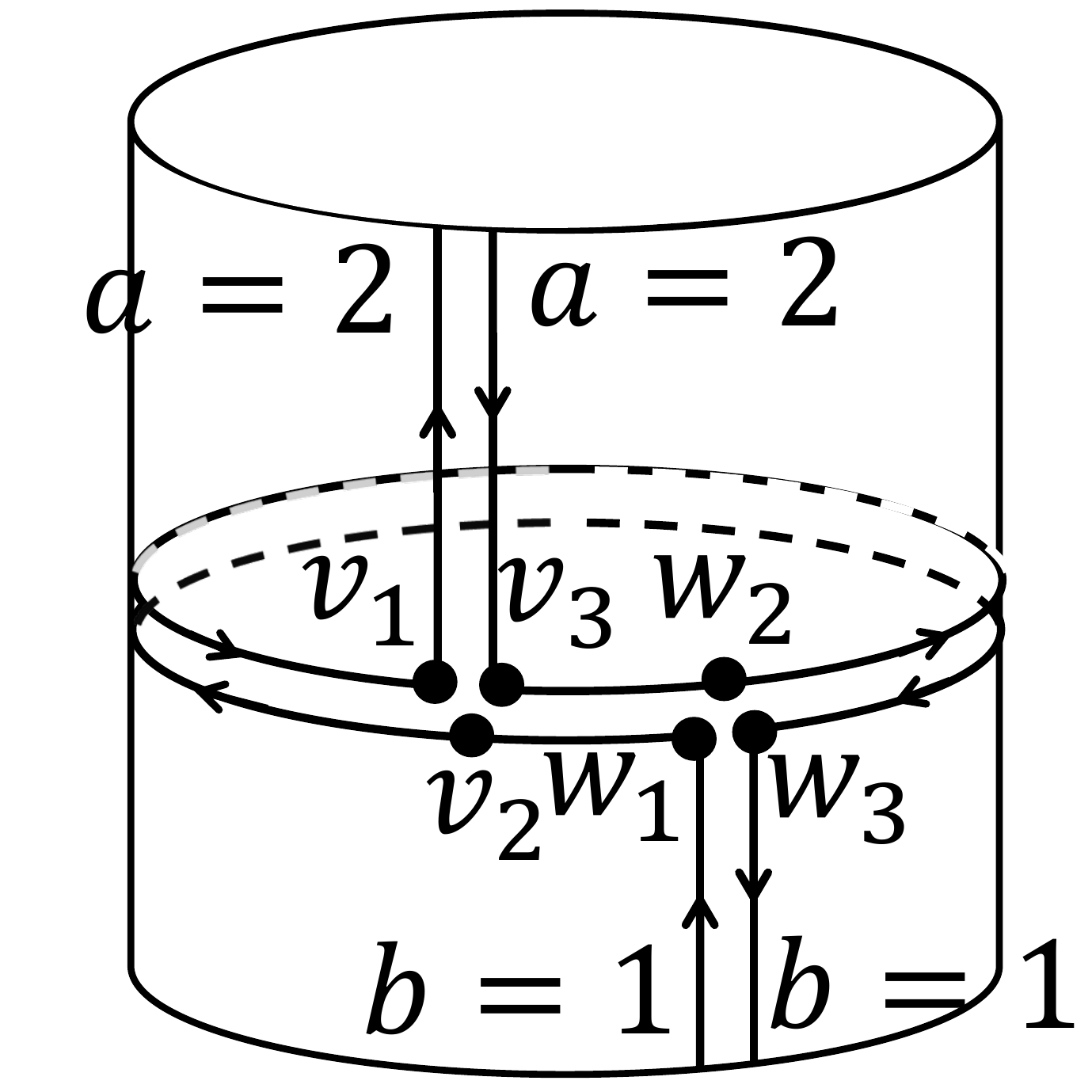}}

\def\p{\includegraphics[width=0.15\textwidth]{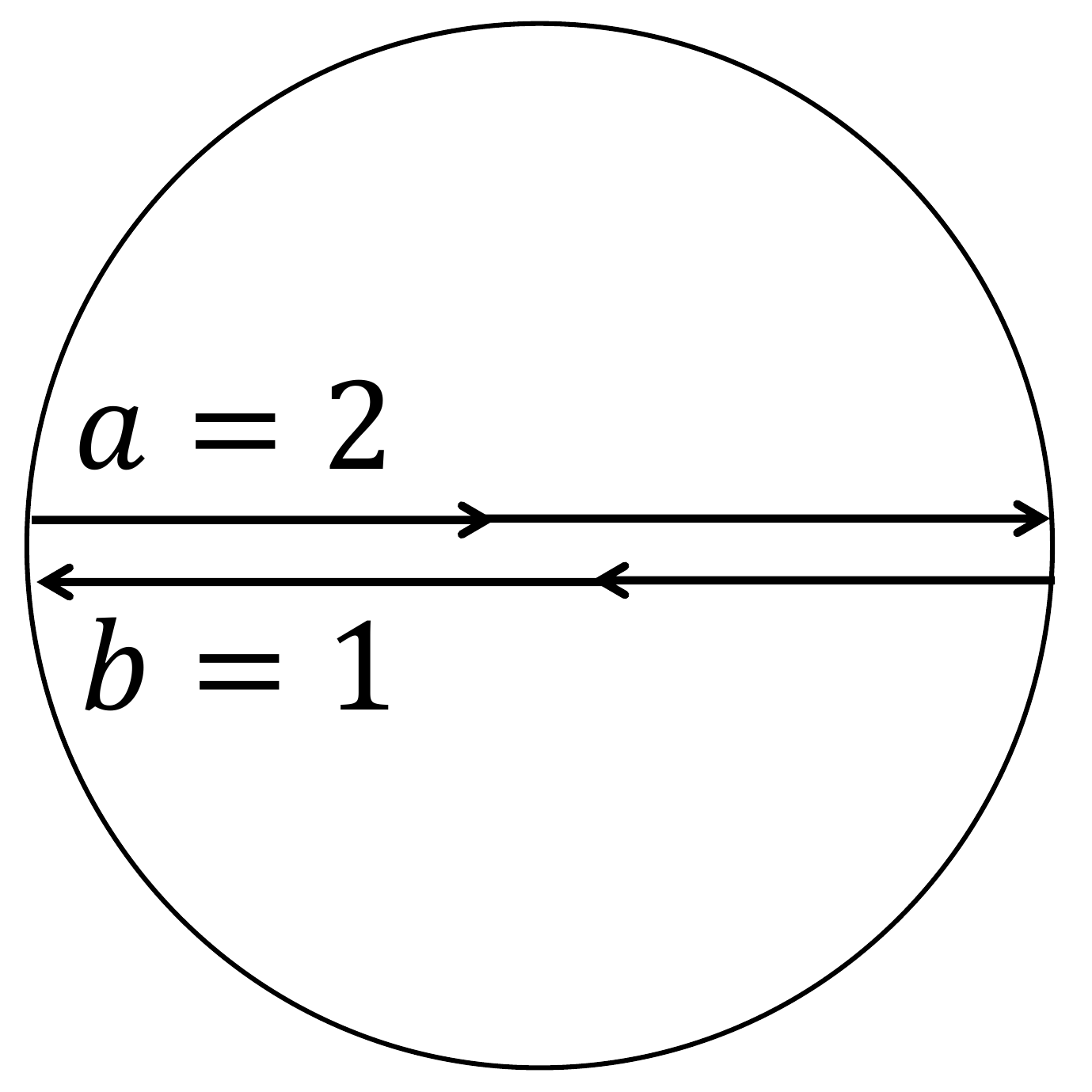}}

\def\1{\includegraphics[width=0.1\textwidth]{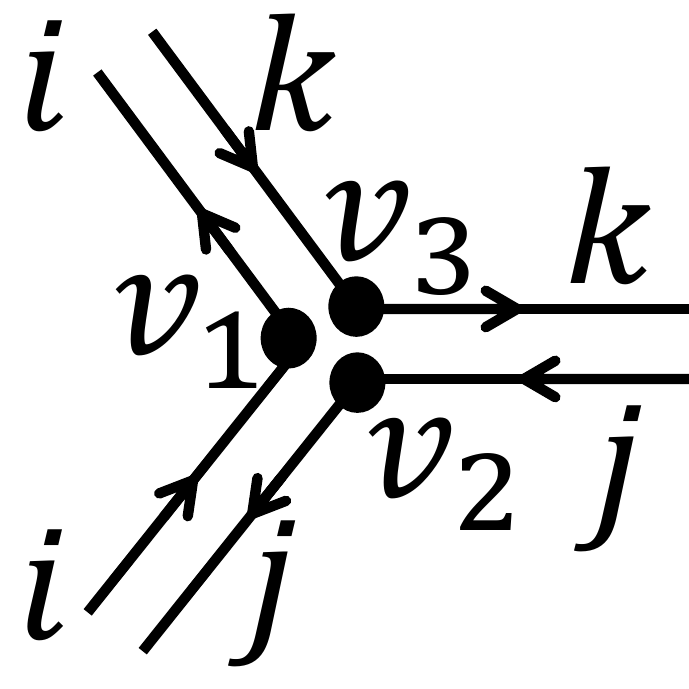}}

\def\2{\includegraphics[width=0.1\textwidth]{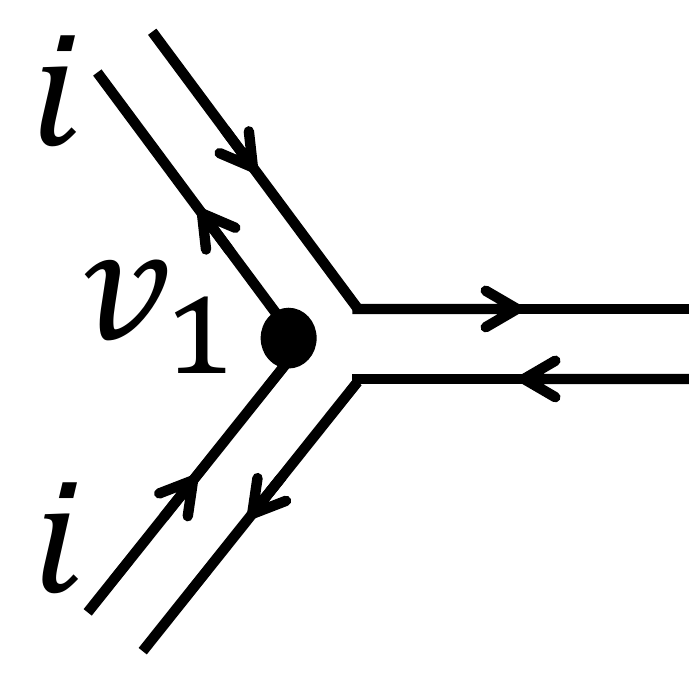}}

\def\3{\includegraphics[width=0.1\textwidth]{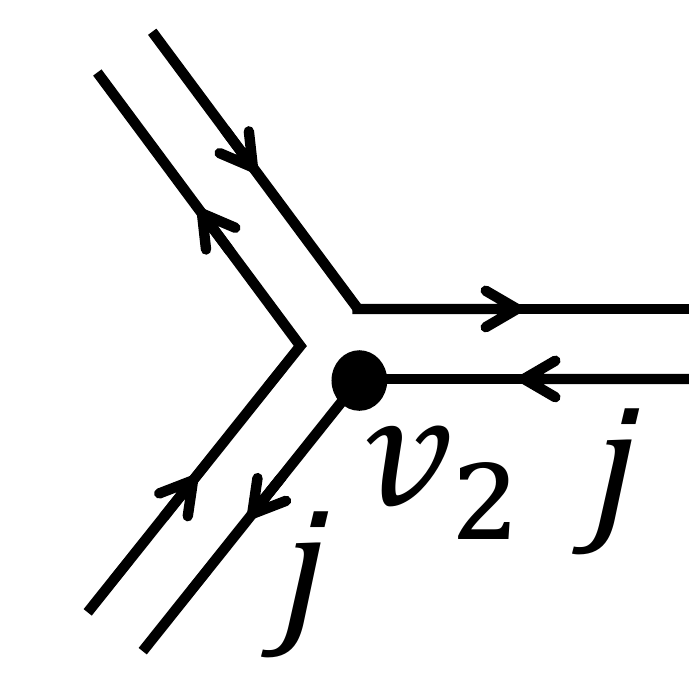}}

\def\4{\includegraphics[width=0.1\textwidth]{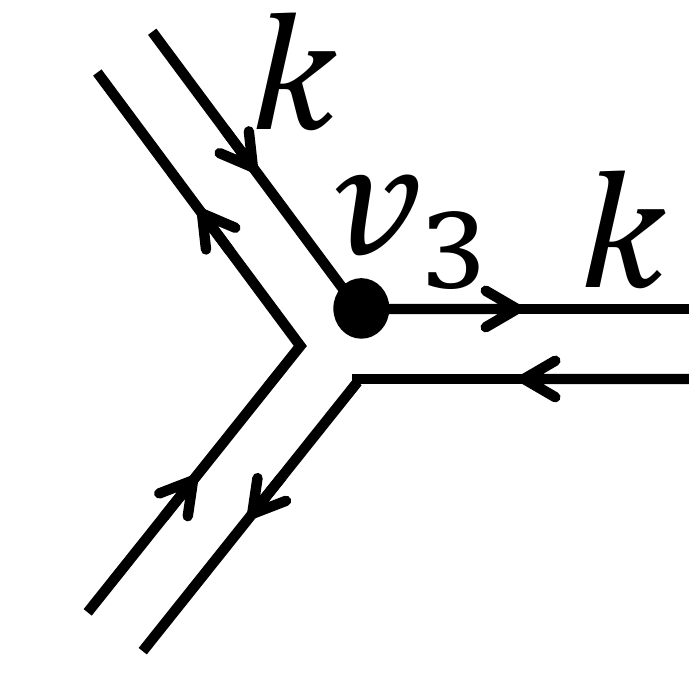}}

\def\5{\includegraphics[width=0.15\textwidth]{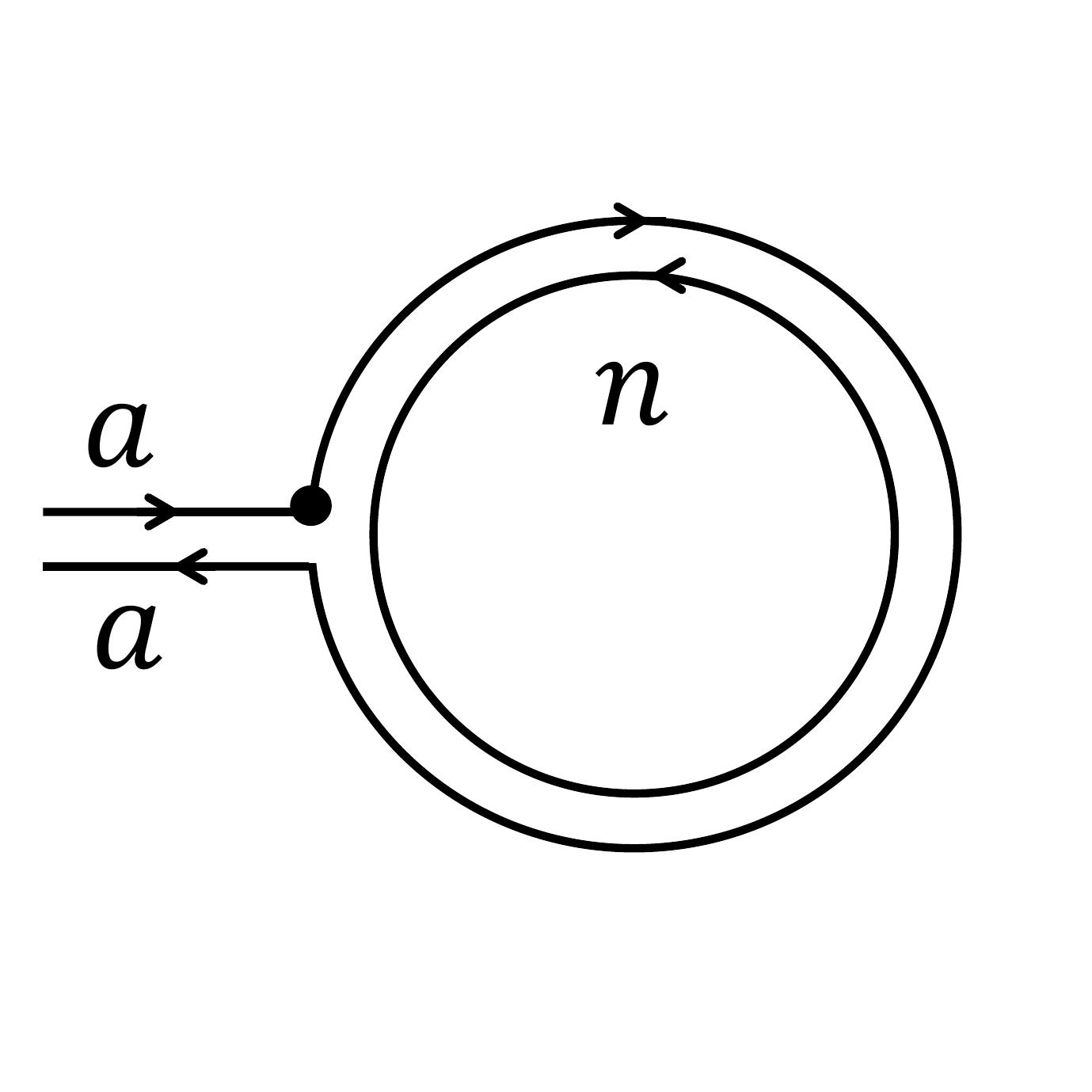}}

\def\6{\includegraphics[width=0.15\textwidth]{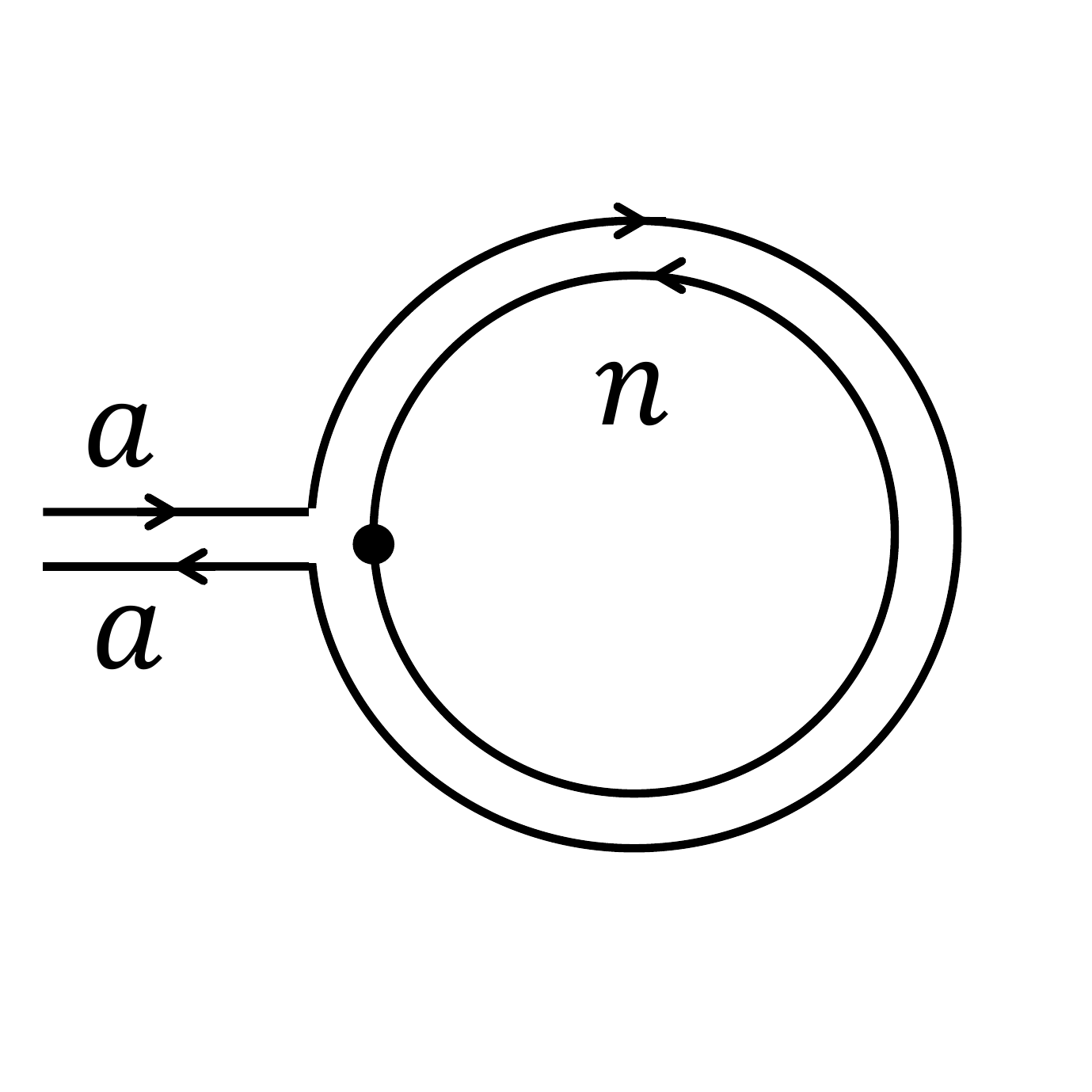}}

\def\7{\includegraphics[width=0.15\textwidth]{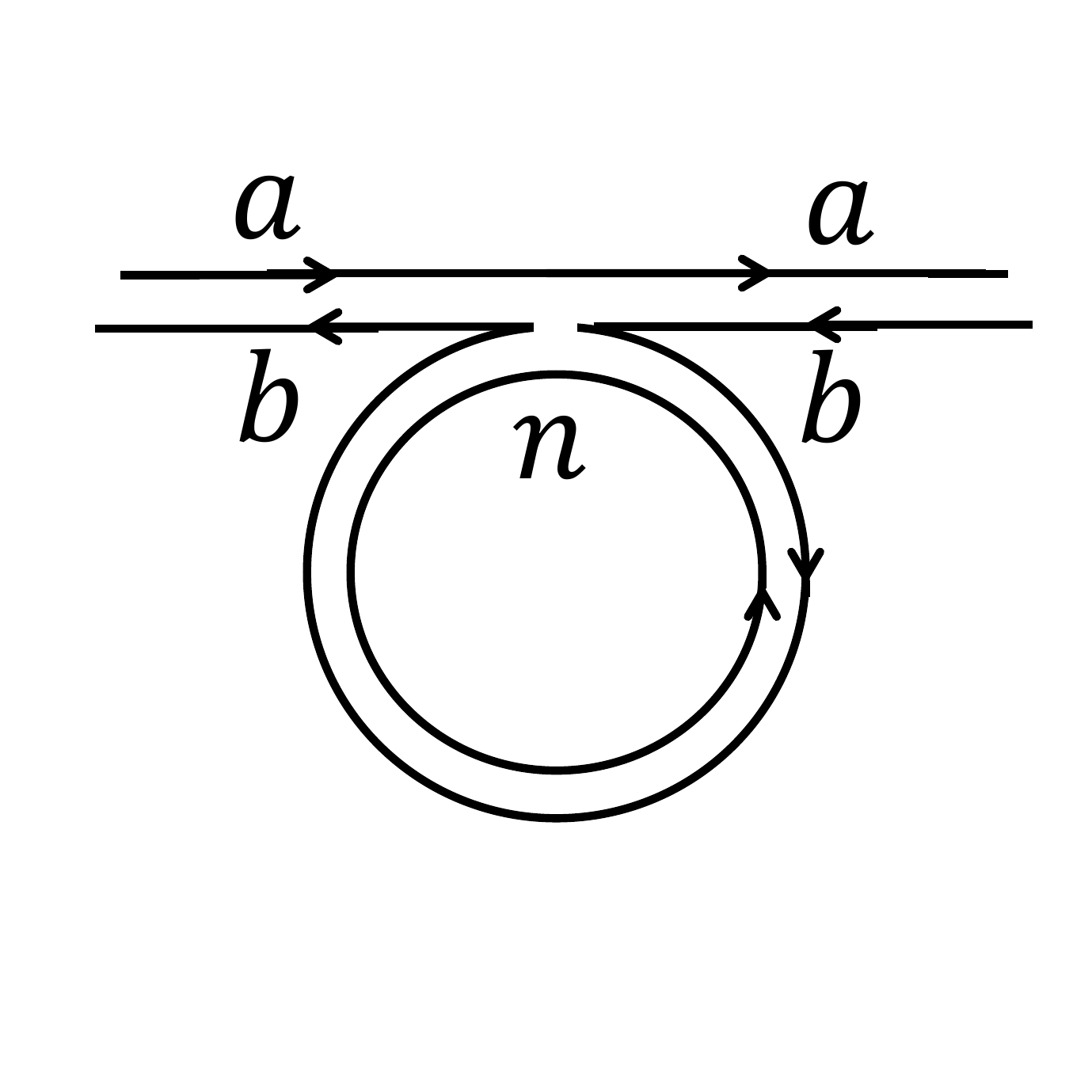}}

\def\8{\includegraphics[width=1.0\textwidth]{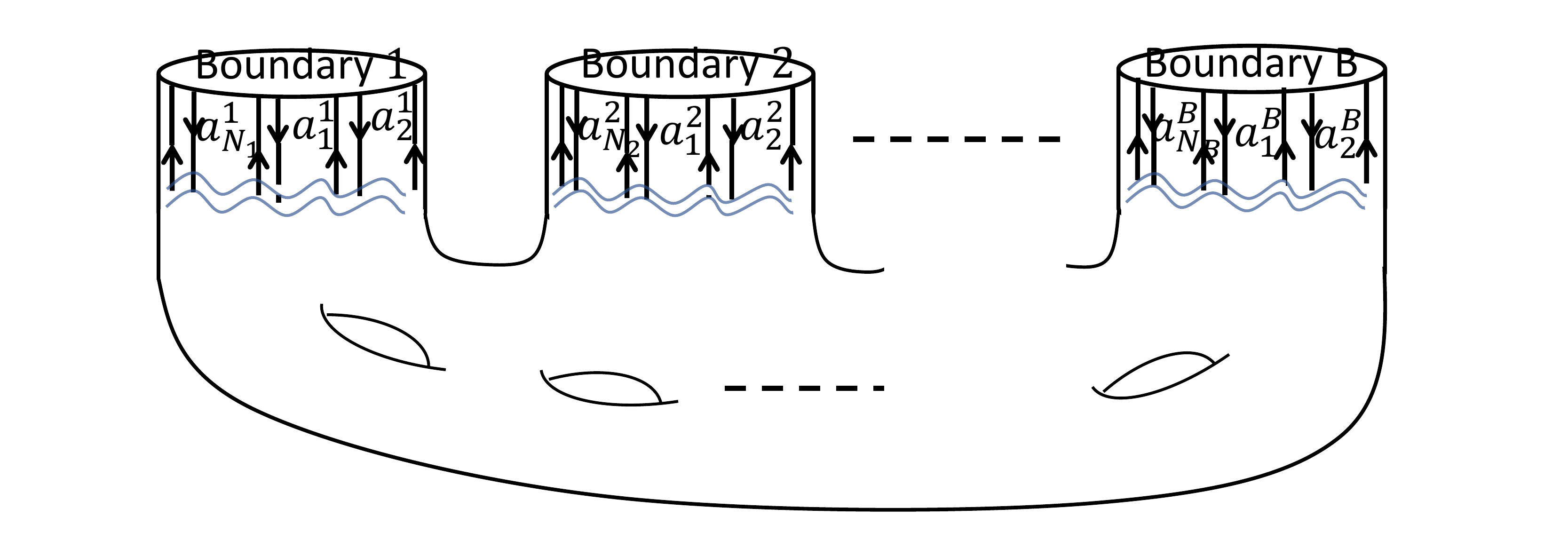}}

\def\9{\includegraphics[width=0.2\textwidth]{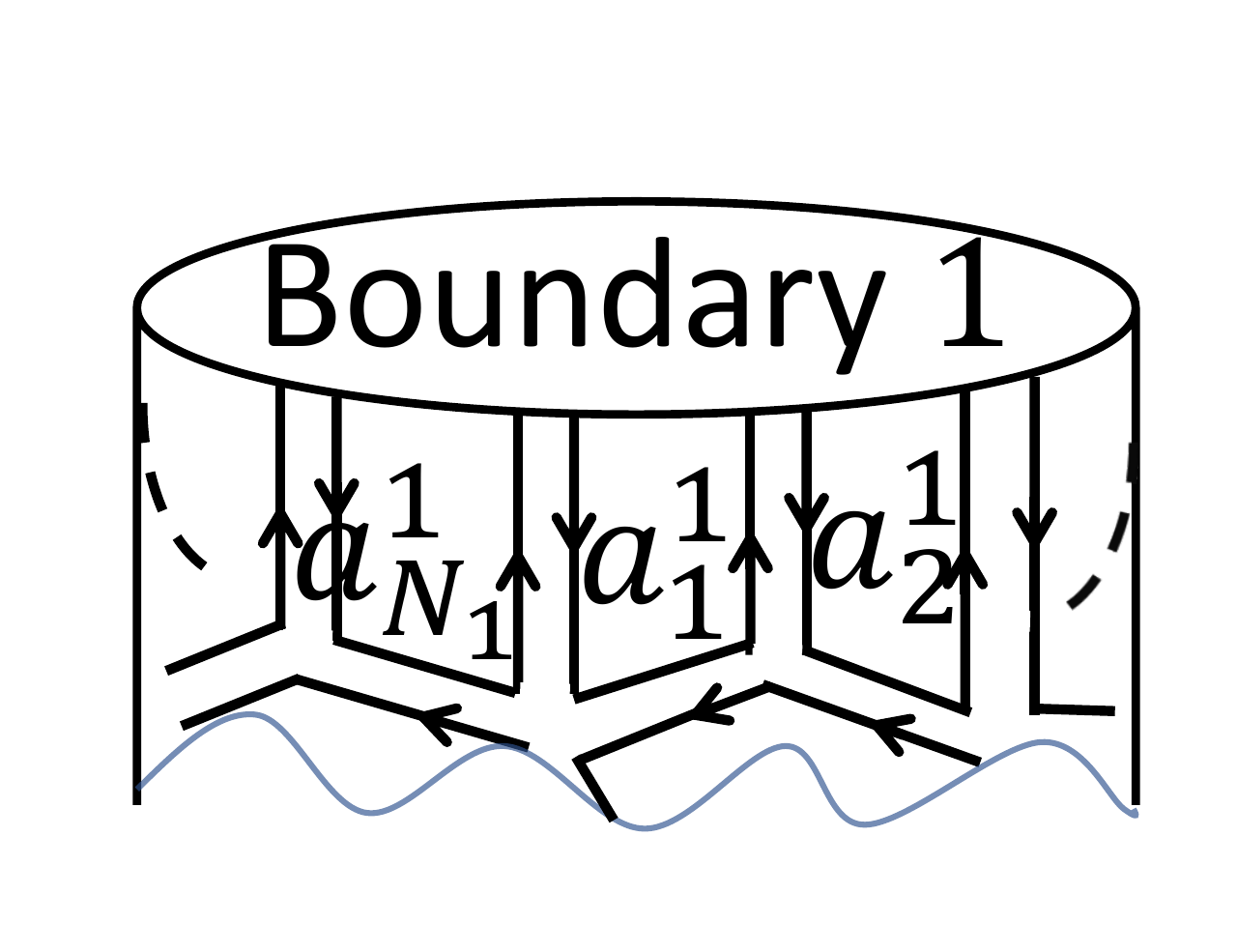}}

\def\thr{\includegraphics[width=0.1\textwidth]{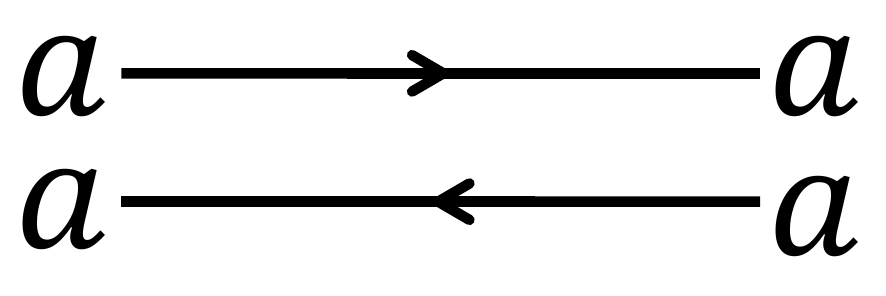}}

\def\eleven{\includegraphics[width=0.1\textwidth]{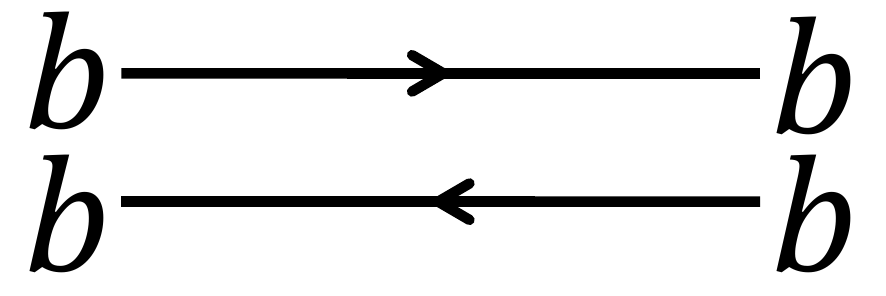}}

\def\twelve{\includegraphics[width=0.15\textwidth]{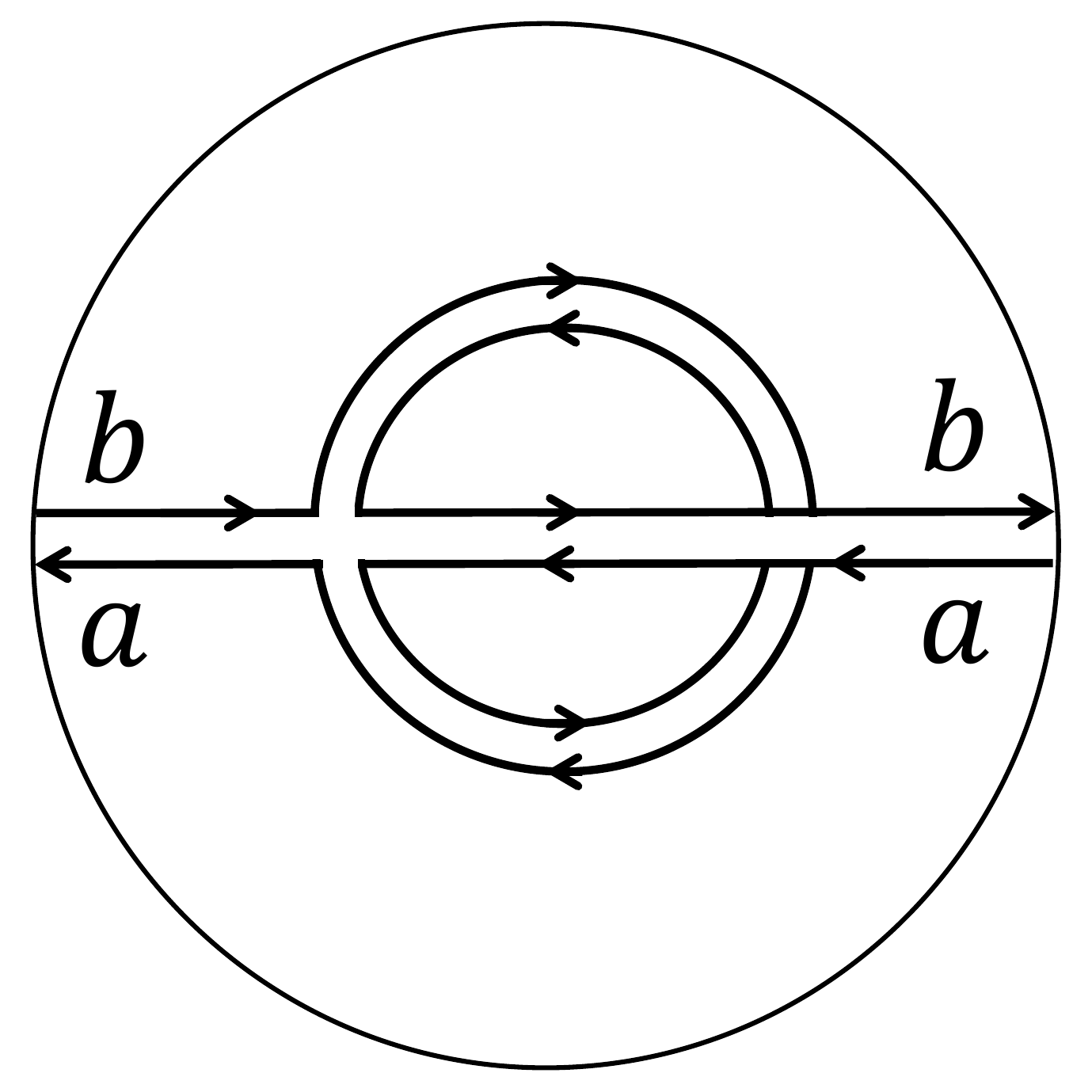}}

\begin{document}
\baselineskip 5mm
\title{Exact Solutions v.s. Perturbative Calculations\\

of Finite $\Phi^{3}$-$\Phi^{4}$ Hybrid-Matrix-Model}
\author{${}^{1,2}$ Naoyuki Kanomata and~ ${}^{1,2}$ Akifumi Sako}
{
${}^1$  Tokyo University of Science,\\ 1-3 Kagurazaka, Shinjuku-ku, Tokyo, 162-8601, Japan\\
${}^2$ Erwin Schr\"odinger International Institute for Mathematics and Physics, University of Vienna\\
Boltzmanngasse 9, 1090 Vienna, Austria
}
\noindent

\vspace{1cm}

\abstract{\vspace{1mm}There is a matrix model corresponding to a scalar field theory on noncommutative spaces called Grosse-Wulkenhaar model ($\Phi^{4}$ matrix model), which is renormalizable by adding a harmonic oscillator potential to scalar $\Phi^{4}$ theory on Moyal spaces. There are more unknowns in $\Phi^{4}$ matrix model than in $\Phi^{3}$ matrix model, for example, in terms of integrability. We then construct a one-matrix model  ($\Phi^{3}$-$\Phi^{4}$ Hybrid-Matrix-Model) with multiple potentials, which is a combination of a $3$-point interaction and a $4$-point interaction, where the $3$-point interaction of $\Phi^{3}$ is multiplied by some positive definite diagonal matrix $M$. This model is solvable due to the effect of this $M$. In particular, the connected $\displaystyle\sum_{i=1}^{B}N_{i}$-point function $G_{|a_{N_{1}}^{1}\cdots a_{N_{1}}^{1}|\cdots|a_{1}^{B}\cdots a_{N_{B}}^{B}|}$ of $\Phi^{3}$-$\Phi^{4}$ Hybrid-Matrix-Model is studied in detail. This $\displaystyle\sum_{i=1}^{B}N_{i}$-point function can be interpreted geometrically and corresponds to the sum over all Feynman diagrams (ribbon graphs) drawn on Riemann surfaces with $B$ boundaries (punctures). Each $|a_{1}^{i}\cdots a_{N_{i}}^{i}|$ represents $N_{i}$ external lines coming from the $i$-th boundary (puncture) in each Feynman diagram. First, we construct Feynman rules for $\Phi^{3}$-$\Phi^{4}$ Hybrid-Matrix-Model and calculate perturbative expansions of some multipoint functions in ordinary methods. Second, we calculate the path integral of the partition function $\mathcal{Z}[J]$ and use the result to compute exact solutions for $1$-point function $G_{|a|}$ with $1$-boundary, $2$-point function $G_{|ab|}$ with $1$-boundary, $2$-point function $G_{|a|b|}$ with $2$-boundaries, and $n$-point function $G_{|a^{1}|a^{2}|\cdots|a^{n}|}$ with $n$-boundaries. They include contributions from Feynman diagrams corresponding to nonplanar Feynman diagrams or higher genus surfaces.}

\allowdisplaybreaks[1]

\section{Introduction}

\subsection{Overview}

The purposes of this paper are to construct a perturbative theory of $\Phi^{3}$-$\Phi^{4}$ Hybrid-Matrix-Model to compute connected multipoint correlation functions and to give their exact solutions by direct calculations of path integrals. Roughly speaking, $\Phi^{3}$-$\Phi^{4}$ Hybrid-Matrix-Model is a $1$-matrix model of a Hermitian matrix $\Phi$ with interactions of $\Phi^{4}$ and $M\Phi^{3}$, where $M$ is some positive definite constant matrix, which is defined in detail in 1.2. Its perturbative theory is also carefully constructed in this paper since its interaction differs from a usual interaction of $\Phi^{3}$ because of $M$. On the other hand, this $M$ makes it possible to solve this theory rigorously.\bigskip

Theories in which several potentials are mixed, such as $\Phi^{3}$-$\Phi^{4}$ Hybrid-Matrix-Model, are often studied in physics. For example, such mixtures of potentials like scalar $\phi^{3}$-$\phi^{4}$ theories appear in cosmology. It is known that the expansion of the universe is accelerating. The energy causing the expansion of the universe is called the cosmological constant. The cosmological constant is thought to correspond to the energy of the vacuum, but this is still unknown. Prior works have considered, for example, a massless minimally coupled quantum scalar field $V(\phi)=\lambda\phi^{4}/4!+\beta\phi^{3}/3!$ with asymmetric self-interactions ($\lambda>0$), in a $(3 + 1)$-dimensional inflationary de Sitter space\cite{Bhattacharya:2022wjl}\cite{Bhattacharya:2022aqi}. It has been suggested that evaluating vacuum expectation values of $V(\phi)=\lambda\phi^{4}/4!+\beta\phi^{3}/3!$ and $\phi$ in this spacetime can lead to screening of the cosmological constant\cite{Bhattacharya:2022wjl}. The renormalized vacuum expectation value of $\phi$ is calculated by computing tadpoles in \cite{Bhattacharya:2022wjl}. The vacuum expectation value of $\phi^{2}$ is also calculated in \cite{Bhattacharya:2022aqi}. The model with $\Phi^{3}$-$\Phi^{4}$ potentials constructed in this paper can be interpreted as quantum field theories on noncommutative spaces after taking a large $N$ limit.

In addition, $\phi^{3}$-$\phi^{4}$ potentials can appear when dealing with matrix models with several potentials mixed in. For example, in a limit where there is an effective potential of a gauge theory on a fuzzy sphere, $\phi^{3}$ and $\phi^{4}$ appear in a mixed form\cite{Delgadillo-Blando:2008cuz}. (Specifically, the phase transition of this matrix model is considered in a study of theoretical predictions and Monte Carlo simulations of the matrix model in \cite{Delgadillo-Blando:2008cuz}.) This matrix model mixed with multiple potentials has also been applied in the phenomenology of particle physics.

Since a matrix model corresponding to a random triangulation of a $2$-dimensional surface has a potential of a form $\displaystyle\sum_{k=1}^{N}\Phi^{k}$ in general, $\Phi^{3}$-$\Phi^{4}$ Hybrid-Matrix-Model can be considered a model whose potential is restricted to some cubic one and quartic one.

Recently, matrix models on noncommutative spaces with several potentials which are different from Grosse-Wulkenhaar model have also been studied. For example, one-matrix multitrace scalar matrix models have been studied in \cite{Ydri:2021cam}. Thus, theories with several potentials have been the subjects of many important studies.\bigskip

One matrix model for noncritical string theories and two-dimensional quantum gravity theories were well-studied in the 1980s and 1990s\cite{DiFrancesco:1993cyw}. Each Feynman diagram in perturbative expansions of the matrix models represents a corresponding simplicial decomposition of a two-dimensional surface. In particular, two-dimensional surfaces are represented by $\Phi^{3}$ matrix model when considered as a graph discretized by a triangulation. The sum over two-dimensional surfaces corresponds to the two-dimensional quantum gravity theories. The proof of the Witten conjecture by Kontsevich can be cited as an example of a proof using $\Phi^{3}$ matrix model. Historically, Fukuma, Kawai, and Nakayama proved that the Virasoro constraint is equivalent to the condition that the solution of the KdV hierarchy satisfies the string equation\cite{Fukuma:1990jw}\cite{Zhou:2013kka}. Witten showed that the Witten-Kontsevich $\tau$-function satisfies the string equation\cite{Zhou:2013kka}\cite{Witten:1990hr}. Furthermore, Witten predicted that the Witten-Kontsevich $\tau$-function is the $\tau$ function of the KdV hierarchy\cite{Zhou:2013kka}\cite{Witten:1990hr}. Finally, Kontsevich proved the Witten conjecture using $\Phi^{3}$ matrix model\cite{Zhou:2013kka}\cite{Kontsevich:1992ti}. The quantum field theories on noncommutative spaces such as Moyal spaces gave a new perspective to matrix models. $\Phi^{3}$ matrix model as a renormalizable quantum field theory on Moyal spaces was first studied by Grosse and Steinacker\cite{Grosse:2005ig}\cite{Grosse:2006qv}\cite{Grosse:2006tc}. The historical background of the studies is that the UV/IR mixing problem occurs and renormalization is not possible when considering quantum field theories on noncommutative spaces. To avoid the UV/IR mixing problem, Grosse and Steinacker adjusted the action of scalar $\Phi^{3}$ theories on Moyal spaces by adding harmonic oscillator potentials. In the following, Grosse-Steinacker $\Phi^{3}$ model is referred to as $\Phi^{3}$ matrix model. This model is renormalized by adding harmonic oscillator potentials to scalar $\Phi^{3}$ theories on Moyal spaces. In particular, all multipoint correlation function of $\Phi^{3}$ matrix model in the large $N,V$ limit was computed by Grosse, Wulkenhaar, and one of the authors by solving exactly Schwinger-Dyson equations\cite{Grosse:2016pob}\cite{Grosse:2016qmk}. The multipoint correlation function in $\Phi^{3}$ matrix model here is the probability amplitude of a particle being observed at $n$-points, a physical quantity in quantum field theories. Subsequently, all the multipoint correlation functions of $\Phi^{3}$ matrix model in the case of finite degrees of freedom were computed by the authors\cite{Kanomata:2022pdo}.

A matrix model as a renormalizable $\Phi^{4}$ theory on Moyal spaces similar to $\Phi^{3}$ matrix model is Grosse-Wulkenhaar $\Phi^{4}$ model\cite{Hock:2020rje}\cite{Grosse:2012uv}. In the following, Grosse-Wulkenhaar $\Phi^{4}$ model is called $\Phi^{4}$ matrix model. This model corresponds to the quantum field theories on noncommutative spaces, which is renormalized by adding harmonic oscillator potentials to scalar $\Phi^{4}$ theories on Moyal spaces. This is how $\Phi^{4}$ matrix model came to be considered. The $2$-point function of $\Phi^{4}$ matrix model whose Feynman diagrams in the perturbative expansion in the large $N,V$ limit can be drawn in the planar diagrams was solved exactly by Grosse, Wulkenhaar, and Hock\cite{Grosse:2019jnv}. The multipoint correlation functions of $\Phi^{4}$ matrix model have been solved by Wulkenhaar and Hock\cite{Hock:2021tbl}. Multipoint correlation function for $\Phi^{4}$ matrix model with nonplanar Feynman diagrams has been studied by Grosse, Wulkenhaar, Hock, and Branahl using a computational method called ``Blobbed Topological Recursion" in \cite{Branahl:2020yru}\cite{Branahl:2021slr}\cite{Branahl:2021uea}\cite{Hock:2023nki}. Using the technology of topological recursion, a hybrid potential theory called $r$-spin model like the model of this paper is also studied in \cite{Belliard}. Recently, Prekrat, Rankovi$\acute{c}$, Todorovi$\acute{c}$-Vasovi$\acute{c}$, Kov$\acute{a}\breve{c}$ik, and Tekel calculated the curvature contribution in Grosse-Wulkenhaar model\cite{Prekrat:2022sir}. Itzykson-Zuber integral is used to approximate the curvature contribution in Grosse-Wulkenhaar model\cite{Prekrat:2022sir}. The multitrace matrix model is approximated using analytical methods, and the multitrace matrix model is also studied numerically by Monte Carlo simulations in \cite{Prekrat:2022sir}.\bigskip

It is a well-known fact that $\Phi^{3}$ matrix model is integrable\cite{Kontsevich:1992ti}. However, the integrability of the $\Phi^{4}$ matrix model as Grosse-Wulkenhaar model is not proved until now. Showing that $\Phi^{3}$-$\Phi^{4}$ Hybrid-Matrix-Model is finite and solvable may provide insight into the solvability of $\Phi^{4}$ matrix model. Therefore, to reveal properties of $\Phi^{3}$-$\Phi^{4}$ Hybrid-Matrix-Model is significance.\bigskip

In this paper, we construct Feynman rules of $\Phi^{3}$-$\Phi^{4}$ Hybrid-Matrix-Model in a well-known way in terms of quantum field theories. The $\Phi^{3}$ interaction causes unconventional Feynman rules because of the insertion of $M$ as $\mathrm{tr}M\Phi^{3}$. Therefore, we will discuss it carefully without omission. Particular attention is paid to the connected $\displaystyle\sum_{i=1}^{B}N_{i}$-point function $G_{|a_{N_{1}}^{1}\cdots a_{N_{1}}^{1}|\cdots|a_{1}^{B}\cdots a_{N_{B}}^{B}|}$. Its details are defined and discussed in 2.2, but this function can be interpreted geometrically and corresponds to the sum over all Feynman diagrams (ribbon graphs) drawn in certain rules on Riemann surfaces with $B$-boundaries (punctures). Each $|a_{1}^{i}\cdots a_{N_{i}}^{i}|$ corresponds to $N_{i}$ external lines coming from the $i$-th boundary (puncture) in the Feynman diagrams (ribbon graphs). First, using the Feynman rules of $\Phi^{3}$-$\Phi^{4}$ Hybrid-Matrix-Model, perturbative expansions for one-point function $G_{|1|}$, two-point function $G_{|21|}$, and two-point function $G_{|2|1|}$ are computed by drawing Feynman diagrams for matrix size $N=2$ case as pedagogical instructions to understand the way of calculations. Next, we perform perturbative calculations for $G_{|a|}$,$G_{|ab|}$, and $G_{|a|b|}$ in the case that the matrix size is any $N$.

On the other hand, the calculation of the partition function $\mathcal{Z}[J]$ in $\Phi^{3}$-$\Phi^{4}$ Hybrid-Matrix-Model can be carried out rigorously. For the computation of the partition function $\mathcal{Z}[J]$, the integral of the off-diagonal elements of the Hermitian matrix is computed using Itzykson-Zuber integral\cite{Itzykson:1979fi}\cite{T.Tao}\cite{Eynard:2015aea}\cite{Zinn-Justin:2002rai}. In contrast, the integral of the diagonal elements of the Hermite matrix is obtained by using a function $P(z)$ that is similar to Pearcey integral. We then use the exact calculated partition function $\mathcal{Z}[J]$ to compute the exact solutions for $G_{|a|}$, $G_{|ab|}$, $G_{|a|b|}$, and $n$-point function $G_{|a^{1}|a^{2}|\cdots|a^{n}|}$  for any $N$ matrix size. We verify that the final results of the perturbative expansions for $N=2$ are in agreement with the saddle point approximation using the results of the exact solutions in $G_{|1|}$, $G_{|21|}$, and $G_{|2|1|}$ by setting $N=2$. Finally, we make remarks about contributions from Feynman diagrams of $\Phi^{3}$-$\Phi^{4}$ Hybrid-Matrix-Model corresponding to nonplanar or higher genus surfaces. \bigskip

This paper is organized as follows. In Section $2$, $\Phi^{3}$-$\Phi^{4}$ Hybrid-Matrix-Model is computed by perturbative expansion methods in ordinary quantum field theories. In Section $3$, we compute the path integral of the partition function $\mathcal{Z}[J]$. In Section $4$, the results of Section $3$ are used to compute exact solutions of one-point function $G_{|a|}$, two-point function $G_{|ab|}$, two-point function $G_{|a|b|}$, and $n$-point function $G_{|a^{1}|a^{2}|\cdots|a^{n}|}$. In Section $5$, the exact solutions of $1$-point function $G_{|1|}$, $2$-point function $G_{|21|}$, and $2$-point function $G_{|2|1|}$ for a matrix of size $N=2$ are approximated using the saddle point method to verify that they are consistent with the perturbative expansion in Section $2$. In Section $6$, we remark contributions from Feynman diagrams of $\Phi^{3}$-$\Phi^{4}$ Hybrid-Matrix-Model corresponding to nonplanar or higher genus surfaces.

\subsection{Setup of $\Phi^{3}$-$\Phi^{4}$ Hybrid-Matrix-Model}
In this subsection, we define $\Phi^{3}$-$\Phi^{4}$ Hybrid-Matrix-Model, and we determine our notations in this paper.

Let $\Phi=(\Phi_{ij})$ be a Hermitian matrix for $i,j=1,2,\ldots,N$ and $E_{m-1}$ be a discretization of a monotonously increasing differentiable function $e$ with $e(0)=0$, 
\begin{align}
E_{m-1}=&\mu^{2}\left(\frac{1}{2}+e\left(\frac{m-1}{\mu^{2}V}\right)\right),
\end{align}
where $\mu^{2}$ is a squared mass, and $V$ is a real constant. Let $E=(E_{m-1}\delta_{mn})$ be a diagonal matrix for $m,n=1,\ldots,N$ and $M=(\sqrt{E_{k-1}}\delta_{kl})$ be a diagonal matrix for $k,l=1,\ldots,N$, i.e. $E=M^{2}$. Let us consider the following action:
\begin{align}
S[\Phi]=V\mathrm{tr}\left(E\Phi^{2}+\kappa\Phi+\frac{1}{2}M\Phi M\Phi+\sqrt{\lambda}M\Phi^{3}+\frac{\lambda}{4}\Phi^{4}\right),
\end{align}
where $\kappa$ is a constant (real), and $\lambda$ is a coupling constant that is non-zero real. To avoid confusion later, $\lambda$ and $V$ are assumed to be positive. When we consider the perturbation theory in Section \ref{sec2}, we put $\kappa=0$.
Let $J=(J_{mn})$ be a Hermitian matrix for $m,n=1,\ldots,N$ as an external field. Let $\mathcal{D}\Phi$ be the integral measure, 
\begin{align}
\displaystyle\mathcal{D}\Phi:=&\prod_{i=1}^{N}d\Phi_{ii}\prod_{1\leq i<j\leq N}d\mathrm{Re}\Phi_{ij}d\mathrm{Im}\Phi_{ij},
\end{align}   
where each variable is divided into real and imaginary parts $\Phi_{ij}=\mathrm{Re}\Phi_{ij}+i\mathrm{Im}\Phi_{ij}$ with $\mathrm{Re}\Phi_{ij}=\mathrm{Re}\Phi_{ji}$ and $\mathrm{Im}\Phi_{ij}=-\mathrm{Im}\Phi_{ji}$. Let us consider the following partition function:
\begin{align}
\mathcal{Z}[J]:=&\int\mathcal{D}\Phi\exp\left(-S[\Phi]+V\mathrm{tr}(J\Phi)\right)\notag\\
=&\int \mathcal{D}\Phi \exp\left(-V\mathrm{tr}\left(E\Phi^{2}+\kappa\Phi+\frac{1}{2}M\Phi M\Phi+\sqrt{\lambda}M\Phi^{3}+\frac{\lambda}{4}\Phi^{4}\right)\right)\exp\left(V\mathrm{tr}\left(J\Phi\right)\right).\label{Z[J]}
\end{align}


\section{Perturbation Theory of $\Phi^{3}$-$\Phi^{4}$ Hybrid-Matrix-Model}\label{sec2}

The aim of this section is to understand $\Phi^{3}$-$\Phi^{4}$ Hybrid-Matrix-Model by usual perturbative methods in field theories. For this purpose, we make its Feynman rules and calculate one-point functions and two types of two-point functions perturbatively. It is made in the same way as Feynman rules for well-known matrix models\cite{Eynard:2015aea}\cite{Hock:2020rje}. However, it differs slightly from the usual one due to the presence of $M$ found in (\ref{Z[J]}). We then construct the perturbative theories a little more carefully for readers who are not familiar with perturbative theories of these matrix models.

\subsection{Feynman Rules of $\Phi^{3}$-$\Phi^{4}$ Hybrid-Matrix-Model $(\kappa=0)$}

We consider the theory of $\displaystyle S_{free}=V\mathrm{tr}\left(E\Phi^{2}+\frac{1}{2}M\Phi M\Phi\right)$, that is no interaction theory, to consider the perturbation theory of $\Phi^{3}$-$\Phi^{4}$ Hybrid-Matrix-Model at first. We calculate $\mathcal{Z}_{free}[J]$;
\begin{align}
\mathcal{Z}_{free}[J]=&\int\mathcal{D}\Phi\exp\left(-V\mathrm{tr}\left(E\Phi^{2}+\frac{1}{2}M\Phi M\Phi\right)\right)\exp\left(V\mathrm{tr}(J\Phi)\right)\notag\\
=&\mathcal{C}'\exp\left(\frac{V}{2}\sum_{n,m=1}^{N}J_{mn}\frac{1}{E_{n-1}+E_{m-1}+\sqrt{E_{n-1}}\sqrt{E_{m-1}}}J_{nm}\right).\label{K}
\end{align}
Here $\displaystyle\mathcal{C}'=\mathcal{Z}_{free}[0]=\left(\prod_{n=1}^{N}\sqrt{\frac{2\pi}{3VE_{n-1}}}\right)\left(\prod_{1\leq n<m\leq N}\frac{\pi}{V(E_{n-1}+E_{m-1}+\sqrt{E_{n-1}E_{m-1}})}\right)$. We introduce the free $n$-point functions:
\begin{align}
\displaystyle\left\langle\prod_{k=1}^{n}\Phi_{i_{k}j_{k}}\right\rangle_{free}:=\frac{1}{\mathcal{Z}_{free}[0]}\int\mathcal{D}\Phi\hspace{1mm}\Phi_{i_{1}j_{1}}\Phi_{i_{2}j_{2}}\cdots\Phi_{i_{n}j_{n}}\exp\left(-V\mathrm{tr}\left(E\Phi^{2}+\frac{1}{2}M\Phi M\Phi\right)\right).
\end{align}

In particular, the propagator is given as
\begin{align}
\langle\Phi_{ba}\Phi_{dc}\rangle_{free}=&\frac{1}{V}\frac{\delta_{ad}\delta_{bc}}{E_{d-1}+E_{c-1}+\sqrt{E_{c-1}}\sqrt{E_{d-1}}}.\label{L}
\end{align}
The Feynman graph of the propagator (ribbon) is then defined as follows:
\begin{align}
\lower1.3ex\hbox{\propagator}=\langle\Phi_{ba}\Phi_{dc}\rangle_{free}=\frac{1}{V}\frac{\delta_{ad}\delta_{bc}}{E_{c-1}+E_{d-1}+\sqrt{E_{c-1}}\sqrt{E_{d-1}}}.
\end{align}
In this paper, we do not distinguish Feynman graphs from the functions (or operations) corresponding to the Feynman graphs, for simplicity. Next, we consider the case of (\ref{Z[J]}) with the condition $\kappa=0$. (\ref{Z[J]}) can be written as follows:

\begin{align}
\mathcal{Z}[J]=&\int\mathcal{D}\Phi \exp\left(-S_{int}[\Phi]\right)\exp\left(-S_{free}[\Phi]\right)\exp\left(+V\mathrm{tr}J\Phi\right).\label{bbb}
\end{align}
Here $\displaystyle S_{int}=V\mathrm{tr}\left(\frac{\lambda}{4}\Phi^{4}+\sqrt{\lambda}M\Phi^{3}\right)$. Using (\ref{bbb}), as in ordinary field theory, if we consider the n-point function,
\begin{align}
\displaystyle\left\langle\prod_{k=1}^{n}\Phi_{i_{k}j_{k}}\right\rangle:=\frac{1}{\mathcal{Z}[0]}\int\mathcal{D}\Phi\hspace{1mm}\Phi_{i_{1}j_{1}}\Phi_{i_{2}j_{2}}\cdots\Phi_{i_{n}j_{n}}\exp\left(-S_{int}\left[\Phi\right]\right)\exp\left(-S_{free}[\Phi]\right),
\end{align}
then 
\begin{align}
\displaystyle\left\langle\prod_{k=1}^{n}\Phi_{i_{k}j_{k}}\right\rangle:=&\frac{1}{\mathcal{Z}[0]}\int\mathcal{D}\Phi\hspace{1mm}\frac{\partial}{\partial J_{j_{1}i_{1}}}\frac{\partial}{\partial J_{j_{2}i_{2}}}\cdots\frac{\partial}{\partial J_{j_{n}i_{n}}}\exp\left(-S_{int}\left[\frac{1}{V}\frac{\partial}{\partial J}\right]\right)\exp\left(-S_{free}[\Phi]\right)\exp\left(+V\mathrm{tr}J\Phi\right)\notag\\
=&\frac{1}{\mathcal{Z}[0]}\frac{\partial}{\partial J_{j_{1}i_{1}}}\frac{\partial}{\partial J_{j_{2}i_{2}}}\cdots\frac{\partial}{\partial J_{j_{n}i_{n}}}\exp\left(-S_{int}\left[\frac{1}{V}\frac{\partial}{\partial J}\right]\right)\mathcal{Z}_{free}[J].
\end{align}
From this, we also obtain the Feynman rule for interactions, which is as follows. First, we consider the three-point interactions. From $\displaystyle-V\mathrm{tr}\sqrt{\lambda}M\Phi^{3}=-V\sqrt{\lambda}\sum_{k,l,m=1}^{N}\sqrt{E_{k-1}}\Phi_{kl}\Phi_{lm}\Phi_{mk}$, the vertex weight of the three-point interaction is determined:
\begin{align}
\lower5ex\hbox{\vertex}\hspace{2mm}=-V\sqrt{\lambda E_{a-1}}.\label{abkde}
\end{align}
The black dot $v$ corresponds to $\sqrt{E_{a-1}}$. Note that this Feynman rule corresponding interaction does not consider statistical factors. In other words, for all Wick contractions with $\mathrm{tr}\sqrt{\lambda}M\Phi^{3}$, we shall add up all graphs with this weight. In this paper, we use the following notation:


\begin{align}
\sum_{v\in\{\{v_{1},v_{2},v_{3}\}\}}\lower5ex\hbox{\1}\hspace{3mm}:=&\sum_{v\in\{\{i,j,k\}\}}\lower5ex\hbox{\1}\notag\\
:=&\hspace{3mm}\lower5ex\hbox{\2}\hspace{3mm}+\lower5ex\hbox{\3}\hspace{3mm}+\lower5ex\hbox{\4},\label{pas}
\end{align}
where $\{\{v_{1},v_{2},v_{3}\}\}$ means multi set. For example, $\displaystyle\sum_{v\in\{\{a,a,n\}\}}v=a+a+n$. So even if the cases $i=j$ and so on, the definition (\ref{pas}) is not changed.

Next, we consider the four-point interactions. From $\displaystyle-V\mathrm{tr}\frac{\lambda}{4}\Phi^{4}$, the vertex weight of the four-point interaction is obtained:
\begin{align}
\lower5ex\hbox{\vertexs}\hspace{2mm}=-\frac{V\lambda}{4}.
\end{align}
Note that this Feynman rule corresponding to this interaction does not consider statistical factors, too. For all Wick contractions with $\displaystyle\mathrm{tr}\frac{\lambda}{4}\Phi^{4}$, we have to sum all terms with this weight. 

For each loop, we add $\displaystyle\sum_{a=1}^{N}$ to sum over all elements. Note that summation $\displaystyle\sum_{a=1}^{N}$ should be carried out after multiplying $\sqrt{E_{a-1}}$ in (\ref{abkde}) for each black dot in any loop. See the following examples:


\begin{align}
\lower8ex\hbox{\5}=-\frac{\sqrt{\lambda}\sqrt{E_{a-1}}}{V3E_{a-1}}\sum_{n=1}^{N}\frac{1}{E_{a-1}+E_{n-1}+\sqrt{E_{a-1}E_{n-1}}},
\end{align}

\begin{align}
\lower8ex\hbox{\6}=-\frac{\sqrt{\lambda}}{V3E_{a-1}}\sum_{n=1}^{N}\frac{\sqrt{E_{n-1}}}{E_{a-1}+E_{n-1}+\sqrt{E_{a-1}E_{n-1}}},
\end{align}

\begin{align}
\lower8ex\hbox{\7}=-\frac{\lambda}{4V^{2}(E_{a-1}+E_{b-1}+\sqrt{E_{a-1}E_{b-1}})^{2}}\sum_{n=1}^{N}\frac{1}{E_{b-1}+E_{n-1}+\sqrt{E_{b-1}E_{n-1}}}.
\end{align}


\subsection{Cumulant of $\Phi^{3}$-$\Phi^{4}$ Hybrid-Matrix-Model $(\kappa=0)$}

\begin{figure}[htbp]
\begin{center}
\includegraphics[width=120mm]{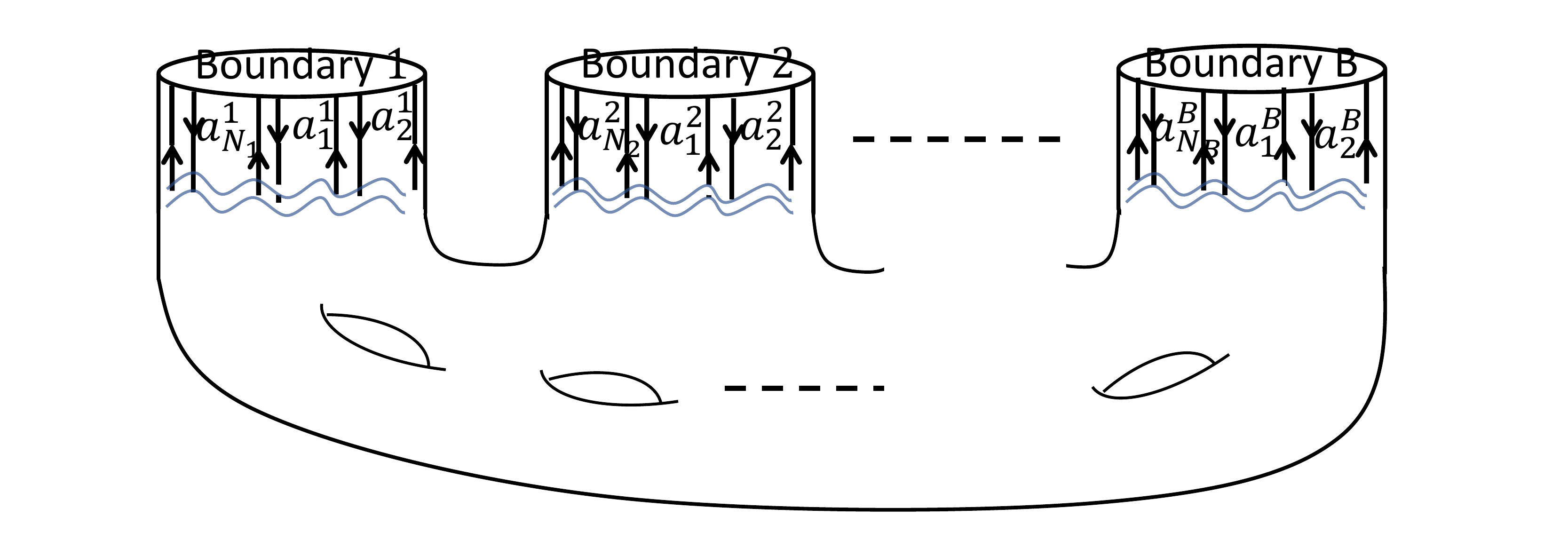}
\caption{The relationship between external ribbons of Feynman diagrams and boundaries as expressed in $G_{|a_{1}^{1}\ldots a_{N_{1}}^{1}|\ldots|a_{1}^{B}\ldots a_{N_{B}}^{B}|}$}\label{qwe}
\end{center}
\end{figure}

Using $\displaystyle\log\frac{ \mathcal{Z}[J]}{\mathcal{Z}[0]}$, the $\displaystyle \sum_{i=1}^{B}N_{i}$-point function $G_{|a_{1}^{1}\ldots a_{N_{1}}^{1}|\ldots|a_{1}^{B}\ldots a_{N_{B}}^{B}|}$ is defined as
\begin{align}
\log\frac{ \mathcal{Z}[J]}{\mathcal{Z}[0]}
:=\sum_{B=1}^\infty \sum_{1\leq N_1 \leq \dots \leq
  N_B}^\infty
\sum_{p_1^1,\dots,p^B_{N_B} =1}^{N} \!\!\!\!
V^{2-B}
&\frac{G_{|p_1^1\dots p_{N_1}^1|\dots|p_1^B\dots p^B_{N_B}|}
}{S_{(N_1,\dots ,N_B)}}
\prod_{\beta=1}^B \frac{\mathbb{J}_{p_1^\beta\dots
    p^\beta_{N_\beta}}}{N_\beta},
\label{logZ}
\end{align}
where $N_{i}$ is the identical valence number for $i=1,\ldots,B$, $\displaystyle\mathbb{J}_{p^{i}_1\dots p^{i}_{N_{i}}}:=\prod_{j=1}^{N_{i}} J_{p^{i}_jp^{i}_{j+1}}$ with $N_{i}+1\equiv 1$, $(N_1,\dots,N_B)=(\underbrace{N'_1,\dots,N'_1}_{\nu_1},\dots,
\underbrace{N'_s,\dots,N'_s}_{\nu_s})$, and $\displaystyle S_{(N_1,\dots ,N_B)}=\prod_{\beta=1}^{s} \nu_{\beta}!$. The $\displaystyle \sum_{i=1}^{B}N_{i}$-point function denoted by $G_{|a_{1}^{1}\ldots a_{N_{1}}^{1}|\ldots|a_{1}^{B}\ldots a_{N_{B}}^{B}|}$ is given by the sum over all Feynman diagrams (ribbon graphs) on Riemann surfaces with $B$-boundaries, and each $|a^{i}_{1}\cdots a^{i}_{N_{i}}|$ corresponds to the Feynman diagrams having $N_{i}$-external ribbons from the $i$-th boundary. (See Figure \ref{qwe}.)

\newpage

\begin{figure}[h]
\begin{center}
\includegraphics[width=50mm]{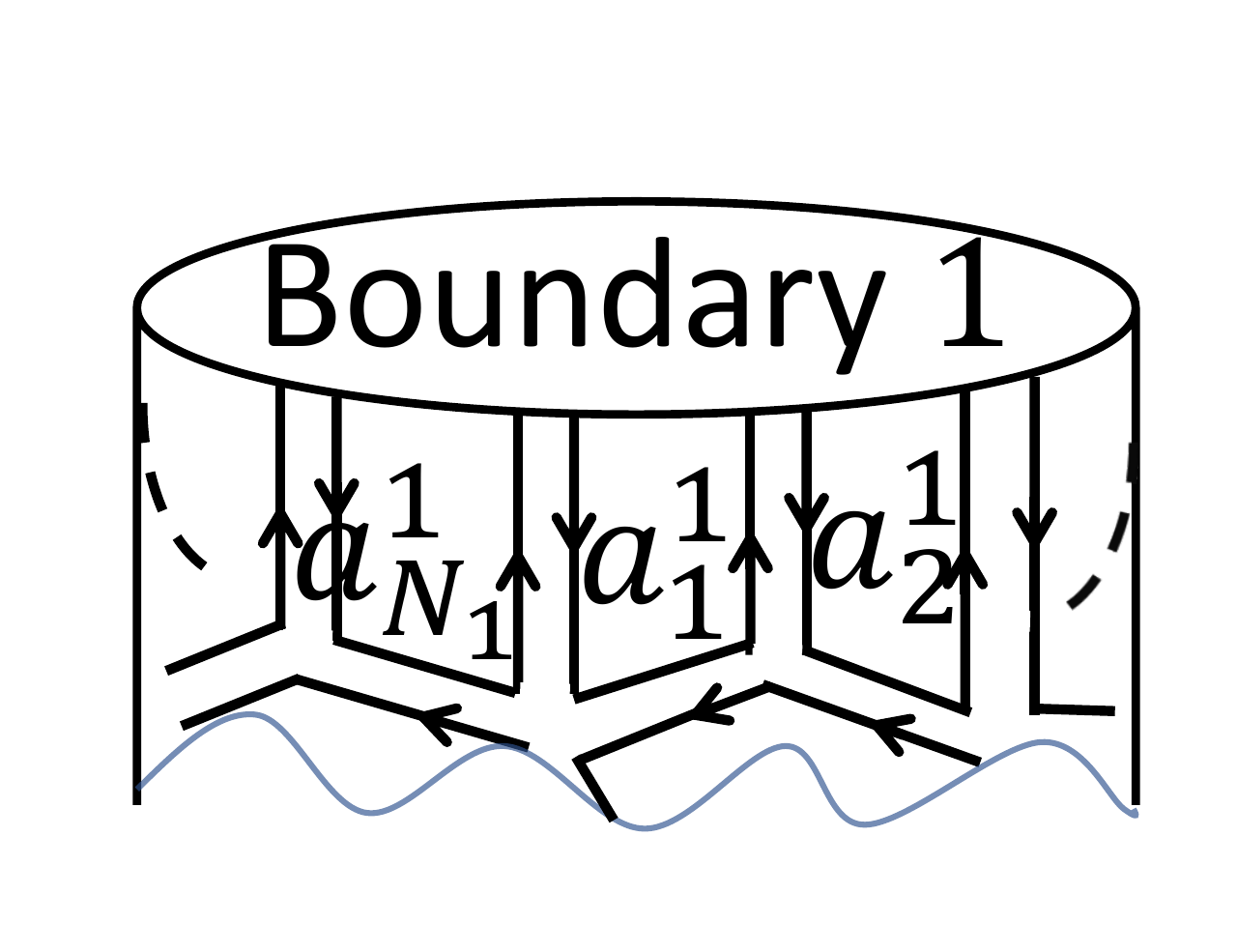}
\caption{The relationship between external ribbons of Feynman diagrams and Boundary $1$ as expressed in $G_{|a_{1}^{1}\ldots a_{N_{1}}^{1}|}$}\label{qwle}
\end{center}
\end{figure}

We give the reason why the Figure \ref{qwe} picture for the Feynman diagram is obtained, in the following. We define a $\displaystyle\sum_{i=1}^{B}N_{i}$-point cumulant which represent contributions of connected Feynman diagrams as
\begin{align}
\langle\Phi_{a_{1}^{1}a_{2}^{1}}\cdots\Phi_{a_{N_{1}}^{1}a_{{1}}^{1}}\Phi_{a_{1}^{2}a_{2}^{2}}\cdots\Phi_{a_{N_{2}}^{2}a_{{1}}^{2}}\cdots\Phi_{a_{1}^{B}a_{2}^{B}}\cdots\Phi_{a_{N_{B}}^{B}a_{{1}}^{B}}\rangle_{c}:=&\left.\frac{1}{V^{N_{1}+\cdots+N_{B}}}\frac{\partial}{\partial J_{a_{2}^{1}a_{1}^{1}}}\cdots\frac{\partial}{\partial J_{a_{1}^{B}a_{N_{B}}^{B}}}\log\mathcal{Z}[J]\right|_{J=0}.\label{kser}
\end{align}
Let us focus on a Feynman diagram with $\mathscr{N}:=\displaystyle\sum_{i=1}^{B}N_{i}$-external ribbons. Let $\Sigma$ be the number of loops contained in the Feynman diagram. Let $k_{3}$ and $k_{4}$ be the number of $V\mathrm{tr}\left(\sqrt{\lambda}M\Phi^{3}\right)$ interactions and the number of $\displaystyle V\mathrm{tr}\left(\frac{\lambda}{4}\Phi^{4}\right)$ interactions in the Feynman diagram, respectively.
The contribution from such Feynman diagram has $V^{k_{3}+k_{4}-\frac{3k_{3}+4k_{4}+\mathscr{N}}{2}}$ since the contribution from vertexes is $V^{k_{3}+k_{4}}$ and the contribution from propagators is $\displaystyle\left(\frac{1}{V}\right)^{\frac{3k_{3}+4k_{4}+\mathscr{N}}{2}}$. Also, the Euler number of a surface with genus $``g"$ and boundaries $``B"$ is $\displaystyle\chi =2-2g-B$. For this Feynman diagram, the corresponding Euler number is given by $\chi=\left(k_{3}+k_{4}+\mathscr{N}\right)-\left(\frac{3k_{3}+4k_{4}+\mathscr{N}}{2}+\mathscr{N}\right)+\left(\mathscr{N}+\sum\right)$. Here $k_{3}+k_{4}+\mathscr{N}$ is the number of vertexes, $\left(\frac{3k_{3}+4k_{4}+\mathscr{N}}{2}+\mathscr{N}\right)$ is the number of the edges, and $\left(\mathscr{N}+\Sigma\right)$ is the number of the faces in the Feynman diagrams. Note that we count one ribbon as one edge, here. Let us see the reason why the last $+\mathscr{N}$ of $\left(\frac{3k_{3}+4k_{4}+\mathscr{N}}{2}+\mathscr{N}\right)$ appears in the number of edge, and $\mathscr{N}$ also represents the number of faces. For example, we see the $i$-th boundary. There are $N_{i}$ faces touching one boundary, since there is $N_{i}$ external ribbons in the Feynman diagram from the term $\displaystyle\prod_{j=1}^{N_{i}}J_{p^{i}_{j}p^{i}_{j+1}}$ with $N_{i}+1\equiv 1$. (See Figure \ref{qwle}.)

\hspace{-5mm}Therefore the number of all surfaces touching the boundary is $\displaystyle\mathscr{N}= \sum_{i=1}^{B}N_{i}$ in this case, and $\mathscr{N}$ edges appear as not ribbons but segments on boundaries. The contribution from the Feynman diagram has $V^{k_{3}+k_{4}-\frac{3k_{3}+4k_{4}+\mathscr{N}}{2}}=V^{\chi-\mathscr{N}-\Sigma}=V^{2-2g-B-\mathscr{N}-\Sigma}$. So, we introduce $G_{|a_{1}^{1}\ldots a_{N_{1}}^{1}|\ldots|a_{1}^{B}\ldots a_{N_{B}}^{B}|}$ for pairwise different $a_{j}^{i}$ $(i=1,\cdots, B,\hspace{2mm} j=1,\cdots, N_{i})$ as in the following equation.
\begin{align}
\langle\Phi_{a_{1}^{1}a_{2}^{1}}\cdots\Phi_{a_{N_{1}}^{1}a_{{1}}^{1}}\Phi_{a_{1}^{2}a_{2}^{2}}\cdots\Phi_{a_{N_{2}}^{2}a_{{1}}^{2}}\cdots\Phi_{a_{1}^{B}a_{2}^{B}}\cdots\Phi_{a_{N_{B}}^{B}a_{{1}}^{B}}\rangle_{c}=&V^{2-\mathscr{N}-B}G_{|a_{1}^{1}\ldots a_{N_{1}}^{1}|\ldots|a_{1}^{B}\ldots a_{N_{B}}^{B}|}.\label{lkjh}
\end{align}
Let us check its consistency with (\ref{logZ}). Note that
\begin{align}
&\frac{1}{V^{\mathscr{N}}}\frac{\partial^{\mathscr{N}}}{\partial J_{a_{2}^{1}a_{1}^{1}}\cdots\partial J_{a_{1}^{B}a_{N_{B}}^{B}}}\sum_{B'=1}^{\infty}\sum_{1\leq\cdots\leq N_{B'}}^{\infty}\sum_{p_1^1,\dots,p^{B'}_{N_{B'}} =1}^{N}\prod_{\beta=1}^{B'} \frac{\mathbb{J}_{p_1^\beta\dots p^\beta_{N_\beta}}}{N_\beta}\left.G_{|p_1^1\dots p_{N_1}^1|\dots|p_1^{B'}\dots p^{B'}_{N_{B'}}|}\right|_{J=0}\notag\\
&=\frac{1}{V^{\mathscr{N}}}G_{|a_{1}^{1}\ldots a_{N_{1}}^{1}|\ldots|a_{1}^{B}\ldots a_{N_{B}}^{B}|}\times S_{(N_1,\dots ,N_B)}.
\end{align}
Then the $\mathscr{N}$-th derivative of the right-hand side of (\ref{logZ}) with respect to $J_{a_{2}^{1}a_{1}^{1}},\cdots,J_{a_{1}^{B}a_{N_{B}}^{B}}$ is given by 
\begin{align}
\frac{1}{V^{\mathscr{N}}}\frac{\partial^{\mathscr{N}}}{\partial J_{a_{2}^{1}a_{1}^{1}}\cdots\partial J_{a_{1}^{B}a_{N_{B}}^{B}}}\left(R.H.S\hspace{1mm}of\hspace{1mm}(\ref{logZ})\right)=&V^{2-\mathscr{N}-B}G_{|a_{1}^{1}\ldots a_{N_{1}}^{1}|\ldots|a_{1}^{B}\ldots a_{N_{B}}^{B}|},
\end{align}
and the corresponding one from the left-hand side of (\ref{logZ}) is given as 

\begin{align}
\frac{1}{V^{\mathscr{N}}}\frac{\partial^{\mathscr{N}}}{\partial J_{a_{2}^{1}a_{1}^{1}}\cdots\partial J_{a_{1}^{B}a_{N_{B}}^{B}}}\left(L.H.S\hspace{1mm}of\hspace{1mm}(\ref{logZ})\right)=&\langle\Phi_{a_{1}^{1}a_{2}^{1}}\cdots\Phi_{a_{N_{1}}^{1}a_{{1}}^{1}}\Phi_{a_{1}^{2}a_{2}^{2}}\cdots\Phi_{a_{N_{2}}^{2}a_{{1}}^{2}}\cdots\Phi_{a_{1}^{B}a_{2}^{B}}\cdots\Phi_{a_{N_{B}}^{B}a_{{1}}^{B}}\rangle_{c}.
\end{align}

Therefore, we found that (\ref{lkjh}) is consistent with (\ref{logZ}) when all $a_{j}^{i}$ are pairwise different. 

If there is no condition that any two indexes do not much, then (\ref{lkjh}) is not necessarily correct. 

\hspace{-5mm}$\langle\Phi_{a_{1}^{1}a_{2}^{1}}\cdots\Phi_{a_{N_{1}}^{1}a_{{1}}^{1}}\Phi_{a_{1}^{2}a_{2}^{2}}\cdots\Phi_{a_{N_{2}}^{2}a_{{1}}^{2}}\cdots\Phi_{a_{1}^{B}a_{2}^{B}}\cdots\Phi_{a_{N_{B}}^{B}a_{{1}}^{B}}\rangle_{c}$ might include contributions from several types of surfaces classified by their boundaries. For example, let us consider $\langle\Phi_{aa}\Phi_{aa}\rangle_{c}$. From (\ref{logZ}), 

\hspace{-5mm}$\displaystyle\langle\Phi_{aa}\Phi_{aa}\rangle_{c}=\frac{1}{V}G_{|aa|}+\frac{1}{V^{2}}G_{|a|a|}$. This means that $\langle\Phi_{aa}\Phi_{aa}\rangle_{c}$ includes contributions from two types of surfaces which are surfaces with one boundary and ones with two boundaries. \bigskip

From these observations, it is concluded that we should prepare a connected oriented surface with $B$ boundaries for drawing each Feynman diagram to calculate $G_{|a_{1}^{1}\ldots a_{N_{1}}^{1}|\ldots|a_{1}^{B}\ldots a_{N_{B}}^{B}|}$. We draw a Feynman diagram with external ribbons with $(a_{1}^{i}a_{2}^{i}),\cdots,(a_{N_{i}}^{i}a_{1}^{i})$ subscripted to each boundary $i$. For any connected segments in a Feynman diagram, both ends are on the same boundary. $G_{|a_{1}^{1}\ldots a_{N_{1}}^{1}|\ldots|a_{1}^{B}\ldots a_{N_{B}}^{B}|}$ is given by the sum over all such Feynman diagrams.

In addition, since it is $V^{k_{3}+k_{4}-\frac{3k_{3}+4k_{4}+\mathscr{N}}{2}}=V^{\chi-\mathscr{N}-\sum}=V^{2-2g-B-\mathscr{N}-\sum}$, we can consider ``genus expansion" of $G_{|a_{1}^{1}\ldots a_{N_{1}}^{1}|\ldots|a_{1}^{B}\ldots a_{N_{B}}^{B}|}$ like \cite{Hock:2019kgb} as
\begin{align}
G_{|a_{1}^{1}\ldots a_{N_{1}}^{1}|\ldots|a_{1}^{B}\ldots a_{N_{B}}^{B}|}=&\sum_{g=0}^{\infty}V^{-2g}G^{(g)}_{|a_{1}^{1}\ldots a_{N_{1}}^{1}|\ldots|a_{1}^{B}\ldots a_{N_{B}}^{B}|}.\label{utgxdsa}
\end{align}

We will discuss contributions  from nontrivial topology surfaces in Section \ref{sec6}.

\subsection{Perturbative Expansion of $1$-Point Function $G_{|1|}$, $2$-Point Function $G_{|21|}$, $2$-Point Function $G_{|2|1|}$\hspace{1mm}($N=2$)}\label{rdx}

In order to familiarize readers with the perturbation calculations for $G_{|a_{1}^{1}\ldots a_{N_{1}}^{1}|\ldots|a_{1}^{B}\ldots a_{N_{B}}^{B}|}$, several specific example calculations are performed in this section. For simple exercises, $N=2$ case is calculated in Subsection \ref{rdx}. In addition, a more simple $N=1$ calculation is included in Appendix \ref{apen1}.

The results obtained in Section \ref{rdx} will be used in Section 5 as a check that the exact solutions are obtained correctly. \bigskip

We calculate the $1$-point function $G_{|1|}$ using perturbative expansion, at first. We compute each term of this expansion by drawing Feynman diagrams on surfaces with one boundary.

\begin{align}
G_{|1|}=&\sum_{n=1}^{2}\sum_{v\in\mathscr{V}_{n}}\lower9ex\hbox{\six}+\mathcal{O}(\lambda\sqrt{\lambda}),
\end{align}
where $\mathscr{V}_{n}=\{\{1,1,n\}\}$. The circle around the Feynman diagram is the boundary. Feynman diagrams and each term of perturbative expansions have a one-to-one correspondence as follows:

\begin{align}
\sum_{n=1}^{2}\sum_{v\in\mathscr{V}_{n}}\lower9ex\hbox{\six}=&-\frac{\sqrt{\lambda}}{3E_{0}V}\sum_{n=1}^{2}\sum_{v\in\{\{1,1,n\}\}}\frac{\sqrt{E_{v-1}}}{E_{n-1}+E_{0}+\sqrt{E_{0}}\sqrt{E_{n-1}}}.
\end{align}
From this, $G_{|1|}$ becomes as follows:

\begin{align}
G_{|1|}=&-\frac{\sqrt{\lambda}}{V}\frac{\sqrt{E_{0}}}{3E_{0}^{2}}-\frac{\sqrt{\lambda}}{V}\frac{\sqrt{E_{0}}}{3E_{0}}\frac{2}{E_{0}+E_{1}+\sqrt{E_{0}}\sqrt{E_{1}}}-\frac{\sqrt{\lambda}}{V}\frac{\sqrt{E_{1}}}{3E_{0}}\frac{1}{E_{0}+E_{1}+\sqrt{E_{0}}\sqrt{E_{1}}}+\mathcal{O}(\lambda\sqrt{\lambda})\label{D}.
\end{align}

Next, we calculate the $2$-point function $G_{|21|}$ on surfaces with one boundary. We compute each term from the expansion of the $2$-point function $G_{|21|}$ by drawing Feynman diagrams.

\begin{align}
G_{|21|}=&V\lower9ex\hbox{\p}+4V\sum_{n=1}^{2}\sum_{(i,j)\in\{(1,2),(2,1)\}}\lower9ex\hbox{\three}\notag\\
&+\frac{V}{2}\sum_{(i,j)\in\{(1,2),(2,1)\}}\sum_{n=1}^{2}\sum_{\omega\in\{\{i,j,n\}\}}\sum_{v\in\{\{i,j,n\}\}}\lower9ex\hbox{\four}\notag\\
&+\frac{V}{2}\times 2\sum_{(i,j)\in\{(1,2),(2,1)\}}\sum_{n=1}^{2}\sum_{\omega\in \{\{i,i,n\}\}}\sum_{v\in\{\{j,i,i\}\}}\lower9ex\hbox{\five}+\mathcal{O}(\lambda^{2}).\label{per}
\end{align}
The first diagram in (\ref{per}) is given as follows:

\begin{align}
V\lower9ex\hbox{\p}=&\frac{1}{E_{0}+E_{1}+\sqrt{E_{0}}\sqrt{E_{1}}}.\label{abcde}
\end{align}
The second term in (\ref{per}) is written as follows:

\begin{align}
&4V\sum_{n=1}^{2}\sum_{(i,j)\in\{(1,2),(2,1)\}}\lower9ex\hbox{\three}\notag\\
=&-\frac{4\lambda}{4V}\sum_{i,j=1,2\hspace{1mm}i\neq j}\sum_{n=1}^{2}\left(\frac{1}{E_{i-1}+E_{j-1}+\sqrt{E_{i-1}}\sqrt{E_{j-1}}}\right)^{2}\left(\frac{1}{E_{i-1}+E_{n-1}+\sqrt{E_{i-1}}\sqrt{E_{n-1}}}\right).
\end{align}
The third term in (\ref{per}) is expressed as follows:

\begin{align}
&\frac{V}{2}\sum_{(i,j)\in\{(1,2),(2,1)\}}\sum_{n=1}^{2}\sum_{\omega\in\{\{i,j,n\}\}}\sum_{v\in\{\{i,j,n\}\}}\lower9ex\hbox{\four}\notag\\
=&+\frac{\lambda}{2V}\sum_{(i,j)\in\{(1,2),(2,1)\}}\sum_{n=1}^{2}\sum_{\omega\in\{\{i,j,n\}\}}\sum_{v\in\{\{i,j,n\}\}}\left(\frac{1}{E_{i-1}+E_{j-1}+\sqrt{E_{i-1}}\sqrt{E_{j-1}}}\right)^{2}\notag\\
&\times\sqrt{E_{v-1}}\sqrt{E_{w-1}}\left(\frac{1}{E_{i-1}+E_{n-1}+\sqrt{E_{i-1}}\sqrt{E_{n-1}}}\right)\left(\frac{1}{E_{j-1}+E_{n-1}+\sqrt{E_{j-1}}\sqrt{E_{n-1}}}\right).
\end{align}
The fourth term in (\ref{per}) is given as follows:

\begin{align}
&2\times\frac{V}{2}\sum_{(i,j)\in\{(1,2),(2,1)\}}\sum_{n=1}^{2}\sum_{\omega\in\{\{i,i,n\}\}}\sum_{v\in\{\{j,i,i\}\}}\lower9ex\hbox{\five}\notag\\
=&\frac{\lambda}{V}\sum_{(i,j)\in\{(1,2),(2,1)\}}\sum_{n=1}^{2}\sum_{\omega\in\{\{i,i,n\}\}}\sum_{v\in\{\{j,i,i\}\}}\left(\frac{1}{3E_{i-1}}\right)\left(\frac{1}{E_{i-1}+E_{j-1}+\sqrt{E_{i-1}}\sqrt{E_{j-1}}}\right)^{2}\sqrt{E_{v-1}}\notag\\
&\times\sqrt{E_{w-1}}\left(\frac{1}{E_{i-1}+E_{n-1}+\sqrt{E_{i-1}}\sqrt{E_{n-1}}}\right).\label{fghij}
\end{align}
From (\ref{abcde})-(\ref{fghij}), $G_{|21|}$ is given as follows:

\begin{align}
G_{|21|}=&\frac{1}{E_{0}+E_{1}+\sqrt{E_{0}}\sqrt{E_{1}}}+\frac{\lambda}{3VE_{0}(E_{0}+E_{1}+\sqrt{E_{0}}\sqrt{E_{1}})^{2}}+\frac{\lambda}{3VE_{1}(E_{0}+E_{1}+\sqrt{E_{0}}\sqrt{E_{1}})^{2}}\notag\\
&+\frac{8\sqrt{E_{0}}\sqrt{E_{1}}\lambda}{3VE_{1}(E_{0}+E_{1}+\sqrt{E_{0}}\sqrt{E_{1}})^{3}}+\frac{8\sqrt{E_{0}}\sqrt{E_{1}}\lambda}{3VE_{0}(E_{0}+E_{1}+\sqrt{E_{0}}\sqrt{E_{1}})^{3}}+\frac{\sqrt{E_{0}}\sqrt{E_{1}}\lambda}{3VE_{1}^{2}(E_{0}+E_{1}+\sqrt{E_{0}}\sqrt{E_{1}})^{2}}\notag\\
&+\frac{\sqrt{E_{0}}\sqrt{E_{1}}\lambda}{3VE_{0}^{2}(E_{0}+E_{1}+\sqrt{E_{0}}\sqrt{E_{1}})^{2}}+\frac{2\lambda E_{0}}{3VE_{1}(E_{0}+E_{1}+\sqrt{E_{0}}\sqrt{E_{1}})^{3}}+\frac{2\lambda E_{1}}{3VE_{0}(E_{0}+E_{1}+\sqrt{E_{0}}\sqrt{E_{1}})^{3}}\notag\\
&+\frac{10\lambda}{3V(E_{0}+E_{1}+\sqrt{E_{0}}\sqrt{E_{1}})^{3}}+\mathcal{O}(\lambda^{2})\label{E}.
\end{align}

Next, let us calculate the $2$-point function $G_{|2|1|}$ that has two boundaries using perturbative expansion. We compute each term of this expansion by drawing Feynman diagrams on surfaces with two boundaries.

\begin{align}
G_{|2|1|}=&4V^{2}\lower9ex\hbox{\eight}+V^{2}\sum_{\omega\in\{\{1,1,2\}\}}\sum_{v\in\{\{2,2,1\}\}}\lower9ex\hbox{\nine}+\mathcal{O}(\lambda^{2})\label{pertur}
\end{align}
There is a one-to-one correspondence between the Feynman diagram and each term in the perturbation expansion. The first diagram in (\ref{pertur}) is given as follows:

\begin{align}
4V^{2}\lower9ex\hbox{\eight}=&-\frac{\lambda}{9E_{0}E_{1}(E_{0}+E_{1}+\sqrt{E_{0}}\sqrt{E_{1}})}\label{adgj}
\end{align}
The second term in (\ref{pertur}) is obtained as follows:

\begin{align}
&V^{2}\sum_{\omega\in\{\{1,1,2\}\}}\sum_{v\in\{\{2,2,1\}\}}\lower9ex\hbox{\nine}\notag\\
=&\frac{5\sqrt{E_{0}}\sqrt{E_{1}}\lambda}{9E_{0}E_{1}(E_{0}+E_{1}+\sqrt{E_{0}}\sqrt{E_{1}})^{2}}+\frac{2\lambda}{9E_{0}(E_{0}+E_{1}+\sqrt{E_{0}}\sqrt{E_{1}})^{2}}+\frac{2\lambda}{9E_{1}(E_{0}+E_{1}+\sqrt{E_{0}}\sqrt{E_{1}})^{2}}\label{lhda}
\end{align}
From (\ref{adgj})-(\ref{lhda}), $G_{|2|1|}$ becomes as follows:

\begin{align}
G_{|2|1|}=&\frac{4\sqrt{E_{0}}\sqrt{E_{1}}\lambda}{9E_{0}E_{1}(E_{0}+E_{1}+\sqrt{E_{0}}\sqrt{E_{1}})^{2}}+\frac{\lambda}{9E_{0}(E_{0}+E_{1}+\sqrt{E_{0}}\sqrt{E_{1}})^{2}}+\frac{\lambda}{9E_{1}(E_{0}+E_{1}+\sqrt{E_{0}}\sqrt{E_{1}})^{2}}+\mathcal{O}(\lambda^{2})\label{1}.
\end{align}

\subsection{Perturbative Expansion of $1$-Point Function $G_{|a|}$, $2$-Point Function $G_{|ab|}$, $2$-Point Function $G_{|a|b|}$}

In Subsection \ref{rdx}, we carried out the calculation for $G_{|1|}$, $G_{|21|}$, $G_{|2|1|}$ perturbatively in the $N=2$ case. In this section, we summarize the similar results for arbitrary $N$ case.\bigskip

At first, we calculate the connected 1-point function $G_{|a|}$ using perturbative expansion. 
\begin{align}
G_{|a|}=&\left.\frac{1}{V}\frac{\partial\log\mathcal{Z}[J]}{\partial J_{aa}}\right|_{J=0}\notag\\
=&\frac{1}{\mathcal{Z}[0]}\int\mathcal{D}\Phi\hspace{1mm}\Phi_{aa}\left(\sum_{k=0}^{\infty}\frac{(-V\lambda)^{k}}{k!4^{k}}\left(\sum_{n_{1},n_{2},n_{3},n_{4}=1}^{N}\Phi_{n_{1}n_{2}}\Phi_{n_{2}n_{3}}\Phi_{n_{3}n_{4}}\Phi_{n_{4}n_{1}}\right)^{k}\right)\notag\\
&\times\left(\sum_{l=0}^{\infty}\frac{(-V\sqrt{\lambda})^{l}}{l!}\left(\sum_{m_{1},m_{3},m_{4}=1}^{N}\sqrt{E_{m_{1}-1}}\Phi_{m_{1}m_{3}}\Phi_{m_{3}m_{4}}\Phi_{m_{4}m_{1}}\right)^{l}\right)\exp\left(-V\mathrm{tr}\left(E\Phi^{2}+\frac{\lambda}{2}M\Phi M\Phi\right)\right)\notag\\
=&-V\sqrt{\lambda}\frac{\mathcal{Z}_{free}[0]}{Z[0]}\sum_{m_{1},m_{3},m_{4}=1}^{N}\sqrt{E_{m_{1}-1}}\langle\Phi_{aa}\Phi_{m_{1}m_{3}}\Phi_{m_{3}m_{4}}\Phi_{m_{4}m_{1}}\rangle_{free}+\mathcal{O}(\lambda\sqrt{\lambda})\label{A}.
\end{align}
We compute each term of this expansion by drawing perturbative expansions of the 1-point function $G_{|a|}$ in Feynman diagrams. The corresponding Feynman diagrams are drawn on oriented surfaces with one boundary with one connected ribbon graph inserted with an index ``$a$".

\begin{align}
G_{|a|}=&\sum_{n=1}^{N}\sum_{v\in\mathscr{V}}\lower9ex\hbox{\one}+\mathcal{O}(\lambda\sqrt{\lambda})\notag\\
=&-\frac{\sqrt{\lambda}}{3E_{a-1}V}\sum_{n=1}^{N}\sum_{v\in\{\{a,a,n\}\}}\frac{\sqrt{E_{v-1}}}{E_{n-1}+E_{a-1}+\sqrt{E_{a-1}E_{n-1}}}+\mathcal{O}(\lambda\sqrt{\lambda})\label{cccccc}.
\end{align}
Here $\mathscr{V}=\{\{v_{1},v_{2},v_{3}\}\}=\{\{a,a,n\}\}$. The circle around the Feynman diagram in (\ref{cccccc}) represents the boundary. So the external lines (Ribbon) grow out of the circle.\bigskip

Next, we calculate the 2-point function $G_{|ab|}(a\neq b)$ using perturbative expansion. 

\begin{align}
G_{|ab|}&=\left.\frac{1}{V}\frac{\partial^{2}\log\mathcal{Z}[J]}{\partial J_{ab}\partial J_{ba}}\right|_{J=0}\notag\\
&=\frac{V}{\mathcal{Z}[0]}\int\mathcal{D}\Phi\hspace{1mm}\Phi_{ab}\Phi_{ba}\left(\sum_{k=0}^{\infty}\frac{(-V\lambda)^{k}}{k!4^{k}}\left(\sum_{n_{1},n_{2},n_{3},n_{4}=1}^{N}\Phi_{n_{1}n_{2}}\Phi_{n_{2}n_{3}}\Phi_{n_{3}n_{4}}\Phi_{n_{4}n_{1}}\right)^{k}\right)\notag\\
&\quad\times\left(\sum_{l=0}^{\infty}\frac{(-V\sqrt{\lambda})^{l}}{l!}\left(\sum_{m_{1},m_{3},m_{4}=1}^{N}\sqrt{E_{m_{1}-1}}\Phi_{m_{1}m_{3}}\Phi_{m_{3}m_{4}}\Phi_{m_{4}m_{1}}\right)^{l}\right)\exp\left(-V\mathrm{tr}\left(E\Phi^{2}+\frac{\lambda}{2}M\Phi M\Phi\right)\right)\notag\\
&=\frac{V\mathcal{Z}_{free}[0]}{Z[0]}\Biggl\{\langle\Phi_{ab}\Phi_{ba}\rangle_{free}+\frac{(-V\lambda)^{1}}{1!4^{1}}\sum_{n_{1},n_{2},n_{3},n_{4}=1}^{N}\langle\Phi_{ab}\Phi_{ba}\Phi_{n_{1}n_{2}}\Phi_{n_{2}n_{3}}\Phi_{n_{3}n_{4}}\Phi_{n_{4}n_{1}}\rangle_{free}\notag\\
&\quad+\frac{(-V\sqrt{\lambda})^{2}}{2!}\sum_{m_{1},m_{3},m_{4},n_{1},n_{3},n_{4}=1}^{N}\sqrt{E_{m_{1}-1}}\sqrt{E_{n_{1}-1}}\langle\Phi_{ab}\Phi_{ba}\Phi_{m_{1}m_{3}}\Phi_{m_{3}m_{4}}\Phi_{m_{4}m_{1}}\Phi_{n_{1}n_{3}}\Phi_{n_{3}n_{4}}\Phi_{n_{4}n_{1}}\rangle_{free}\Biggl\}\notag\\
&\quad+\mathcal{O}(\lambda^{2})\label{osef}.
\end{align}
We compute each term of (\ref{osef}) by drawing Feynman diagrams. The corresponding Feynman diagram should be a connected graph with two ribbon graphs with indices $a$ and $b$ inserted at two points from a single boundary:
\begin{align}
G_{|ab|}=&V\lower9ex\hbox{\two}+4V\sum_{(i,j)\in\{(a,b),(b,a)\}}\sum_{n=1}^{N}\lower9ex\hbox{\three}\notag\\
&+\frac{V}{2!}\sum_{(i,j)\in\{(a,b),(b,a)\}}\sum_{n=1}^{N}\sum_{\omega\in \mathscr{W}}\sum_{v\in\mathscr{V}}\lower9ex\hbox{\four}+2\times\frac{V}{2!}\sum_{(i,j)\in\{(a,b),(b,a)\}}\sum_{n=1}^{N}\sum_{\omega\in \mathscr{W}}\sum_{v\in\mathscr{V}}\lower9ex\hbox{\five}\notag\\
&+\mathcal{O}(\lambda^{2})\label{pgfewq}\notag\\
\end{align}
Note that $``4"$ in the $2$nd term and $``2"$ in the $4$th term are the statistical factors.

\begin{align}
G_{|ab|}=&\frac{1}{E_{a-1}+E_{b-1}+\sqrt{E_{a-1}}\sqrt{E_{b-1}}}\notag\\
&-\frac{\lambda}{V}\sum_{(i,j)\in\{(a,b),(b,a)\}}\sum_{n=1}^{N}\left(\frac{1}{E_{i-1}+E_{j-1}+\sqrt{E_{i-1}}\sqrt{E_{j-1}}}\right)^{2}\left(\frac{1}{E_{i-1}+E_{n-1}+\sqrt{E_{i-1}}\sqrt{E_{n-1}}}\right)\notag\\
&+\frac{\lambda}{2V}\sum_{(i,j)\in\{(a,b),(b,a)\}}\sum_{n=1}^{N}\sum_{\omega\in\{\{i,j,n\}\}}\sum_{v\in\{\{i,j,n\}\}}\left(\frac{1}{E_{i-1}+E_{j-1}+\sqrt{E_{i-1}}\sqrt{E_{j-1}}}\right)^{2}\notag\\
&\times\sqrt{E_{v-1}}\sqrt{E_{w-1}}\left(\frac{1}{E_{i-1}+E_{n-1}+\sqrt{E_{i-1}}\sqrt{E_{n-1}}}\right)\left(\frac{1}{E_{j-1}+E_{n-1}+\sqrt{E_{j-1}}\sqrt{E_{n-1}}}\right)\notag\\
&+2\times\frac{\lambda}{2V}\sum_{(i,j)\in\{(a,b),(b,a)\}}\sum_{n=1}^{N}\sum_{\omega\in\{\{i,i,n\}\}}\sum_{v\in\{\{i,i,j\}\}}\left(\frac{1}{3E_{i-1}}\right)\left(\frac{1}{E_{i-1}+E_{j-1}+\sqrt{E_{i-1}}\sqrt{E_{j-1}}}\right)^{2}\sqrt{E_{v-1}}\notag\\
&\times\sqrt{E_{w-1}}\left(\frac{1}{E_{i-1}+E_{n-1}+\sqrt{E_{i-1}}\sqrt{E_{n-1}}}\right)+\mathcal{O}(\lambda^{2}).\label{ddddds}
\end{align}

\begin{rem}

We give reasons why a nonplanar Feynman diagram does not appear in (\ref{pgfewq}). You might think that the Wick expansion of $\displaystyle\sum_{n_{1},n_{2},n_{3},n_{4}=1}^{N}\langle\Phi_{ab}\Phi_{ba}\Phi_{n_{1}n_{2}}\Phi_{n_{2}n_{3}}\Phi_{n_{3}n_{4}}\Phi_{n_{4}n_{1}}\rangle$ would yield a term like $\displaystyle\sum_{n_{1},n_{2},n_{3},n_{4}=1}^{N}\langle\Phi_{ab}\Phi_{n_{1}n_{2}}\rangle\langle\Phi_{ba}\Phi_{n_{3}n_{4}}\rangle\langle\Phi_{n_{2}n_{3}}\Phi_{n_{4}n_{1}}\rangle$. In the corresponding Feynman diagram, it is like (\ref{dfk}), but since $b$ and $a$ are connected by a line, $\delta_{ab}$ is generated and $0$ is obtained from $a\neq b$.

\begin{align}
\lower8ex\hbox{\twelve}\lower-1ex\hbox{$\displaystyle=0$}\label{dfk}
\end{align}

\end{rem}
\bigskip
As the third example, we calculate the 2-point function $G_{|a|b|}(a\neq b)$ using perturbative expansion. In this case, corresponding Feynman diagrams are drawn on surfaces with two boundaries like a cylinder. The external lines $\lower2ex\hbox{\thr}$ and $\lower2ex\hbox{\eleven}$ are grown out from different boundaries, respectively, and the two boundaries are not connected by any line in any non-zero Feynman diagram.

\begin{align}
G_{|a|b|}=&\left.\frac{\partial^{2}\log\mathcal{Z}[J]}{\partial J_{aa}\partial J_{bb}}\right|_{J=0}\notag\\
&=-\frac{1}{\mathcal{Z}[0]^{2}}\left.\frac{\partial\mathcal{Z}[J]}{\partial J_{aa}}\right|_{J=0}\left.\frac{\partial\mathcal{Z}[J]}{\partial J_{bb}}\right|_{J=0}+\left.\frac{1}{\mathcal{Z}[0]}\frac{\partial^{2}\mathcal{Z}[J]}{\partial J_{aa}\partial J_{bb}}\right|_{J=0}\notag\\
&=\frac{-V^{2}}{\mathcal{Z}[0]^{2}}\int\mathcal{D}\Phi\hspace{1mm}\Phi_{aa}\left(\sum_{k=0}^{\infty}\frac{(-V\lambda)^{k}}{k!4^{k}}\left(\sum_{n_{1},n_{2},n_{3},n_{4}=1}^{N}\Phi_{n_{1}n_{2}}\Phi_{n_{2}n_{3}}\Phi_{n_{3}n_{4}}\Phi_{n_{4}n_{1}}\right)^{k}\right)\notag\\
&\quad\times\left(\sum_{l=0}^{\infty}\frac{(-V\sqrt{\lambda})^{l}}{l!}\left(\sum_{m_{1},m_{3},m_{4}=1}^{N}\sqrt{E_{m_{1}-1}}\Phi_{m_{1}m_{3}}\Phi_{m_{3}m_{4}}\Phi_{m_{4}m_{1}}\right)^{l}\right)\exp\left(-V\mathrm{tr}\left(E\Phi^{2}+\frac{\lambda}{2}M\Phi M\Phi\right)\right)\notag\\
&\quad\times\int\mathcal{D}\Phi\hspace{1mm}\Phi_{bb}\left(\sum_{k=0}^{\infty}\frac{(-V\lambda)^{k}}{k!4^{k}}\left(\sum_{n_{1},n_{2},n_{3},n_{4}=1}^{N}\Phi_{n_{1}n_{2}}\Phi_{n_{2}n_{3}}\Phi_{n_{3}n_{4}}\Phi_{n_{4}n_{1}}\right)^{k}\right)\notag\\
&\quad\times\left(\sum_{l=0}^{\infty}\frac{(-V\sqrt{\lambda})^{l}}{l!}\left(\sum_{m_{1},m_{3},m_{4}=1}^{N}\sqrt{E_{m_{1}-1}}\Phi_{m_{1}m_{3}}\Phi_{m_{3}m_{4}}\Phi_{m_{4}m_{1}}\right)^{l}\right)\exp\left(-V\mathrm{tr}\left(E\Phi^{2}+\frac{\lambda}{2}M\Phi M\Phi\right)\right)\notag\\
&\quad+\frac{V^{2}}{\mathcal{Z}[0]}\int\mathcal{D}\Phi\hspace{1mm}\Phi_{aa}\Phi_{bb}\left(\sum_{k=0}^{\infty}\frac{(-V\lambda)^{k}}{k!4^{k}}\left(\sum_{n_{1},n_{2},n_{3},n_{4}=1}^{N}\Phi_{n_{1}n_{2}}\Phi_{n_{2}n_{3}}\Phi_{n_{3}n_{4}}\Phi_{n_{4}n_{1}}\right)^{k}\right)\notag\\
&\quad\times\left(\sum_{l=0}^{\infty}\frac{(-V\sqrt{\lambda})^{l}}{l!}\left(\sum_{m_{1},m_{3},m_{4}=1}^{N}\sqrt{E_{m_{1}-1}}\Phi_{m_{1}m_{3}}\Phi_{m_{3}m_{4}}\Phi_{m_{4}m_{1}}\right)^{l}\right)\exp\left(-V\mathrm{tr}\left(E\Phi^{2}+\frac{\lambda}{2}M\Phi M\Phi\right)\right).
\end{align}
So, we estimate the following.
\begin{align}
G_{|a|b|}=&\frac{-V^{2}\mathcal{Z}_{free}[0]^{2}}{\mathcal{Z}[0]^{2}}\times\left((-V\sqrt{\lambda})^{1}\sum_{m_{1},m_{3},m_{4}=1}^{N}\sqrt{E_{m_{1}-1}}\langle\Phi_{aa}\Phi_{m_{1}m_{3}}\Phi_{m_{3}m_{4}}\Phi_{m_{4}m_{1}}\rangle_{free}\right)\notag\\
&\quad\times\left((-V\sqrt{\lambda})^{1}\sum_{m_{1},m_{3},m_{4}=1}^{N}\sqrt{E_{m_{1}-1}}\langle\Phi_{bb}\Phi_{m_{1}m_{3}}\Phi_{m_{3}m_{4}}\Phi_{m_{4}m_{1}}\rangle_{free}\right)\notag\\
&\quad+\frac{V^{2}\mathcal{Z}_{free}[0]}{Z[0]}\Biggl\{\langle\Phi_{aa}\Phi_{bb}\rangle_{free}+\frac{(-V\lambda)^{1}}{1!4^{1}}\sum_{n_{1},n_{2},n_{3},n_{4}=1}^{N}\langle\Phi_{aa}\Phi_{bb}\Phi_{n_{1}n_{2}}\Phi_{n_{2}n_{3}}\Phi_{n_{3}n_{4}}\Phi_{n_{4}n_{1}}\rangle_{free}\notag\\
&\quad+\frac{(-V\sqrt{\lambda})^{2}}{2!}\sum_{m_{1},m_{3},m_{4},n_{1},n_{3},n_{4}=1}^{N}\sqrt{E_{m_{1}-1}}\sqrt{E_{n_{1}-1}}\langle\Phi_{aa}\Phi_{bb}\Phi_{m_{1}m_{3}}\Phi_{m_{3}m_{4}}\Phi_{m_{4}m_{1}}\Phi_{n_{1}n_{3}}\Phi_{n_{3}n_{4}}\Phi_{n_{4}n_{1}}\rangle_{free}\Biggl\}\notag\\
&\quad+\mathcal{O}(\lambda^{2})\label{H}.
\end{align}
We compute each term of this expansion by drawing Feynman diagrams.

\begin{align}
G_{|a|b|}=&4V^{2}\lower9ex\hbox{\ten}+V^{2}\sum_{\omega\in\mathscr{W}}\sum_{v\in\mathscr{V}}\lower9ex\hbox{\seven}+\mathcal{O}(\lambda^{2})\notag\\
=&-\frac{\lambda}{9E_{a-1}E_{b-1}\left(E_{a-1}+E_{b-1}+\sqrt{E_{a-1}}\sqrt{E_{b-1}}\right)}+\sum_{\omega\in\mathscr{W}}\sum_{v\in\mathscr{V}}\frac{\sqrt{E_{v-1}}\sqrt{E_{w-1}}\lambda}{9E_{a-1}E_{b-1}\left(E_{a-1}+E_{b-1}+\sqrt{E_{a-1}}\sqrt{E_{b-1}}\right)^{2}}\notag\\
&+\mathcal{O}(\lambda^{2}),
\end{align}
where $\mathscr{V}=\{\{a,a,b\}\}$ and $\mathscr{W}=\{\{b,b,a\}\}$. More explicitly, this is rewritten as 

\begin{align}
G_{|a|b|}=&-\frac{\lambda}{9E_{a-1}E_{b-1}(E_{a-1}+E_{b-1}+\sqrt{E_{a-1}}\sqrt{E_{b-1}})}+\frac{5\sqrt{E_{a-1}}\sqrt{E_{b-1}}\lambda}{9E_{a-1}E_{b-1}(E_{a-1}+E_{b-1}+\sqrt{E_{a-1}}\sqrt{E_{b-1}})^{2}}\notag\\
&+\frac{2\lambda}{9E_{a-1}(E_{a-1}+E_{b-1}+\sqrt{E_{a-1}}\sqrt{E_{b-1}})^{2}}+\frac{2\lambda}{9E_{b-1}(E_{a-1}+E_{b-1}+\sqrt{E_{a-1}}\sqrt{E_{b-1}})^{2}}+\mathcal{O}(\lambda^{2}).
\end{align}


\section{Exact Calculation of Partition Function $\mathcal{Z}[J]$}\label{sec3}
In this section, the calculation of the partition function\footnote{For the case with $J=0$ and $\kappa=0$, the partition function of this model derives a higher KdV hierarchy. See for example\cite{Adler}\cite{Itzykson:1992ya}\cite{Kontsevich:1992ti}\cite{Witten}.} is carried out rigorously for any $N$. The flow of computations is similar to that of \cite{Kanomata:2022pdo}. \bigskip

We introduce a new variable $X$ by $\displaystyle\Phi=X-\frac{1}{\sqrt{\lambda}}M$. Here $X=(X_{mn})$ is a Hermitian matrix, too. We do a change of variables of the integral measure $\mathcal{D}\Phi$ as $\displaystyle d\Phi_{ij}=\sum_{m,n=1}^{N}\frac{\partial\Phi_{ij}}{\partial X_{mn}}dX_{mn}=dX_{ij}$. Then $\mathcal{Z}[J]$ is given as
\begin{align}
\mathcal{Z}[J]=&\int \mathcal{D}\Phi \exp\left(-V\mathrm{tr}\left(E\Phi^{2}+\kappa\Phi+\frac{\lambda}{4}\Phi^{4}+\sqrt{\lambda}M\Phi^{3}+\frac{1}{2}M\Phi M\Phi\right)\right)\exp\left(V\mathrm{tr}\left(J\Phi\right)\right)\notag\\
=&\exp\left(-V\mathrm{tr}\left(\frac{3}{4\lambda}M^{3}-\frac{\kappa}{\sqrt{\lambda}}I+\frac{1}{\sqrt{\lambda}}J\right)M\right)\notag\\
&\int\mathcal{D}X\exp\left(-\frac{\lambda V}{4}\mathrm{tr}(X^4)\right)\exp\left(V\mathrm{tr}\left\{\left(\frac{1}{\sqrt{\lambda}}M^{3}-\kappa I+J\right)X\right\}\right).\label{XY}
\end{align}
Here $I$ is the unit matrix. Note that 
\begin{align*}
\mathcal{D}X=&\left(\prod_{i=1}^{N}dx_{i}\right)\left(\prod_{1\leq k<l\leq N}(x_{l}-x_{k})^{2}\right)dU,
\end{align*}
where $x_{i}$ is the eigenvalues of $X$ for $i=1,\cdots,N$, $dU$ is the Haar probability measure of the unitary group $U(N)$, and $U$ is the unitary matrix which diagonalize $X$\cite{Eynard:2015aea}. Then (\ref{XY}) can be rewritten as the following:
\begin{align}
\mathcal{Z}[J]=&\exp\left(-V\mathrm{tr}\left(\frac{3}{4\lambda}M^{3}-\frac{\kappa}{\sqrt{\lambda}}I+\frac{1}{\sqrt{\lambda}}J\right)M\right)\notag\\
&\int\left(\prod_{i=1}^{N}dx_{i}\exp\left(-\frac{\lambda V}{4}x^{4}_{i}\right)\right)\left(\prod_{1\leq k<l\leq N}(x_{l}-x_{k})^{2}\right)\notag\\
&\int_{U(N)}dU\exp\left(V\mathrm{tr}\left\{\left(\frac{1}{\sqrt{\lambda}}M^{3}-\kappa I+J\right)U\widetilde{X}U^{*}\right\}\right),\label{YX}
\end{align}
where $\widetilde{X}$ is the diagonal matrix $\widetilde{X}=U^{*}XU$. We use the following formula.\bigskip

The Harish-Chandra-Itzykson-Zuber integral \cite{Itzykson:1979fi},\cite{T.Tao},\cite{Zinn-Justin:2002rai} for the unitary group $U(n)$ is
\begin{align}
\int_{U(n)}\exp\left(t\mathrm{tr}\left(AUBU^{*}\right)\right)dU=&c_{n}\frac{\displaystyle\det_{1\leq i,j\leq n}\left(\exp\left(t\lambda_{i}(A)\lambda_{j}(B)\right)\right)}{t^{\frac{(n^{2}-n)}{2}}\displaystyle\Delta(\lambda(A))\displaystyle\Delta(\lambda(B))}.\label{Itzykson}
\end{align}
Here $A=(A_{ij})$, and $B=(B_{ij})$ are some Hermitian matrices whose eigenvalues denoted by $\lambda_{i}(A)$ and $\lambda_{i}(B)$ $(i=1,\cdots,n)$, respectively. $t$ is the non-zero complex parameter, $\displaystyle\Delta(\lambda(A)):=\prod_{1\leq i<j\leq n}(\lambda_{j}(A)-\lambda_{i}(A))$ is the Vandermonde determinant, and $\displaystyle c_{n}:=\left(\prod_{i=1}^{n-1}i!\right)\times\pi^{\frac{n(n-1)}{2}}$ is the constant. $\left(\exp\left(t\lambda_{i}(A)\lambda_{j}(B)\right)\right)$ is the $n\times n$ matrix with the $i$-th row and the $j$-th column being $\exp\left(t\lambda_{i}(A)\lambda_{j}(B)\right)$.

Applying the Harish-Chandra-Itzykson-Zuber integral (\ref{Itzykson}) to $\displaystyle\int dU\exp\left(V\mathrm{tr}\left\{\left(\frac{1}{\sqrt{\lambda}}M^{3}-\kappa I+J\right)U\widetilde{X}U^{*}\right\}\right)$ in (\ref{YX}), the result is
\begin{align}
\displaystyle\int_{U(N)}dU\exp\left(V\mathrm{tr}\left\{\left(\frac{1}{\sqrt{\lambda}}M^{3}-\kappa I+J\right)U\widetilde{X}U^{*}\right\}\right)=&\frac{C}{N!}\frac{\displaystyle\det_{1\leq i,j\leq N}\exp\left(Vx_{i}s_{j}\right)}{\displaystyle\prod_{i<j}(x_{j}-x_{i})\prod_{i<j}(s_{j}-s_{i})},
\end{align}
where $s_{t}$ is the eigenvalues of the matrix $\displaystyle\frac{1}{\sqrt{\lambda}}M^{3}-\kappa I+J$ for $t=1,\cdots,N$ and $\displaystyle C=\left(\prod_{p=1}^{N}p!\right)\times\left(\frac{\pi}{V}\right)^{\frac{N(N-1)}{2}}$. $\left(\exp\left(Vx_{i}s_{j}\right)\right)$ denotes the $N\times N$ matrix with the $i$-th row and the $j$-th column being $\exp\left(Vx_{i}s_{j}\right)$. Then the partition function $\mathcal{Z}[J]$ is described as
\begin{align}
\mathcal{Z}[J]=&\frac{C}{N!}\exp\left(-V\mathrm{tr}\left(\frac{3}{4\lambda}M^{3}-\frac{\kappa}{\sqrt{\lambda}}I+\frac{1}{\sqrt{\lambda}}J\right)M\right)\frac{1}{\displaystyle\prod_{1\leq t<u\leq N}(s_{u}-s_{t})}\notag\\
&\int\left(\prod_{i=1}^{N}dx_{i}\exp\left(-\frac{\lambda V}{4}x^{4}_{i}\right)\right)\left(\prod_{1\leq k<l\leq N}(x_{l}-x_{k})\right)\displaystyle\det_{1\leq m,n\leq N}\exp\left(Vx_{m}s_{n}\right).\label{aaaa}
\end{align}
Let us transform the part of the $x_{i}$ integrations in (\ref{aaaa}). By definition of the determinant,
\begin{align}
&\int\left(\prod_{i=1}^{N}dx_{i}\exp\left(-\frac{\lambda V}{4}x^{4}_{i}\right)\right)\left(\prod_{1\leq k<l\leq N}(x_{l}-x_{k})\right)\displaystyle\det_{1\leq i,j\leq N}\exp\left(Vx_{i}s_{j}\right)\notag\\
=&\sum_{\sigma\in S_{N}}\int\left(\prod_{i=1}^{N}dx_{i}\exp\left(-\frac{\lambda V}{4}x^{4}_{i}\right)\right)\left(\prod_{1\leq k<l\leq N}(x_{l}-x_{k})\right)(-1)^{\sigma}\left(\prod_{j=1}^{N}e^{Vx_{\sigma(j)}s_{j}}\right).\notag\\
\end{align}
Here $S_{N}$ is a symmetric group. Next we changed variables as $x_{\sigma(i)}\mapsto x_{i}~(i=1,\cdots,N)$. Note that the Vandermonde determinant $\det\left(x_{\sigma(i)}^{j-1}\right)=(-1)^{\sigma}\det x_{i}^{j-1}$. The above is written as

\begin{align}
&\sum_{\sigma\in S_{N}}\int\left(\prod_{i=1}^{N}dx_{i}\exp\left(-\frac{\lambda V}{4}x^{4}_{i}\right)\right)\left(\prod_{1\leq k<l\leq N}(x_{l}-x_{k})\right)(-1)^{\sigma}(-1)^{\sigma}\left(\prod_{j=1}^{N}e^{Vx_{j}s_{j}}\right)\notag\\
=&N!\int\left(\prod_{i=1}^{N}dx_{i}\exp\left(-\frac{\lambda  V}{4}x^{4}_{i}\right)\displaystyle\exp\left(Vx_{i}s_{i}\right)\right)\prod_{1\leq k<l\leq N}(x_{l}-x_{k}).
\end{align}

From this, the partition function $\mathcal{Z}[J]$ becomes as follows:

\begin{align}
\mathcal{Z}[J]=&C\exp\left(-V\mathrm{tr}\left(\frac{3}{4\lambda}M^{3}-\frac{\kappa}{\sqrt{\lambda}}I+\frac{1}{\sqrt{\lambda}}J\right)M\right)\frac{1}{\displaystyle\prod_{1\leq t<u\leq N}(s_{u}-s_{t})}\notag\\
&\int\left(\prod_{i=1}^{N}dx_{i}\exp\left(-\frac{\lambda  V}{4}x^{4}_{i}\right)\displaystyle\exp\left(Vx_{i}s_{i}\right)\right)\prod_{1\leq k<l\leq N}(x_{l}-x_{k}).
\end{align}

Using $\displaystyle\prod_{1\leq k<l\leq N}(x_{l}-x_{k})=\det_{1\leq k,l\leq N}\left(x_{k}^{l-1}\right)$, we calculate the remaining integral in the right-hand side in (\ref{aaaa}) as 
\begin{align}
&\int_{-\infty}^{\infty}\left(\prod_{i=1}^{N}dx_{i}\exp\left(-\frac{\lambda  V}{4}x^{4}_{i}\right)\displaystyle\exp\left(Vx_{i}s_{i}\right)\right)\det_{1\leq k,l\leq\mathrm{N}}\left(x^{k-1}_{l}\right)\notag\\
=&\sum_{\sigma\in S_{\mathrm{N}}}\mathrm{sgn}\sigma\prod_{i=1}^{N}\phi_{\sigma(i)}(s_{i})\notag\\
=&\det_{1\leq i,j\leq N}\left(\phi_{i}(s_{j})\right),\label{aaa}
\end{align}
where $\displaystyle\phi_{k}(z)$ is defined by
\begin{align}
\displaystyle\phi_{k}(z)=&\int_{-\infty}^{\infty}dx\hspace{2mm}x^{k-1}\exp\left(-\frac{\lambda V}{4}x^4+Vxz\right),
\end{align}
and $(\phi_{i}(s_{j}))$ is the $N\times N$ matrix with the $i$-th row and the $j$-th column being $\phi_{i}(s_{j})$. Summarizing the results (\ref{aaaa}) and (\ref{aaa}), we obtain the following:

\begin{proposition}\label{Pro3.1}

Let $\mathcal{Z}[J]$ be the partition function of $\Phi^{3}$-$\Phi^{4}$ Hybrid-Matrix-Model given by (\ref{Z[J]}). Then, $\mathcal{Z}[J]$ is given as
\begin{align*}
\mathcal{Z}[J]=&C\exp\left(-V\mathrm{tr}\left(\frac{3}{4\lambda}M^{3}-\frac{\kappa}{\sqrt{\lambda}}I+\frac{1}{\sqrt{\lambda}}J\right)M\right)\frac{\displaystyle\det_{1\leq i,j\leq N}\left(\phi_{i}(s_{j})\right)}{\displaystyle\prod_{1\leq t<u\leq\mathrm{N}}(s_{u}-s_{t})}.
\end{align*}
\end{proposition}
Note that $\displaystyle\phi_{k}(z)$ is expressed as 
\begin{align}
\displaystyle\phi_{k}(z)=&\left(\frac{1}{V}\right)^{k-1}\left(\frac{d}{dz}\right)^{k-1}\int_{-\infty}^{\infty}dx\exp\left(-\frac{\lambda V}{4}x^4+Vxz\right)\label{b}.
\end{align}
We use 
\begin{align}
P(z)=&\int_{-\infty}^{\infty}dx\exp\left(-\frac{\lambda V}{4}x^4+Vxz\right).\label{a}
\end{align}

If $V$ is a pure imaginary number, this $P(z)$ is a special case of the following:

\begin{align}
P(x,y):=&\int_{-\infty}^{\infty}dt\exp\left(i\left(t^{4}+xt^{2}+yt\right)\right).
\end{align}

This is called Pearcey integral\cite{Pedro}, where $0\leq \arg x\leq \pi$ and $y\in\mathbb{R}$. Substituting (\ref{a}) for (\ref{b}), $\displaystyle\phi_{k}(z)$ is calculated as follows:
\begin{align}
\displaystyle\phi_{k}(z)=&\left(\frac{1}{V}\right)^{k-1}\left(\frac{d}{dz}\right)^{k-1}P(z).
\end{align}

\begin{proposition}\label{Pro3.2}
Let $\left(P^{(j-1)}(s_{i})\right)$ be the $N\times N$ matrix with the $i$-th row and the $j$-th column being $\displaystyle P^{(j-1)}(s_{i})=\left(\frac{d}{ds_{i}}\right)^{j-1}\!\!\!\!\!\!P(s_{i})$. We then obtain the following:
\begin{align}
\det\left(P^{(j-1)}(s_{i})\right)=&\left(\prod_{1\leq i<j\leq N}\left(\partial_{s_{i}}-\partial_{s_{j}}\right)\right)P(s_{1})\cdots P(s_{N}).\notag
\end{align}
\end{proposition}
\hspace{-5mm}This proposition is identical to Proposition \ref{Pro3.2} in \cite{Kanomata:2022pdo} and its proof is also given in \cite{Kanomata:2022pdo}. We introduce 
\begin{align}
P_{N}(s_{1},\cdots,s_{N})=\left(\displaystyle\prod_{1\leq i< j\leq N}(\partial_{s_{i}}-\partial_{s_{j}})\right)P(s_{1})\cdots P(s_{N})
=\det
\begin{pmatrix}
P(s_{1})&\cdots&P(s_{N})\\
P^{(1)}(s_{1})&\cdots&P^{(1)}(s_{N})\\
\vdots&    &\vdots\\
P^{(N-1)}(s_{1})&\cdots&P^{(N-1)}(s_{N})
\end{pmatrix}
.
\end{align}
From this, $\displaystyle\det_{1\leq i,j\leq N}(\phi_{i}(s_{j}))$ is calculated as follows:
\begin{align}
\displaystyle\det_{1\leq i,j\leq N}(\phi_{i}(s_{j}))=&\frac{1}{V^{\frac{N(N-1)}{2}}}P_{N}(s_{1},\cdots,s_{N}).
\end{align}

Summarizing the above results, we obtain the following:

\begin{theorem}

Let $\mathcal{Z}[J]$ be the partition function of $\Phi^{3}$-$\Phi^{4}$ Hybrid-Matrix-Model given by (\ref{Z[J]}). Then, $\mathcal{Z}[J]$ is given as
\begin{align}
\mathcal{Z}[J]=&\int \mathcal{D}\Phi \exp\left(-V\mathrm{tr}\left(E\Phi^{2}+\kappa\Phi+\frac{\lambda}{4}\Phi^{4}+\sqrt{\lambda}M\Phi^{3}+\frac{1}{2}M\Phi M\Phi\right)\right)\exp\left(V\mathrm{tr}\left(J\Phi\right)\right)\notag\\
=&C'\frac{e^{\frac{-V}{\sqrt{\lambda}}\mathrm{tr}(JM)}P_{N}(s_{1},\cdots,s_{N})}{\displaystyle\prod_{1\leq t<u\leq N}(s_{u}-s_{t})}.\label{ZZ}
\end{align}
Here $\displaystyle C'=\exp\left(-V\mathrm{tr}\left(\frac{3}{4\lambda}M^{3}-\frac{\kappa}{\sqrt{\lambda}}I\right)M\right)\left(\prod_{p=1}^{N}p!\right)\frac{\pi^{\frac{N(N-1)}{2}}}{V^{N(N-1)}}$, and $s_{t}$ is the eigenvalues of the matrix $\displaystyle\frac{1}{\sqrt{\lambda}}M^{3}-\kappa I+J$ for $t=1,\cdots,N$.

\end{theorem}


\section{Exact Calculations of $1$-Point Function $G_{|a|}$, $2$-Point Function $G_{|ab|}$, $2$-Point Function $G_{|a|b|}$, and $G_{|a_{1}|a_{2}|\cdots|a_{n}|}$\hspace{1mm}$(3\leq n)$}\label{sec4}

In this section, $G_{|a|}$, $G_{|ab|}$, $G_{|a|b|}$, and $G_{|a_{1}|a_{2}|\cdots|a_{n}|}$ are calculated exactly, by using Theorem 3.3.

\subsection{$1$-Point Function $G_{|a|}$}

In the calculation of the 1-point function $G_{|a|}$, the external field $J$ can be treated as the diagonal matrix $J=diag\left(J_{11},\cdots,J_{NN}\right)$. Then the eigenvalues $s_{t}$ in (\ref{ZZ}) are given $\displaystyle s_{t}=\frac{E_{t-1}\sqrt{E_{t-1}}}{\sqrt{\lambda}}+J_{tt}-\kappa$. From Theorem 3.3, the $1$-point function $G_{|a|}$ is calculated as follows:
\begin{align}
G_{|a|}=\left.\frac{1}{V}\frac{\partial\log\mathcal{Z}[J]}{\partial J_{aa}}\right|_{J=0}=\displaystyle\frac{\displaystyle\frac{1}{V}\displaystyle\frac{\partial}{\partial J_{aa}}\Biggl(\frac{e^{\frac{-V}{\sqrt{\lambda}}\mathrm{tr}(JM)}\displaystyle P_{N}(s_{1},\cdots,s_{N})}{\displaystyle\prod_{1\leq t<u\leq N}\left(s_{u}-s_{t}\right)}\Biggl)\Biggl|_{J=0}}{\frac{\displaystyle P_{N}(s_{1},\cdots,s_{N})\Biggl|_{J=0}}{\displaystyle\left.\prod_{1\leq p<q\leq N}\left(s_{q}-s_{p}\right)\right|_{J=0}}}.\label{d}
\end{align}
Note that
\begin{align}
&\frac{\partial}{\partial J_{aa}}\Biggl\{e^{\frac{-V}{\sqrt{\lambda}}\mathrm{tr}(\mathrm{J}M)}\displaystyle P_{N}(s_{1},\cdots,s_{N})\Biggl\}\notag\\
&=-\frac{V}{\sqrt{\lambda}}\sqrt{E_{a-1}}e^{\frac{-V}{\sqrt{\lambda}}\mathrm{tr}(JM)}\displaystyle P_{N}(s_{1},\cdots,s_{N})+e^{\frac{-V}{\sqrt{\lambda}}\mathrm{tr}(JM)}\left(\partial_{a}\displaystyle P_{N}(s_{1},\cdots,s_{N})\right),\label{ykkkk}
\end{align}
where $\displaystyle\partial_{a}P_{N}(s_{1},\cdots,s_{N})=\frac{\partial}{\partial s_{a}}P_{N}(s_{1},\cdots,s_{N})$. Next, we use the following formula. Let $\displaystyle\Delta=\Delta(\vec{x}_{n})=\det_{1\leq i,j\leq n}\left((x_{j})^{i-1}\right)$ be the Vandermonde determinant for $\vec{x}_{n}=(x_{1},\cdots,x_{n})\in\mathbb{R}^{n}$. For any $1\leq k\leq n$
\begin{align}
\frac{\partial M_{n}}{\partial x_{k}}=&\sum_{i=1,i\neq k}^{n}\frac{M_{n}(\vec{x}_{n})}{x_{k}-x_{i}}.\label{www}
\end{align}
Using this formula, we get
\begin{align}
\left.\frac{\partial}{\partial J_{aa}}\prod_{1\leq i,j\leq N}(s_{j}-s_{i})\right|_{J=0}=&\left.\frac{\partial}{\partial J_{aa}}\left\{\displaystyle\det_{1\leq i,j\leq N}\left(\Biggl(\frac{E_{j-1}\sqrt{E_{j-1}}}{\sqrt{\lambda}}+J_{jj}-\kappa\Biggl)^{i-1}\right)\right\}\right|_{J=0}\notag\\
&=\displaystyle\sum_{i=1,i\neq a}^{N}\frac{\displaystyle\sqrt{\lambda}\det_{1\leq i,j\leq N}\left(\Biggl(\frac{E_{j-1}\sqrt{E_{j-1}}}{\sqrt{\lambda}}-\kappa\Biggl)^{i-1}\right)}{E_{a-1}\sqrt{E_{a-1}}-E_{i-1}\sqrt{E_{i-1}}}.\label{xkkkk}
\end{align}
Substituting (\ref{xkkkk}) into (\ref{d}), finally $G_{|a|}$ is expressed as
\begin{align}
G_{|a|}=&-\frac{\sqrt{E_{a-1}}}{\sqrt{\lambda}}-\frac{1}{V}\sum_{i=1,i\neq a}^{N}\frac{\sqrt{\lambda}}{E_{a-1}\sqrt{E_{a-1}}-E_{i-1}\sqrt{E_{i-1}}}+\frac{1}{V}\partial_{a}\log P_{N}(z_{1},\cdots,z_{N}),\label{h}
\end{align}
where $\displaystyle z_{j}=\frac{E_{j-1}\sqrt{E_{j-1}}}{\sqrt{\lambda}}-\kappa$ for $j=1,\ldots,N$, and $\displaystyle\partial_{a}=\frac{\partial}{\partial z_{a}}$.


\subsection{$2$-Point Function $G_{|ab|}$}

Let us consider $2$-point function $G_{|ab|}$ ($a\neq b,\hspace{1mm} a,b\in\{1,2,\cdots,N\}$). For the calculation, we put $J$ as all components without $J_{ab},J_{ba}$ are zero. Note that $\mathrm{tr}JM=0$ for this $J$. 

At first, we estimate eigenvalues $s_{t}$ for $t=1,\ldots,N$ of the matrix $\displaystyle\frac{1}{\sqrt{\lambda}}M^{3}-\kappa I+J$. The eigenequation is 
\begin{align}
&0=\det\left(sI-\left(\frac{1}{\sqrt{\lambda}}M^{3}-\kappa I+J\right)\right)\notag\\
&=\left(\prod_{i=1, i\neq a, i\neq b}^{N}\left(s-\frac{1}{\sqrt{\lambda}}E_{i-1}\sqrt{E_{i-1}}+\kappa\right)\right)\Biggl\{s^{2}+\left(-\frac{1}{\sqrt{\lambda}}E_{b-1}\sqrt{E_{b-1}}-\frac{1}{\sqrt{\lambda}}E_{a-1}\sqrt{E_{a-1}}+2\kappa\right)s\notag\\
&\hspace{3mm}-\frac{1}{\sqrt{\lambda}}E_{a-1}\sqrt{E_{a-1}}\kappa-\frac{1}{\sqrt{\lambda}}E_{b-1}\sqrt{E_{b-1}}\kappa+\frac{1}{\lambda}E_{a-1}E_{b-1}\sqrt{E_{a-1}}\sqrt{E_{b-1}}+\kappa^{2}-J_{ab}J_{ba}\Biggl\}.\notag\\
\end{align}
We label the eigenvalues as $\displaystyle s_{t}=\frac{1}{\sqrt{\lambda}}E_{t-1}\sqrt{E_{t-1}}-\kappa$ for $t\neq a,b$, 
\begin{align}
\displaystyle s_{a}=\frac{\displaystyle\frac{1}{\sqrt{\lambda}}E_{a-1}\sqrt{E_{a-1}}+\frac{1}{\sqrt{\lambda}}E_{b-1}\sqrt{E_{b-1}}-2\kappa+\sqrt{\left(\frac{1}{\sqrt{\lambda}}E_{a-1}\sqrt{E_{a-1}}-\frac{1}{\sqrt{\lambda}}E_{b-1}\sqrt{E_{b-1}}\right)^{2}+4J_{ab}J_{ba}}}{2},\label{sa}
\end{align}
and 
\begin{align}
\displaystyle s_{b}=\frac{\displaystyle\frac{1}{\sqrt{\lambda}}E_{a-1}\sqrt{E_{a-1}}+\frac{1}{\sqrt{\lambda}}E_{b-1}\sqrt{E_{b-1}}-2\kappa-\sqrt{\left(\frac{1}{\sqrt{\lambda}}E_{a-1}\sqrt{E_{a-1}}-\frac{1}{\sqrt{\lambda}}E_{b-1}\sqrt{E_{b-1}}\right)^{2}+4J_{ab}J_{ba}}}{2}.\label{sb}
\end{align}

Let us calculate $G_{|ab|}$ by using these $s_{t}$. From Theorem 3.3,
\begin{align}
G_{|ab|}=&\left.\frac{1}{V}\frac{\partial^{2}\log\mathcal{Z}[J]}{\partial J_{ab}\partial J_{ba}}\right|_{J=0}\notag\\
=&\left.\frac{1}{V}\frac{\partial^{2}}{\partial J_{ab}\partial J_{ba}}\left\{\log P_{N}\left(s_{1},\cdots,s_{N}\right)-\log\prod_{1\leq t<u\leq N}(s_{u}-s_{t})\right\}\right|_{J=0}\notag\\
=&\frac{1}{V}\left\{\displaystyle\left.\frac{\displaystyle\frac{\displaystyle\partial^{2}}{\partial J_{ab}\partial J_{ba}}P_{N}(s_{1},\cdots,s_{N})}{\displaystyle P_{N}(s_{1},\cdots,s_{N})}\right|_{J=0}-\left.\frac{\displaystyle\frac{\displaystyle\partial^{2}}{\partial J_{ab}\partial J_{ba}}\Biggl\{\prod_{1\leq t<u\leq N}(s_{u}-s_{t})\Biggl\}}{\displaystyle\prod_{1\leq t<u\leq N}(s_{u}-s_{t})}\right|_{J=0}\right\}.\label{e}
\end{align}
Here we use $\left.\displaystyle\frac{\partial P_{N}(s_{1},\cdots,s_{N})}{\partial J_{ab}}\right|_{J=0}=\displaystyle\left.\frac{\displaystyle\partial\det_{1\leq k,l\leq N}\left((s_{l})^{k-1}\right)}{\partial J_{ab}}\right|_{J=0}=0$, since $s_{a}$ and $s_{b}$ are functions of $(J_{ab}J_{ba})$ as we see in (\ref{sa}) and (\ref{sb}), then $\displaystyle\frac{\partial P_{N}(s_{1},\cdots,s_{N})}{\partial J_{ab}}$ and $\displaystyle\frac{\displaystyle\partial\det_{1\leq k,l\leq N}\left((s_{l})^{k-1}\right)}{\partial J_{ab}}$ are of the form $J_{ba}\times(\cdots)$. 

\hspace{-5mm}Using the fact that $\displaystyle\left.\frac{\partial s_{k}}{\partial J_{ab}}\right|_{J=0}=0$ and  

$\displaystyle\left.\frac{\partial^{2} s_{a}}{\partial J_{ba}\partial J_{ab}}\right|_{J=0}=\frac{\sqrt{\lambda}}{\displaystyle\left|E_{a-1}\sqrt{E_{a-1}}-E_{b-1}\sqrt{E_{b-1}}\right|}=-\left.\frac{\partial^{2}s_{b}}{\partial J_{ab}\partial J_{ba}}\right|_{J=0}$, we obtain
\begin{align}
\displaystyle\left.\frac{\displaystyle\partial^{2}}{\partial J_{ab}\partial J_{ba}}P_{N}(s_{1},\cdots,s_{N})\right|_{J=0}=&\frac{\sqrt{\lambda}}{\displaystyle\left|E_{a-1}\sqrt{E_{a-1}}-E_{b-1}\sqrt{E_{b-1}}\right|}\left(\partial_{a}P_{N}\left(z_{1},\ldots,z_{N}\right)-\partial_{b}P_{N}\left(z_{1},\ldots,z_{N}\right)\right),\label{rrr}
\end{align}
where $\displaystyle z_{j}=\frac{E_{j-1}\sqrt{E_{j-1}}}{\sqrt{\lambda}}-\kappa$ for $j=1,\ldots,N$. Similarly, we get 
\begin{align}
\displaystyle\left.\frac{\displaystyle\partial^{2}}{\partial J_{ab}\partial J_{ba}}\Biggl\{\prod_{1\leq t<u\leq N}(s_{u}-s_{t})\Biggl\}\right|_{J=0}=&\frac{(\sqrt{\lambda})^{2}}{\displaystyle\left|E_{a-1}\sqrt{E_{a-1}}-E_{b-1}\sqrt{E_{b-1}}\right|}\det_{1\leq k,l\leq N}\left((s_{k})^{l-1}\right)\notag\\
&\hspace{-2cm}\times\left(\sum_{i=1,i\neq a}^{N}\frac{1}{E_{a-1}\sqrt{E_{a-1}}-E_{i-1}\sqrt{E_{i-1}}}-\sum_{i=1,i\neq b}^{N}\frac{1}{E_{b-1}\sqrt{E_{b-1}}-E_{i-1}\sqrt{E_{i-1}}}\right),\label{rrrr}
\end{align}
where we use the formula (\ref{www}), again. Substituting (\ref{rrr}) and (\ref{rrrr}) into (\ref{e}), $G_{|ab|}$\hspace{2mm}($b<a$, i.e.$E_{b}<E_{a}$) is finally obtained as
\begin{align}
G_{|ab|}=&\frac{\sqrt{\lambda}}{\displaystyle V(E_{a-1}\sqrt{E_{a-1}}-E_{b-1}\sqrt{E_{b-1}})}\Biggl\{\left(\frac{\partial_{a}P_{N}(z_{1},\cdots,z_{N})}{P_{N}(z_{1},\cdots,z_{N})}-\frac{\partial_{b}P_{N}(z_{1},\cdots,z_{N})}{P_{N}(z_{1},\cdots,z_{N})}\right)\notag\\
&-\sqrt{\lambda}\left(\sum_{i=1,i\neq a}^{N}\frac{1}{E_{a-1}\sqrt{E_{a-1}}-E_{i-1}\sqrt{E_{i-1}}}-\sum_{i=1,i\neq b}^{N}\frac{1}{E_{b-1}\sqrt{E_{b-1}}-E_{i-1}\sqrt{E_{i-1}}}\right)\Biggl\}.\label{f}
\end{align}


\subsection{$2$-Point Function $G_{|a|b|}$}

In the calculation of the 2-point functions $G_{|a|b|}$, the external field $J$ can be treated as the diagonal matrix $J=diag\left(J_{11},\cdots,J_{NN}\right)$. Then the eigenvalues $s_{t}$ in (\ref{ZZ}) are given $\displaystyle s_{t}=\frac{E_{t-1}\sqrt{E_{t-1}}}{\sqrt{\lambda}}+J_{tt}-\kappa$ for $t=1,\cdots,N$. Then, the $2$-point function $G_{|a|b|}$ is calculated as follows:
\begin{align}
G_{|a|b|}=&\left.\frac{\partial^{2}\log\mathcal{Z}[J]}{\partial J_{aa}\partial J_{bb}}\right|_{J=0}\notag\\
=&\left.\frac{\partial^{2}}{\partial J_{aa}\partial J_{bb}}\left\{\log P_{N}\left(s_{1},\cdots,s_{N}\right)-\log\prod_{1\leq t<u\leq N}(s_{u}-s_{t})\right\}\right|_{J=0}\notag\\
=&-\left.\frac{\partial_{a}P_{N}(s_{1},\cdots,s_{N})}{P_{N}(s_{1},\cdots,s_{N})}\right|_{J=0}\left.\frac{\partial_{b}P_{N}(s_{1},\cdots,s_{N})}{P_{N}(s_{1},\cdots,s_{N})}\right|_{J=0}+\left.\frac{\partial_{a}\partial_{b}P_{N}(s_{1},\cdots,s_{N})}{P_{N}(s_{1},\cdots,s_{N})}\right|_{J=0}\notag\\
&-\frac{\lambda}{(E_{a-1}\sqrt{E_{a-1}}-E_{b-1}\sqrt{E_{b-1}})^{2}}.
\end{align}
Finally $G_{|a|b|}$ is expressed as
\begin{align}
G_{|a|b|}=&-\frac{\partial_{a}P_{N}(z_{1},\cdots,z_{N})}{P_{N}(z_{1},\cdots,z_{N})}\frac{\partial_{b}P_{N}(z_{1},\cdots,z_{N})}{P_{N}(z_{1},\cdots,z_{N})}+\frac{\partial_{a}\partial_{b}P_{N}(z_{1},\cdots,z_{N})}{P_{N}(z_{1},\cdots,z_{N})}\notag\\
&-\frac{\lambda}{(E_{a-1}\sqrt{E_{a-1}}-E_{b-1}\sqrt{E_{b-1}})^{2}},\label{yuewq}
\end{align}
where $\displaystyle z_{j}=\frac{E_{j-1}\sqrt{E_{j-1}}}{\sqrt{\lambda}}-\kappa$ for $j=1,\ldots,N$,$\displaystyle\partial_{a}=\frac{\partial}{\partial z_{a}}$, and $\displaystyle\partial_{b}=\frac{\partial}{\partial z_{b}}$.

\subsection{$n$-point function $G_{|a^{1}|a^{2}|\cdots|a^{n}|}$\hspace{1mm}$(3\leq n)$}

Let us calculate $G_{|a^{1}|a^{2}|\cdots|a^{n}|}$ for $3\leq n$. Here $a^{\beta}$ is the pairwise different indices for $\beta=1,\ldots,n$. To calculate $G_{|a^{1}|a^{2}|\cdots|a^{n}|}$, it is enough to take $J$ as a diagonal matrix $J=diag(J_{11},\cdots, J_{NN})$. Then the eigenvalues $s_{t}$ in (\ref{ZZ}) are given $\displaystyle s_{t}=\frac{E_{t-1}\sqrt{E_{t-1}}}{\sqrt{\lambda}}+J_{tt}-\kappa$ for $t=1,2,\cdots, N$. From the definition in (\ref{logZ}), the $n$-point function $G_{|a^{1}|a^{2}|\cdots|a^{n}|}$ is given by

\begin{align}
G_{|a^{1}|a^{2}|\cdots|a^{n}|}=&V^{n-2}\frac{\partial^{n}}{\partial J_{a^{1}a^{1}}\cdots\partial J_{a^{n}a^{n}}}\left.\log\frac{\mathcal{Z}[J]}{\mathcal{Z}[0]}\right|_{J=0}\notag\\
=&V^{n-2}\frac{\partial^{n}}{\partial J_{a^{1}a^{1}}\cdots\partial J_{a^{n}a^{n}}}\left.\log\left(\frac{e^{\frac{-V}{\sqrt{\lambda}}\mathrm{tr}(JM)}P_{N}(s_{1},\cdots,s_{N})}{\displaystyle\prod_{1\leq t<u\leq N}(s_{u}-s_{t})}\right)\right|_{J=0}.
\end{align}

Since $\displaystyle s_{t}=\frac{E_{t-1}\sqrt{E_{t-1}}}{\sqrt{\lambda}}+J_{tt}-\kappa$, $\displaystyle\left.\frac{\partial^{3}}{\partial J_{a^{1}a^{1}}\partial J_{a^{2}a^{2}}\partial J_{a^{3}a^{3}}}\left(\sum_{1\leq t<u\leq N}\log(s_{u}-s_{t})\right)\right|_{J=0}=0$. In addition 

\begin{align}
\frac{\partial^{2}}{\partial J_{a^{1}a^{1}}\partial J_{a^{2}a^{2}}}\log\exp\left(-\frac{V}{\sqrt{\lambda}}\mathrm{tr}JM\right)=&-\frac{\partial^{2}}{\partial J_{a^{1}a^{1}}\partial J_{a^{2}a^{2}}}\frac{V}{\sqrt{\lambda}}\sum_{i=1}^{N}J_{ii}\sqrt{E_{i-1}}=0.
\end{align}

Then, the $n$-point function $G_{|a^{1}|a^{2}|\cdots|a^{n}|}$\hspace{1mm}$(3\leq n)$ is obtained as follows:

\begin{align}
G_{|a^{1}|a^{2}|\cdots|a^{n}|}=&V^{n-2}\frac{\partial^{n}}{\partial z_{a^{1}}\cdots\partial z_{a^{n}}}\log P_{N}(z_{1},\cdots,z_{N}),\label{prbgd}
\end{align}
where $\displaystyle z_{i}=\frac{E_{i-1}\sqrt{E_{i-1}}}{\sqrt{\lambda}}-\kappa$.

\section{Approximations from Exact Solutions by Saddle point method}\label{sec5}

To ensure that perturbative calculations in Section 2 and exact results in Section 4 are consistent, we shall reproduce the contents of Section $2$ by approximating the results of Section $4$. Thereafter, the calculations are performed with $\kappa=0$. 

By the change of variable $\displaystyle x=\frac{1}{\displaystyle\left(V\frac{\lambda}{4}\right)^{\frac{1}{4}}}k$, $\displaystyle P(z)=\displaystyle\int_{-\infty}^{\infty}dx\exp\left(-V\frac{\lambda}{4}x^{4}+Vxz\right)$ is transformed into
\begin{align}
P\left(\frac{z\lambda^{\frac{1}{4}}}{V^{\frac{3}{4}}\sqrt{2}}\right)=&\frac{1}{\displaystyle\left(V\frac{\lambda}{4}\right)^{\frac{1}{4}}}\displaystyle\int_{-\infty}^{\infty}dk\exp\left(-k^{4}+zk\right).\label{J}
\end{align}
The function under integration has three saddle points. We choose an integral path through one of them, $\displaystyle k=\frac{z^{\frac{1}{3}}}{2^{\frac{2}{3}}}$. To estimate the integral around the neighborhood of $\displaystyle k=\frac{z^{\frac{1}{3}}}{2^{\frac{2}{3}}}$ we put $\displaystyle k=\frac{z^{\frac{1}{3}}}{2^{\frac{2}{3}}}+\xi$. Then (\ref{J}) can be evaluated as 

\begin{align}
P\left(\frac{z\lambda^{\frac{1}{4}}}{V^{\frac{3}{4}}\sqrt{2}}\right)=&\frac{1}{\displaystyle\left(V\frac{\lambda}{4}\right)^{\frac{1}{4}}}\displaystyle\int_{-\infty}^{\infty}d\xi\exp\left(-\left(\frac{z^{\frac{1}{3}}}{2^{\frac{2}{3}}}+\xi\right)^{4}+z\left(\frac{z^{\frac{1}{3}}}{2^{\frac{2}{3}}}+\xi\right)\right)\notag\\
=&\mathcal{C}(z)\left(\frac{\sqrt{2}}{\lambda^{\frac{1}{4}}V^{\frac{1}{4}}}+\frac{7}{9\times 2^{\frac{5}{6}}V^{\frac{1}{4}}\lambda^{\frac{1}{4}}z^{\frac{4}{3}}}+\frac{385}{648\times 2^{\frac{1}{6}}\lambda^{\frac{1}{4}}V^{\frac{1}{4}}z^{\frac{8}{3}}}\right)+\mathcal{O}\left(\frac{1}{z^{4}}\right).
\end{align}
Here $\displaystyle \mathcal{C}(z)=\frac{2^{\frac{1}{6}}\sqrt{\pi}}{\sqrt{3}z^{\frac{1}{3}}}\exp\left(-\frac{z^{\frac{4}{3}}}{2^{\frac{8}{3}}}+\frac{z^{\frac{4}{3}}}{2^{\frac{2}{3}}}\right)$. To evaluate $n$-point functions, $\displaystyle z=\frac{\sqrt{2}V^{\frac{3}{4}}E_{i-1}^{\frac{3}{2}}}{\lambda^{\frac{3}{4}}}$ cases are used.  For the case $\displaystyle z_{i}=\frac{\sqrt{2}V^{\frac{3}{4}}E_{i-1}^{\frac{3}{2}}}{\lambda^{\frac{3}{4}}}$,

\begin{align}
P\left(\frac{E_{i-1}\sqrt{E_{i-1}}}{\sqrt{\lambda}}\right)=&\mathcal{C}'(E_{i})\left(\frac{\sqrt{2}}{\lambda^{\frac{1}{4}}V^{\frac{1}{4}}}+\frac{7\lambda^{\frac{3}{4}}}{18\sqrt{2}E_{i-1}^{2}V^{\frac{5}{4}}}+\frac{385\lambda^{\frac{7}{4}}}{1296\sqrt{2}E_{i-1}^{4}V^{\frac{9}{4}}}\right)+\mathcal{O}(\lambda^{\frac{11}{4}})\label{Q},
\end{align}
where $\displaystyle\mathcal{C}'(E_{i})=\sqrt{\frac{\pi\lambda^{\frac{1}{2}}}{3V^{\frac{1}{2}}E_{i-1}}}\exp\left(\frac{3VE_{i-1}^{2}}{4\lambda}\right)$. 

Next, we estimate $\partial P(z)$, similarly. After changing variable as $\displaystyle x=\frac{1}{\displaystyle\left(V\frac{\lambda}{4}\right)^{\frac{1}{4}}}k$,  

\hspace{-5mm}$\displaystyle\partial P(z)=\displaystyle V\int_{-\infty}^{\infty}dx\hspace{1mm}x\exp\left(-V\frac{\lambda}{4}x^{4}+Vxz\right)$ is transformed into
\begin{align}
\left(\partial P\right)\left(\frac{z\lambda^{\frac{1}{4}}}{V^{\frac{3}{4}}\sqrt{2}}\right)=&\frac{V}{\displaystyle\left(V\frac{\lambda}{4}\right)^{\frac{1}{2}}}\displaystyle\int_{\infty}^{\infty}dk\hspace{1mm}k\exp\left(-k^{4}+zk\right).\label{O}
\end{align}
The saddle point on the integral path is $\displaystyle k=\frac{z^{\frac{1}{3}}}{2^{\frac{2}{3}}}$. In the neighborhood of the saddle point $\displaystyle k=\frac{z^{\frac{1}{3}}}{2^{\frac{2}{3}}}$, we put $\displaystyle k=\frac{z^{\frac{1}{3}}}{2^{\frac{2}{3}}}+\xi$, then (\ref{O}) can be evaluated as 

\begin{align}
\left(\partial P\right)\left(\frac{z\lambda^{\frac{1}{4}}}{V^{\frac{3}{4}}\sqrt{2}}\right)=&\frac{V}{\displaystyle\left(V\frac{\lambda}{4}\right)^{\frac{1}{2}}}\displaystyle\int_{-\infty}^{\infty}d\xi\left(\frac{z^{\frac{1}{3}}}{2^{\frac{2}{3}}}+\xi\right)\exp\left(-\left(\frac{z^{\frac{1}{3}}}{2^{\frac{2}{3}}}+\xi\right)^{4}+z\left(\frac{z^{\frac{1}{3}}}{2^{\frac{2}{3}}}+\xi\right)\right)\notag\\
=&V\mathcal{C}(z)\left(\frac{2^{\frac{1}{3}}z^{\frac{1}{3}}}{\sqrt{\lambda}\sqrt{V}}-\frac{5}{18\sqrt{\lambda}\sqrt{V}z}-\frac{455}{648\times 2^{\frac{1}{3}}\sqrt{\lambda}\sqrt{V}z^{\frac{7}{3}}}\right)+\mathcal{O}\left(\frac{1}{z^{\frac{11}{3}}}\right).
\end{align}
For the case $\displaystyle z_{i}=\frac{\sqrt{2}V^{\frac{3}{4}}E_{i-1}^{\frac{3}{2}}}{\lambda^{\frac{3}{4}}}$,

\begin{align}
\left(\partial P\right)\left(\frac{E_{i-1}\sqrt{E_{i-1}}}{\sqrt{\lambda}}\right)=&V\mathcal{C}'(E_{i})\left(\frac{\sqrt{2}\sqrt{E_{i-1}}}{\lambda^{\frac{3}{4}}V^{\frac{1}{4}}}-\frac{5\lambda^{\frac{1}{4}}}{18\sqrt{2}E_{i-1}^{\frac{3}{2}}V^{\frac{5}{4}}}-\frac{455\lambda^{\frac{5}{4}}}{1296\sqrt{2}E_{i-1}^{\frac{7}{2}}V^{\frac{9}{4}}}\right)+\mathcal{O}(\lambda^{\frac{9}{4}})\label{R}.
\end{align}
Next, we consider $\partial^{2}P(z)$. $\displaystyle\partial^{2}P(z)=\displaystyle V^{2}\int_{-\infty}^{\infty}dx\hspace{1mm}x^{2}\exp\left(-V\frac{\lambda}{4}x^{4}+Vxz\right)$ is transformed similarly into 
\begin{align}
\left(\partial^{2}P\right)\left(\frac{z\lambda^{\frac{1}{4}}}{V^{\frac{3}{4}}\sqrt{2}}\right)=&\frac{V^{2}}{\displaystyle\left(V\frac{\lambda}{4}\right)^{\frac{3}{4}}}\displaystyle\int_{\infty}^{\infty}dk\hspace{1mm}k^{2}\exp\left(-k^{4}+zk\right).\label{P}
\end{align}
In the neighborhood of $\displaystyle k=\frac{z^{\frac{1}{3}}}{2^{\frac{2}{3}}}$, (\ref{P}) can be evaluated as 

\begin{align}
\left(\partial^{2}P\right)\left(\frac{z\lambda^{\frac{1}{4}}}{V^{\frac{3}{4}}\sqrt{2}}\right)=&\frac{V^{2}}{\displaystyle\left(V\frac{\lambda}{4}\right)^{\frac{3}{4}}}\displaystyle\int_{-\infty}^{\infty}d\xi\left(\frac{z^{\frac{1}{3}}}{2^{\frac{2}{3}}}+\xi\right)^{2}\exp\left(-\left(\frac{z^{\frac{1}{3}}}{2^{\frac{2}{3}}}+\xi\right)^{4}+z\left(\frac{z^{\frac{1}{3}}}{2^{\frac{2}{3}}}+\xi\right)\right)\notag\\
=&V^{2}\mathcal{C}(z)\left(\frac{2^{\frac{1}{6}}z^{\frac{2}{3}}}{\lambda^{\frac{3}{4}}V^{\frac{3}{4}}}-\frac{2^{\frac{5}{6}}}{3\lambda^{\frac{3}{4}}V^{\frac{3}{4}}z^{\frac{2}{3}}}+\frac{7}{18\times 2^{\frac{1}{6}}\lambda^{\frac{3}{4}}V^{\frac{3}{4}}z^{\frac{2}{3}}}-\frac{35\sqrt{2}}{27\lambda^{\frac{3}{4}}V^{\frac{3}{4}}z^{2}}+\frac{1705}{648\sqrt{2}\lambda^{\frac{3}{4}}V^{\frac{3}{4}}z^{2}}\right)\notag\\
&+\mathcal{O}\left(\frac{1}{z^{\frac{10}{3}}}\right).
\end{align}
For the case $\displaystyle z_{i}=\frac{\sqrt{2}V^{\frac{3}{4}}E_{i-1}^{\frac{3}{2}}}{\lambda^{\frac{3}{4}}}$,

\begin{align}
\left(\partial^{2}P\right)\left(\frac{E_{i-1}\sqrt{E_{i-1}}}{\sqrt{\lambda}}\right)=&V^{2}\mathcal{C}'(E_{i})\left(\frac{\sqrt{2}E_{i-1}}{\lambda^{\frac{5}{4}}V^{\frac{1}{4}}}-\frac{5}{18\sqrt{2}\lambda^{\frac{1}{4}}E_{i-1}V^{\frac{5}{4}}}+\frac{25\lambda^{\frac{3}{4}}}{1296\sqrt{2}E_{i-1}^{3}V^{\frac{9}{4}}}\right)+\mathcal{O}(\lambda^{\frac{7}{4}})\label{S}.
\end{align}
We use these approximate quantities in the following subsections.


\subsection{Approximation of $1$-Point Function $G_{|1|}$ by Saddle Point Method ($N=2$)}

We consider the $1$-point function $G_{|1|}$ in the case of (\ref{h}) in $N=2$.

\begin{align}
G_{|1|}=&-\frac{\sqrt{E_{0}}}{\sqrt{\lambda}}+\frac{1}{V}\partial_{1}\log P_{2}(z_{1},z_{2})-\frac{1}{V}\frac{\sqrt{\lambda}}{E_{0}\sqrt{E_{0}}-E_{1}\sqrt{E_{1}}},\notag\\
\end{align}
where $\displaystyle z_{1}=\frac{\sqrt{2}V^{\frac{3}{4}}E_{0}^{\frac{3}{2}}}{\lambda^{\frac{3}{4}}}$, and $\displaystyle z_{2}=\frac{\sqrt{2}V^{\frac{3}{4}}E_{1}^{\frac{3}{2}}}{\lambda^{\frac{3}{4}}}$.
Using (\ref{Q}), (\ref{R}) and (\ref{S}), we approximate $1$-Point Function $G_{|1|}$ ($N=2$) as

\begin{align}
G_{|1|}=&-\frac{\sqrt{E_{0}}}{\sqrt{\lambda}}+\frac{1}{V}\frac{\partial_{1}P_{2}(z_{1},z_{2})}{P_{2}(z_{1},z_{2})}-\frac{1}{V}\frac{\sqrt{\lambda}}{E_{0}\sqrt{E_{0}}-E_{1}\sqrt{E_{1}}}\notag\\
=&-\frac{\sqrt{\lambda}}{V}\frac{\sqrt{E_{0}}}{3E_{0}^{2}}-\frac{\sqrt{\lambda}}{V}\frac{\sqrt{E_{0}}}{3E_{0}}\frac{2}{E_{0}+E_{1}+\sqrt{E_{0}}\sqrt{E_{1}}}-\frac{\sqrt{\lambda}}{V}\frac{\sqrt{E_{1}}}{3E_{0}}\frac{1}{E_{0}+E_{1}+\sqrt{E_{0}}\sqrt{E_{1}}}+\mathcal{O}(\lambda\sqrt{\lambda})\label{F}.
\end{align}
This is consistent with the calculation of the 1-point function $G_{|1|}$ ($N=2$) using perturbative expansion as $(\ref{D})=(\ref{F})$.


\subsection{Approximation of $2$-Point Function $G_{|21|}$ by Saddle Point Method ($N=2$)}

We consider the $2$-point function $G_{|21|}$ in the case of (\ref{f}) in $N=2$.

\begin{align}
G_{|21|}=&-\frac{1}{V}\frac{\sqrt{\lambda}}{E_{1}\sqrt{E_{1}}-E_{0}\sqrt{E_{0}}}\frac{\partial_{1}P_{2}(z_{1},z_{2})}{P_{2}(z_{1},z_{2})}+\frac{1}{V}\frac{\sqrt{\lambda}}{E_{1}\sqrt{E_{1}}-E_{0}\sqrt{E_{0}}}\frac{\partial_{2}P_{2}(z_{1},z_{2})}{P_{2}(z_{1},z_{2})}-\frac{2}{V}\frac{\lambda}{(E_{1}\sqrt{E_{1}}-E_{0}\sqrt{E_{0}})^{2}},
\end{align}
where $\displaystyle z_{1}=\frac{\sqrt{2}V^{\frac{3}{4}}E_{0}^{\frac{3}{2}}}{\lambda^{\frac{3}{4}}}$, and $\displaystyle z_{2}=\frac{\sqrt{2}V^{\frac{3}{4}}E_{1}^{\frac{3}{2}}}{\lambda^{\frac{3}{4}}}$.
Using (\ref{Q}), (\ref{R}) and (\ref{S}), we approximate $2$-Point Function $G_{|21|}$ ($N=2$) as

\begin{align}
G_{|21|}=&\frac{1}{E_{0}+E_{1}+\sqrt{E_{0}}\sqrt{E_{1}}}+\frac{\sqrt{E_{1}}\lambda}{3V(\sqrt{E_{1}}-\sqrt{E_{0}})E_{0}\sqrt{E_{0}}(E_{1}\sqrt{E_{1}}-E_{0}\sqrt{E_{0}})}\notag\\
&+\frac{\sqrt{E_{0}}\lambda}{3V(\sqrt{E_{1}}-\sqrt{E_{0}})E_{1}\sqrt{E_{1}}(E_{1}\sqrt{E_{1}}-E_{0}\sqrt{E_{0}})}-\frac{2}{V}\frac{\lambda}{(E_{1}\sqrt{E_{1}}-E_{0}\sqrt{E_{0}})^{2}}+\mathcal{O}(\lambda^{2})\notag\\
=&\frac{1}{E_{0}+E_{1}+\sqrt{E_{0}}\sqrt{E_{1}}}+\frac{\lambda}{3VE_{0}(E_{0}+E_{1}+\sqrt{E_{0}}\sqrt{E_{1}})^{2}}+\frac{\lambda}{3VE_{1}(E_{0}+E_{1}+\sqrt{E_{0}}\sqrt{E_{1}})^{2}}\notag\\
&+\frac{8\sqrt{E_{0}}\sqrt{E_{1}}\lambda}{3VE_{1}(E_{0}+E_{1}+\sqrt{E_{0}}\sqrt{E_{1}})^{3}}+\frac{8\sqrt{E_{0}}\sqrt{E_{1}}\lambda}{3VE_{0}(E_{0}+E_{1}+\sqrt{E_{0}}\sqrt{E_{1}})^{3}}+\frac{\sqrt{E_{0}}\sqrt{E_{1}}\lambda}{3VE_{1}^{2}(E_{0}+E_{1}+\sqrt{E_{0}}\sqrt{E_{1}})^{2}}\notag\\
&+\frac{\sqrt{E_{0}}\sqrt{E_{1}}\lambda}{3VE_{0}^{2}(E_{0}+E_{1}+\sqrt{E_{0}}\sqrt{E_{1}})^{2}}+\frac{2\lambda E_{0}}{3VE_{1}(E_{0}+E_{1}+\sqrt{E_{0}}\sqrt{E_{1}})^{3}}+\frac{2\lambda E_{1}}{3VE_{0}(E_{0}+E_{1}+\sqrt{E_{0}}\sqrt{E_{1}})^{3}}\notag\\
&+\frac{10\lambda}{3V(E_{0}+E_{1}+\sqrt{E_{0}}\sqrt{E_{1}})^{3}}+\mathcal{O}(\lambda^{2})\label{G}.
\end{align}
This is consistent with the result of the 2-point function $G_{|21|}$ ($N=2$) using perturbative expansion i.e. $(\ref{E})=(\ref{G})$.


\subsection{Approximation of $2$-Point Function $G_{|2|1|}$ by the saddle point method ($N=2$)}

We consider the $2$-point function $G_{|2|1|}$ in the case of (\ref{1}) in $N=2$.

\begin{align}
G_{|2|1|}=&-\frac{\partial_{1}P_{2}(z_{1},z_{2})}{P_{2}(z_{1},z_{2})}\frac{\partial_{2}P_{2}(z_{1},z_{2})}{P_{2}(z_{1},z_{2})}+\frac{\partial_{1}\partial_{2}P_{2}(z_{1},z_{2})}{P_{2}(z_{1},z_{2})}\notag\\
&-\frac{\lambda}{E_{0}^{3}+E_{1}^{3}-2E_{0}\sqrt{E_{0}}E_{1}\sqrt{E_{1}}},
\end{align}
where $\displaystyle z_{1}=\frac{\sqrt{2}V^{\frac{3}{4}}E_{0}^{\frac{3}{2}}}{\lambda^{\frac{3}{4}}}$, and $\displaystyle z_{2}=\frac{\sqrt{2}V^{\frac{3}{4}}E_{1}^{\frac{3}{2}}}{\lambda^{\frac{3}{4}}}$. Using (\ref{Q}), (\ref{R}) and (\ref{S}), we approximate $2$-Point Function $G_{|21|}$ ($N=2$) as

\begin{align}
G_{|2|1|}=&\frac{4\sqrt{E_{0}}\sqrt{E_{1}}\lambda}{9E_{0}E_{1}(E_{0}+E_{1}+\sqrt{E_{0}}\sqrt{E_{1}})^{2}}+\frac{\lambda}{9E_{0}(E_{0}+E_{1}+\sqrt{E_{0}}\sqrt{E_{1}})^{2}}+\frac{\lambda}{9E_{1}(E_{0}+E_{1}+\sqrt{E_{0}}\sqrt{E_{1}})^{2}}+\mathcal{O}(\lambda^{2}).\label{2}
\end{align}
Then, we verified the consistency between the exact solution and the perturbative calculation for $G_{|2|1|}$ by $(\ref{1})=(\ref{2})$.


\section{Contributions from Non-trivial Topology Surfaces}\label{sec6}

\begin{figure}[h]
\begin{center}
\includegraphics[width=30mm]{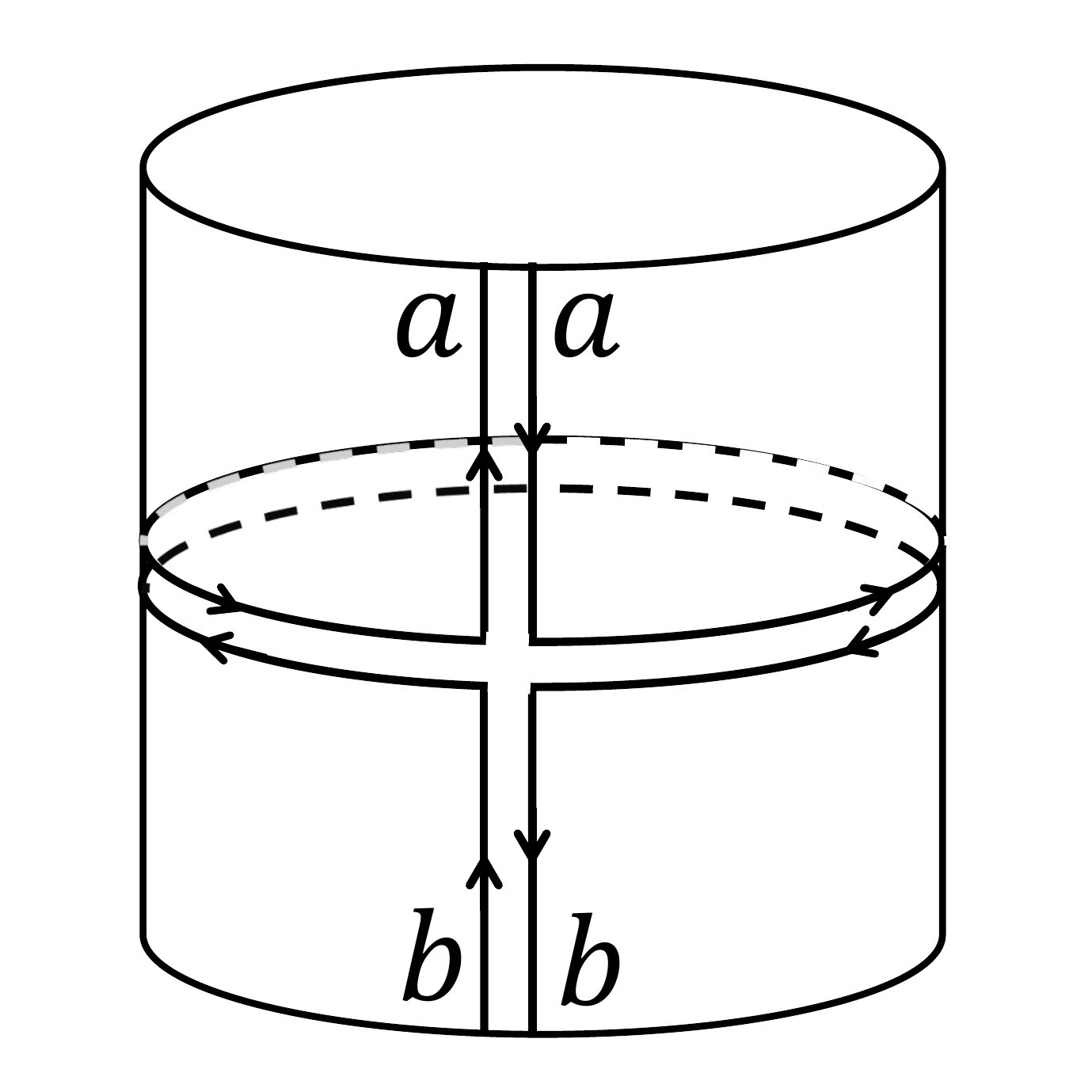}\lower-8ex\hbox{$\displaystyle=-\frac{\lambda}{36V^{2}E_{a-1}E_{b-1}(E_{a-1}+E_{b-1}+\sqrt{E_{a-1}}\sqrt{E_{b-1}})}$}
\caption{The surface of $2$-point function $G_{|a|b|}$ with two boundaries}\label{qwert}
\end{center}
\end{figure}

In this section, we make remarks about contributions from Feynman diagrams corresponding to nonplanar or higher genus surfaces. As we saw in Section 2, perturbative expansions of $\Phi^{3}$-$\Phi^{4}$ Hybrid-Matrix-Model are given by the sum over not only planar but also non-planar Feynman diagrams. Nonplanar graphs are diagrams that cannot be drawn on a plane. For example, a nonplanar graph in Figure \ref{qwert} appeared when the $2$-point function $G_{|a|b|}$ was calculated.

Calculations in this paper are carried out for finite $N$, so nonplanar Feynman diagrams are taken into account.\bigskip

\begin{figure}[h]
\begin{center}
\includegraphics[width=30mm]{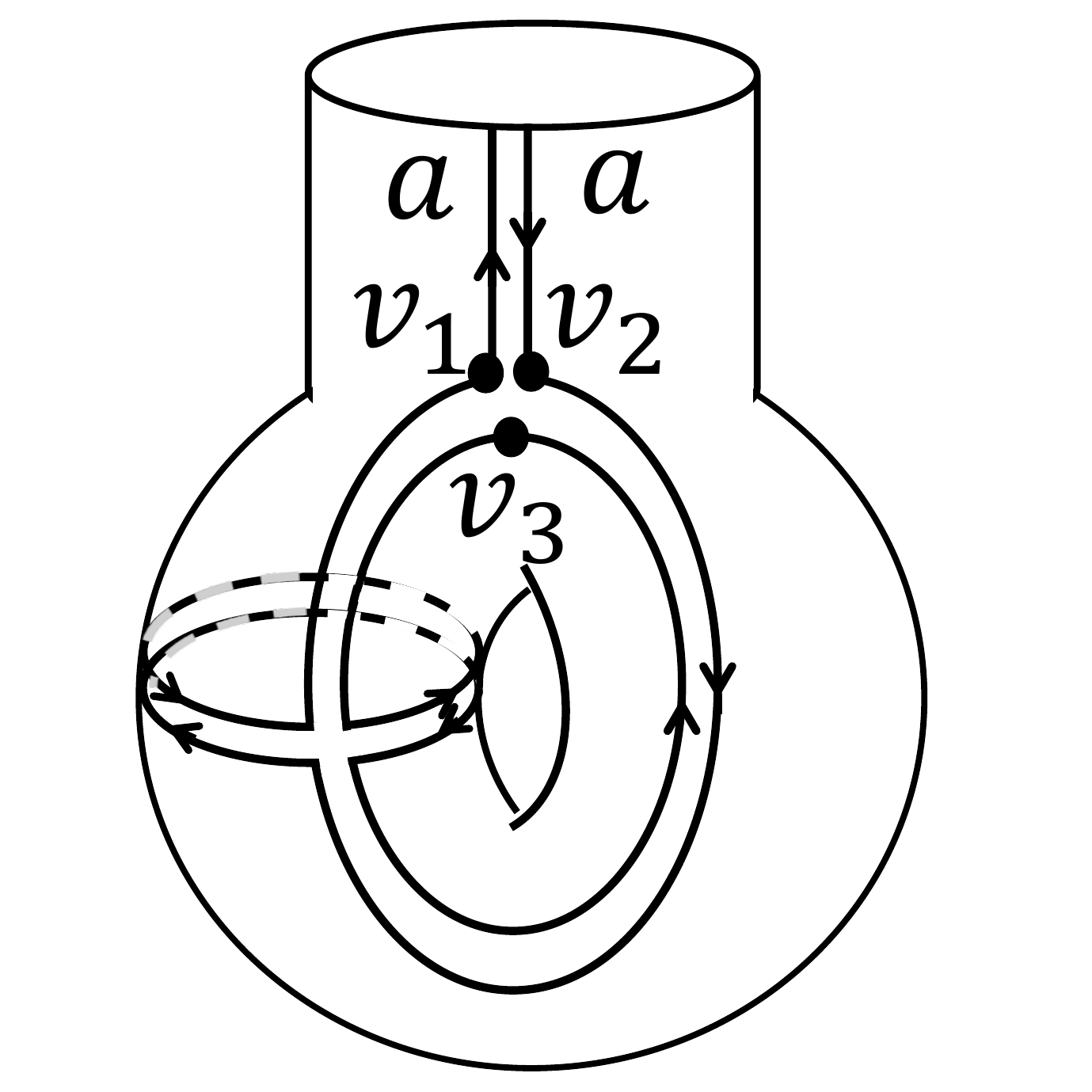}\lower-8ex\hbox{$\displaystyle=+\frac{\lambda\sqrt{\lambda}}{324V^{2}E_{a-1}^{4}}\sum_{v\in\{\{a,a,a\}\}}\sqrt{E_{v-1}}$}
\caption{A surface for $1$-point function $G_{|a|}$ with one boundary and one genus}\label{qou}
\end{center}
\end{figure}

Next, let us consider contributions from higher genus surfaces. Specifically, we consider ``genus expansion" in one-point function $G_{|a|}$. First, we consider the contribution to the one-point function $G_{|a|}$ from a surface of one genus. (See Figure \ref{qou}.)
\newpage

\begin{figure}[h]
\begin{center}
\includegraphics[width=70mm]{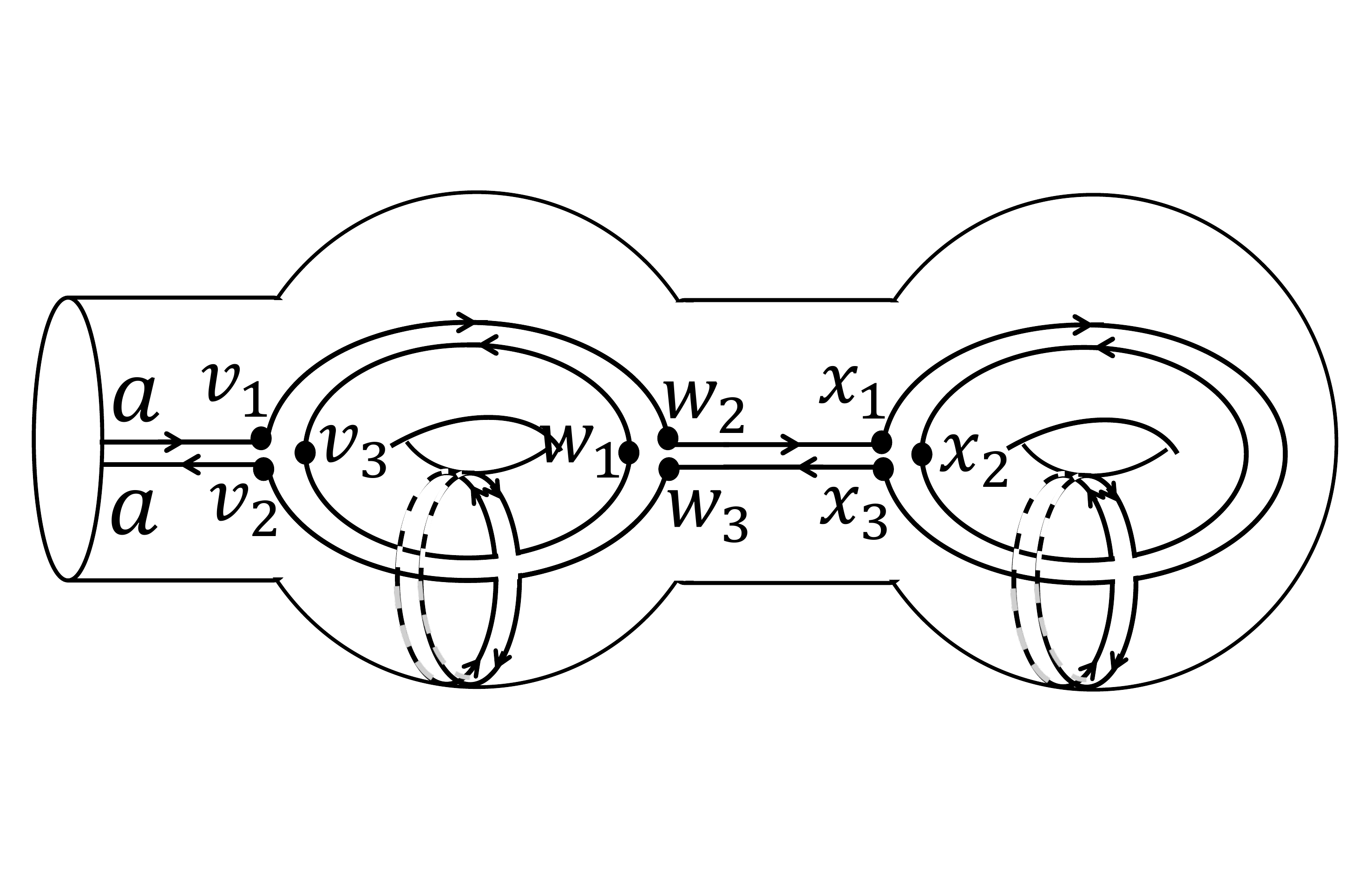}\lower-15ex\hbox{$\displaystyle=-\frac{\lambda^{3}\sqrt{\lambda}}{16\times 3^{9}E_{a-1}^{9}V^{4}}\sum_{\substack{v,w,x\in\{\{a,a,a\}\}}}\sqrt{E_{v-1}}\sqrt{E_{w-1}}\sqrt{E_{x-1}}$}
\caption{The surface for $1$-point function $G_{|a|}$ with one boundary and two genus}\label{quo}
\end{center}
\end{figure}

This diagram is constructed by using $\Sigma=0$, $k_{3}=1$, $k_{4}=1$, and $\mathscr{N}=1$, where we use the notation in Subsection 2.2. The contribution from the Feynman diagram has $V^{k_{3}+k_{4}-\frac{3k_{3}+4k_{4}+\mathscr{N}}{2}}$ as we saw in Subsection 2.2. From this formula, $V^{k_{3}+k_{4}-\frac{3k_{3}+4k_{4}+\mathscr{N}}{2}}=V^{1+1-\frac{3+4+1}{2}}=V^{-2}$. Also $V^{k_{3}+k_{4}-\frac{3k_{3}+4k_{4}+\mathscr{N}}{2}}=V^{\chi-\mathscr{N}-\Sigma}=V^{2-2g-B-\mathscr{N}-\Sigma}$. From this, $V^{2-2g-1-1-0}=V^{-2}$. It is consistent with Figure \ref{qou} that the genus of the surface is $g=1$ . Second, we consider the contribution for $G_{|a|}$ from a two genus surface. (See Figure \ref{quo}.)

From this diagram, we find that $\Sigma=0$, $k_{3}=3$, $k_{4}=2$, and $\mathscr{N}=1$. From them, $V^{k_{3}+k_{4}-\frac{3k_{3}+4k_{4}+\mathscr{N}}{2}}=V^{3+2-\frac{9+8+1}{2}}=V^{-4}$. Also $V^{k_{3}+k_{4}-\frac{3k_{3}+4k_{4}+\mathscr{N}}{2}}=V^{\chi-\mathscr{N}-\Sigma}=V^{2-2g-B-\mathscr{N}-\Sigma}$. On the other hand, Figure \ref{quo} implies $V^{2-2g-1-1-0}=V^{-4}$. It is consistent with Figure \ref{quo} that the genus of the surface is $g=2$. These observations show that our calculations in Section \ref{sec4} took into account any contributions from higher genus surfaces.

In other words, if we expand (\ref{h}),(\ref{f}),(\ref{yuewq}),(\ref{prbgd}), and so on as (\ref{utgxdsa}), then each $G^{(g)}_{|a_{1}^{1}\ldots a_{N_{1}}^{1}|\ldots|a_{1}^{B}\ldots a_{N_{B}}^{B}|}$ is obtained as the contribution from the fixed genus $g$.\\

\bigskip


{\bf Acknowledgement}\\
{A.S. was supported by JSPS KAKENHI Grant Number 21K03258. This study was supported by Erwin Schr\"odinger International Institute for Mathematics and Physics (ESI) through the project ``Research in Teams Project Integrability".}


\appendix
\section{Appendix}\label{apen}

\subsection{Calculation of Perturbative Expansion ($N=1$)}\label{apen1}
For $N=1$, we calculate the $1$-point function $G_{|1|}$ using perturbative expansion. For $N=1$, (\ref{A}) is described as
\newpage
\begin{align}
G_{|1|}=&\left.\frac{1}{V}\frac{\partial\log\mathcal{Z}[J]}{\partial J}\right|_{J=0}\notag\\
=&\frac{1}{\mathcal{Z}[0]}\int_{-\infty}^{\infty}dx\hspace{1mm}x\left(\sum_{k=0}^{\infty}\frac{(-V\lambda)^{k}}{k!4^{k}}x^{4k}\right)\left(\sum_{l=0}^{\infty}\frac{(-V)^{l}}{l!}(\sqrt{\lambda})^{l}(\sqrt{E_{0}})^{l}x^{3l}\right)\exp\left(-V\frac{3}{2}E_{0}x^{2}\right)\notag\\
=&\frac{1}{\mathcal{Z}[0]}\int_{-\infty}^{\infty}\left(\frac{-V\sqrt{\lambda}\sqrt{E_{0}}}{1!}\right)x^{4}\exp\left(-V\frac{3}{2}E_{0}x^{2}\right)dx+\frac{V^{2}\lambda\sqrt{\lambda}\sqrt{E_{0}}}{4\mathcal{Z}[0]}\int_{-\infty}^{\infty}x^{8}\exp\left(-V\frac{3}{2}E_{0}x^{2}\right)dx\notag\\
&+\frac{1}{\mathcal{Z}[0]}\int_{-\infty}^{\infty}\frac{(-V\sqrt{\lambda}\sqrt{E_{0}})^{3}}{3!}x^{10}\exp\left(-V\frac{3}{2}E_{0}x^{2}\right)dx+\mathcal{O}(\lambda^{2}\sqrt{\lambda})\label{B}.
\end{align}
(\ref{B}) is calculated directly as follows:
\begin{align}
G_{|1|}=&(\ref{B})\notag\\
=&\left\{-\frac{\sqrt{E_{0}}\sqrt{\lambda}}{3VE_{0}^{2}}-\frac{\sqrt{E_{0}}\lambda\sqrt{\lambda}}{36V^{2}E_{0}^{4}}+\frac{5\sqrt{E_{0}}\lambda\sqrt{\lambda}}{54V^{2}E_{0}^{4}}\right\}+\frac{35\sqrt{E_{0}}\lambda\sqrt{\lambda}}{108V^{2}E_{0}^{4}}-\frac{35\sqrt{E_{0}}\lambda\sqrt{\lambda}}{54V^{2}E_{0}^{4}}+\mathcal{O}(\lambda^{2}\sqrt{\lambda})\notag\\
=&-\frac{\sqrt{\lambda}}{3VE_{0}\sqrt{E_{0}}}-\frac{7\lambda\sqrt{\lambda}}{27V^{2}E_{0}^{\frac{7}{2}}}+\mathcal{O}(\lambda^{2}\sqrt{\lambda}).\label{X}
\end{align}
Similar to Section \ref{sec5} this result is verified from the exact solution of $G_{|1|}$ for $N=1$ that is given in \ref{A.2}.

\subsection{Approximation of $1$-Point Function $G_{|1|}$ by Saddle Point Method ($N=1$)}\label{A.2}
We consider the $1$-point function $G_{|1|}$ in the case of (\ref{h}) in $N=1$.
\begin{align}
G_{|1|}=&-\frac{\sqrt{E_{0}}}{\sqrt{\lambda}}+\frac{1}{V}\frac{\partial_{1}P_{1}(z_{1})}{P_{1}(z_{1})},
\end{align}
where $\displaystyle z_{1}=\frac{\sqrt{2}V^{\frac{3}{4}}E_{0}^{\frac{3}{2}}}{\lambda^{\frac{3}{4}}}$, and $\displaystyle z_{2}=\frac{\sqrt{2}V^{\frac{3}{4}}E_{1}^{\frac{3}{2}}}{\lambda^{\frac{3}{4}}}$. Using (\ref{Q}) and (\ref{R}), we approximate $1$-Point Function $G_{|1|}$ ($N=1$) as
\begin{align}
G_{|1|}=&-\frac{\sqrt{E_{0}}}{\sqrt{\lambda}}+\frac{\sqrt{E_{0}}}{\sqrt{\lambda}}-\frac{\sqrt{\lambda}}{12VE_{0}\sqrt{E_{0}}}-\frac{\sqrt{\lambda}}{3VE_{0}\sqrt{E_{0}}}+\frac{5\sqrt{\lambda}}{18VE_{0}\sqrt{E_{0}}}+\frac{35\lambda\sqrt{\lambda}}{54\times 2^{4}\times E_{0}^{\frac{7}{2}}V^{2}}\notag\\
&+\frac{35\lambda\sqrt{\lambda}}{27\times 2^{2}\times E_{0}^{\frac{7}{2}}V^{2}}-\frac{35\lambda\sqrt{\lambda}}{V^{2}\times 2^{2}\times 18\times E_{0}^{\frac{7}{2}}}-\frac{35\lambda\sqrt{\lambda}}{V^{2}\times 54\times E_{0}^{\frac{7}{2}}}+\frac{11\times 35\lambda\sqrt{\lambda}}{V^{2}\times 2^{2}\times E_{0}^{\frac{7}{2}}\times 2\times 3^4}\notag\\
&+\frac{\sqrt{E_{0}}}{\sqrt{\lambda}}\left\{\frac{\lambda}{12E_{0}^{2}V}-\frac{5\lambda}{18E_{0}^{2}V}-\frac{35\lambda^{2}}{V^{2}\times 216\times 4E_{0}^{4}}+\frac{35\lambda^{2}}{72V^{2}E_{0}^{4}}-\frac{11\times 35\lambda^{2}}{V^{2}\times 8\times 3^{4}E_{0}^{4}}\right\}\notag\\
&-\frac{\sqrt{\lambda}}{12VE_{0}\sqrt{E_{0}}}\left\{\frac{\lambda}{12E_{0}^{2}V}-\frac{5\lambda}{18E_{0}^{2}V}\right\}-\frac{\sqrt{\lambda}}{3VE_{0}\sqrt{E_{0}}}\left\{\frac{\lambda}{12E_{0}^{2}V}-\frac{5\lambda}{18E_{0}^{2}V}\right\}\notag\\
&+\frac{5\sqrt{\lambda}}{18VE_{0}\sqrt{E_{0}}}\left\{\frac{\lambda}{12E_{0}^{2}V}-\frac{5\lambda}{18VE_{0}^{2}}\right\}+\frac{\sqrt{E_{0}}}{\sqrt{\lambda}}\left\{\frac{\lambda^{2}}{144E_{0}^{4}V^{2}}+\frac{25\lambda^{2}}{18^{2}E_{0}^{4}V^{2}}-\frac{10\lambda^{2}}{12\times 18E_{0}^{4}V^{2}}\right\}+\mathcal{O}(\lambda^{2}\sqrt{\lambda})\notag\\
=&-\frac{\sqrt{\lambda}}{3VE_{0}\sqrt{E_{0}}}-\frac{7\lambda\sqrt{\lambda}}{27V^{2}E_{0}^{\frac{7}{2}}}+\mathcal{O}(\lambda^{2}\sqrt{\lambda})\label{C}.
\end{align}
This is consistent with the calculation of 1-point function $G_{|1|}$ ($N=1$) using perturbative expansion i.e. $(\ref{X})=(\ref{C})$.\\



\begin{thebibliography}{999}

\bibitem{Adler}
M.~Adler and P.~van Moerbeke, 
``A matrix integral solution to two-dimensional $W_{p}$-gravity,"
Commun. Math. Phys. \textbf{147}, $25$-$56$ (1992) doi:10.1007/BF02099527

\bibitem{Belliard}
R.~Belliard, S.~Charbonnier, B.~Eynard and E.~Garcia-Failde,
``Topological recursion for generalised Kontsevich graphs and $r$-spin intersection numbers,"
https://doi.org/10.48550/arXiv.2105.08035, arXiv.2105.08035, 2021.


\bibitem{Bhattacharya:2022wjl}
S.~Bhattacharya and N.~Joshi,
``Non-perturbative analysis for a massless minimal quantum scalar with $V(\phi)=\lambda \phi^4/4!+\beta \phi^3/3!$ in the inflationary de Sitter spacetime,''
[arXiv:2211.12027 [hep-th]].

\bibitem{Bhattacharya:2022aqi}
S.~Bhattacharya,
``Massless minimal quantum scalar field with an asymmetric self interaction in de Sitter spacetime,''
JCAP \textbf{09}, 041 (2022)
doi:10.1088/1475-7516/2022/09/041
[arXiv:2202.01593 [hep-th]].

\bibitem{Branahl:2021slr}
J.~Branahl, H.~Grosse, A.~Hock and R.~Wulkenhaar,
``From scalar fields on quantum spaces to blobbed topological recursion,''
J. Phys. A \textbf{55}, no.42, 423001 (2022)
doi:10.1088/1751-8121/ac9260
[arXiv:2110.11789 [hep-th]].


\bibitem{Branahl:2020yru}
J.~Branahl, A.~Hock and R.~Wulkenhaar, Blobbed Topological Recursion of the Quartic Kontsevich Model I: Loop Equations and Conjectures. Commun. Math. Phys. (2022). 

https://doi.org/10.1007/s00220-022-04392-z.


\bibitem{Branahl:2021uea}
J.~Branahl and A.~Hock,
``Genus one free energy contribution to the quartic Kontsevich model,''
[arXiv:2111.05411 [math-ph]].


\bibitem{Delgadillo-Blando:2008cuz}
R.~Delgadillo-Blando, D.~O'Connor and B.~Ydri,
``Matrix Models, Gauge Theory and Emergent Geometry,''
JHEP \textbf{05}, 049 (2009)
doi:10.1088/1126-6708/2009/05/049
[arXiv:0806.0558 [hep-th]].


\bibitem{Eynard:2015aea}
B. Eynard, T. Kimura, S. Ribault, Random matrices, https://doi.org/10.48550/arXiv.1510.04430, arXiv.1510.04430,
2015.


\bibitem{DiFrancesco:1993cyw}
P.~Di Francesco, P.~H.~Ginsparg and J.~Zinn-Justin,
2-D Gravity and random matrices,
Phys. Rept. \textbf{254}, 1-133 (1995)
doi:10.1016/0370-1573(94)00084-G
[arXiv:hep-th/9306153 [hep-th]].

\bibitem{Fukuma:1990jw}
M.~Fukuma, H.~Kawai and R.~Nakayama,
Continuum Schwinger-dyson Equations and Universal Structures in Two-dimensional Quantum Gravity,
Int. J. Mod. Phys. A \textbf{6}, 1385-1406 (1991)
doi:10.1142/S0217751X91000733


\bibitem{Grosse:2005ig}
H.~Grosse and H.~Steinacker,
Renormalization of the noncommutative $\phi^{3}$ model through the Kontsevich model,
Nucl. Phys. B \textbf{746}, 202-226 (2006)
doi:10.1016/j.nuclphysb.2006.04.007
[arXiv:hep-th/0512203 [hep-th]].

\bibitem{Grosse:2006qv}
H.~Grosse and H.~Steinacker,
A Nontrivial solvable noncommutative $\phi^{3}$ model in 4 dimensions,
JHEP \textbf{08}, 008 (2006)
doi:10.1088/1126-6708/2006/08/008
[arXiv:hep-th/0603052 [hep-th]].

\bibitem{Grosse:2006tc}
H.~Grosse and H.~Steinacker,
Exact renormalization of a noncommutative $\phi^{3}$ model in 6 dimensions,
Adv. Theor. Math. Phys. \textbf{12}, no.3, 605-639 (2008)
doi:10.4310/ATMP.2008.v12.n3.a4
[arXiv:hep-th/0607235 [hep-th]].


\bibitem{Grosse:2012uv}
H.~Grosse and R.~Wulkenhaar,
Self-Dual Noncommutative $\phi^4$ -Theory in Four Dimensions is a Non-Perturbatively Solvable and Non-Trivial Quantum Field Theory,
Commun. Math. Phys. \textbf{329}, 1069-1130 (2014)
doi:10.1007/s00220-014-1906-3
[arXiv:1205.0465 [math-ph]].


\bibitem{Grosse:2016pob}
H.~Grosse, A.~Sako and R.~Wulkenhaar,
Exact solution of matricial $\Phi^3_2$ quantum field theory,
Nucl. Phys. B \textbf{925}, 319-347 (2017)
doi:10.1016/j.nuclphysb.2017.10.010
[arXiv:1610.00526 [math-ph]].

\bibitem{Grosse:2016qmk}
H.~Grosse, A.~Sako and R.~Wulkenhaar,
The $\Phi^3_4$ and $\Phi^3_6$ matricial QFT models have reflection positive two-point function,
Nucl. Phys. B \textbf{926}, 20-48 (2018)
doi:10.1016/j.nuclphysb.2017.10.022
[arXiv:1612.07584 [math-ph]].

\bibitem{Grosse:2019jnv}
H.~Grosse, A.~Hock and R.~Wulkenhaar,
Solution of all quartic matrix models,
[arXiv:1906.04600 [math-ph]].

\bibitem{Hock:2020rje}
A.~Hock, Matrix field theory, Ph.D. Thesis, WWU M\"{u}nster, 2020, arXiv:2005.07525.

\bibitem{Hock:2019kgb}
A.~Hock, H.~Grosse and R.~Wulkenhaar,
``A Laplacian to Compute Intersection Numbers on $\overline{{{\mathcal {M}}}}_{g,n}$ and Correlation Functions in NCQFT,''
Commun. Math. Phys. \textbf{399}, no.1, 481-517 (2023)
doi:10.1007/s00220-022-04557-w
[arXiv:1903.12526 [math-ph]].

\bibitem{Hock:2021tbl}
A.~Hock and R.~Wulkenhaar,
Blobbed topological recursion of the quartic Kontsevich model II: Genus=0,
[arXiv:2103.13271 [math-ph]].

\bibitem{Hock:2023nki}
A.~Hock and R.~Wulkenhaar,
``Blobbed topological recursion from extended loop equations,''
[arXiv:2301.04068 [math-ph]].

\bibitem{Itzykson:1979fi}
C.~Itzykson and J.~B.~Zuber,
The Planar Approximation. 2.,
J. Math. Phys. \textbf{21}, 411 (1980)
doi:10.1063/1.524438

\bibitem{Itzykson:1992ya}
C.~Itzykson and J.~B.~Zuber,
``Combinatorics of the Modular Group II The Kontsevich integrals,''
Int. J. Mod. Phys. A \textbf{7}, 5661-5705 (1992)
doi:10.1142/S0217751X92002581
[arXiv:hep-th/9201001 [hep-th]].

\bibitem{Kanomata:2022pdo}
N.~Kanomata and A.~Sako,
Exact solution of the $\Phi_{2}^{3}$ finite matrix model,
Nucl. Phys. B \textbf{982}, 115892 (2022)
doi:10.1016/j.nuclphysb.2022.115892
[arXiv:2205.15798 [hep-th]].

\bibitem{Kontsevich:1992ti}
M.~Kontsevich,
Intersection theory on the moduli space of curves and the matrix Airy function,
Commun. Math. Phys. \textbf{147}, 1-23 (1992)
doi:10.1007/BF02099526

\bibitem{Pedro}
José L. López, Pedro J. Pagola,
Convergent and asymptotic expansions of the Pearcey integral,
Journal of Mathematical Analysis and Applications,
Volume 430, Issue 1,
2015,
Pages 181-192,
ISSN 0022-247X,
https://doi.org/10.1016/j.jmaa.2015.04.078.

\bibitem{Prekrat:2022sir}
D.~Prekrat, D.~Rankovi\'c, N.~K.~Todorovi\'c-Vasovi\'c, S.~Kov\'a\v{c}ik and J.~Tekel,
``Approximate treatment of noncommutative curvature in quartic matrix model,''
JHEP \textbf{01}, 109 (2023)
doi:10.1007/JHEP01(2023)109
[arXiv:2209.00592 [hep-th]].


\bibitem{T.Tao}
  T.~Tao,\\
http://terrytao.wordpress.com/2013/02/08/the-harish-chandra-itzykson-zuber-integral-formula/.


\bibitem{Witten}
E.~Witten,
Algebraic geometry associated with matrix models of two-dimensional gravity, Topological methods in Modern Mathematics, 1993.

\bibitem{Witten:1990hr}
E.~Witten,
Two-dimensional gravity and intersection theory on moduli space,
Surveys Diff. Geom. \textbf{1}, 243-310 (1991)
doi:10.4310/SDG.1990.v1.n1.a5

\bibitem{Ydri:2021cam}
B.~Ydri, R.~Khaled and C.~Soudani,
``Quantized noncommutative geometry from multitrace matrix models,''
Int. J. Mod. Phys. A \textbf{37}, no.10, 2250052 (2022)
doi:10.1142/S0217751X2250052X
[arXiv:2110.06677 [hep-th]].

\bibitem{Zhou:2013kka}
J.~Zhou,
Explicit Formula for Witten-Kontsevich Tau-Function,
[arXiv:1306.5429 [math.AG]].

\bibitem{Zinn-Justin:2002rai}
P.~Zinn-Justin and J.~B.~Zuber,
On some integrals over the U(N) unitary group and their large N limit,
J. Phys. A \textbf{36}, 3173-3194 (2003)
doi:10.1088/0305-4470/36/12/318
[arXiv:math-ph/0209019 [math-ph]].


\end{thebibliography}
\end{document}